\documentclass[12pt]{article}
\newcommand{\FlipTR}{UCR-TR-2021-FLIP-MSE-6}

\usepackage{amsmath}
\usepackage{amssymb}
\usepackage{amsfonts}
\usepackage{graphicx}
\usepackage[utf8]{inputenc}     
\usepackage{aas_macros}				  
\usepackage{bm}                 
\usepackage{amsthm}

\usepackage{slashed}				
\usepackage{mathrsfs}				
\usepackage{bbm}					
\usepackage{cancel}					

\usepackage[dvipsnames]{xcolor}
\usepackage[hang,flushmargin]{footmisc} 

\usepackage{fancyhdr}		
\usepackage{lipsum}			
\usepackage{framed}			
\usepackage{subcaption}	
\usepackage{cite}			  
\usepackage{xspace}			

\usepackage{booktabs}		

\usepackage[font=small]{caption} 	
\usepackage{float}         			  

\usepackage[margin=2.5cm]{geometry}   
\graphicspath{{figures/}}			
\numberwithin{equation}{section}    

\usepackage{multicol}
\usepackage{etoolbox}
\usepackage{relsize}
\patchcmd{\thebibliography}
  {\list}
  {\begin{multicols}{2}\smaller\list}
  {}
  {}
\appto{\endthebibliography}{\end{multicols}}

\let\oldenumerate\enumerate
\renewcommand{\enumerate}{
  \oldenumerate
  \setlength{\itemsep}{1pt}
  \setlength{\parskip}{0pt}
  \setlength{\parsep}{0pt}
}

\let\olditemize\itemize
\renewcommand{\itemize}{
  \olditemize
  \setlength{\itemsep}{1pt}
  \setlength{\parskip}{0pt}
  \setlength{\parsep}{0pt}
}

\usepackage{lineno}

\usepackage{scalefnt} 
\newcommand\acro[2][0.9]{{\scalefont{#1}#2}} 

\renewcommand{\tilde}{\widetilde}   
\renewcommand{\text}{\textnormal}	
\renewcommand{\vec}[1]{\mathbf{#1}} 

\newcommand{\email}[1]{\href{mailto:#1}{#1}}
\newenvironment{institutions}[1][2em]{\begin{list}{}{\setlength\leftmargin{#1}\setlength\rightmargin{#1}}\item[]}{\end{list}}

\providecommand{\DIFadd}[1]{{\protect\color{blue}#1}} 
\providecommand{\DIFdel}[1]{{\protect\color{red}\protect\scriptsize{#1}}}

\usepackage[
	colorlinks=true,
	citecolor=green!50!black,
	linkcolor=NavyBlue!75!black,
	urlcolor=green!50!black,
	hypertexnames=false]{hyperref}

\fancypagestyle{firststyle}{
	\rhead{\footnotesize%
	\texttt{\FlipTR}%
	}}

  \usepackage{tocloft}

\usepackage{etoc}

\providecommand{\DIFadd}[1]{{\protect\color{blue}%
\ifmmode\hbox{\ul{{\mbox{$#1$}}}}\else\ul{#1}\fi}} 

\providecommand{\DIFdel}[1]{{\protect\color{red}\protect\scriptsize{#1}
\ifmmode\hbox{\ul{{\mbox{$#1$}}}}\else\ul{#1}\fi
}}

\begin{document}

\thispagestyle{firststyle}  

\begin{center}
    \vspace*{2cm}
    {\huge \textbf{Continuum-Mediated\\Self-Interacting Dark Matter} \par} 
    \vspace{.7cm}

   { \bf
    Ian Chaffey$^{a}$,
    Sylvain~Fichet$^{b}$,
    and
    Philip Tanedo$^{a}$
    }
   \\
   \vspace{-.2em}
   { \tt \footnotesize
    \email{ichaf001@ucr.edu},
    \email{sfichet@caltech.edu},
  \email{flip.tanedo@ucr.edu}
   }

   \begin{institutions}[1.7cm]
   \footnotesize
   $^{a}$
   {
      \textit{Department of Physics \& Astronomy,
      University of California, Riverside,}
      \acro{CA} \textit{92521}
      }
  \\
  \vspace*{0.05cm}
  $^{b}$
  {
        {\scalefont{.9}ICTP SAIFR}
        \textit{\&}
        {\scalefont{.9}IFT-UNESP}
        \textit{R.~Dr.~Bento Teobaldo Ferraz 271, S\~ao Paulo, Brazil}
        }
   \end{institutions}

\end{center}


\vspace*{1cm}

\begin{abstract}
\noindent 

Dark matter may self-interact through a continuum of low-mass states. 
This happens if dark matter couples to a strongly-coupled nearly-conformal hidden sector.
This type of theory is holographically described by brane-localized dark matter interacting with bulk fields in a slice of \acro{5D} anti-de~Sitter space.
The long-range potential in this scenario depends on a non-integer power of the spatial separation, in contrast to the Yukawa potential generated by the exchange of a single \acro{4D} mediator. 
The resulting self-interaction cross section scales like a non-integer power of velocity. 
We identify the Born, classical and resonant regimes
 and investigate them using state-of-the-art numerical methods. We  demonstrate the viability of our continuum-mediated framework to address the astrophysical small-scale structure anomalies.
Investigating the continuum-mediated Sommerfeld enhancement, we demonstrate that a pattern of resonances can occur depending on the non-integer power.
We conclude that continuum mediators introduce novel power-law scalings which open  new possibilities for dark matter self-interaction phenomenology.

\end{abstract}

\newpage

\setcounter{tocdepth}{2} 
\tableofcontents
\normalsize
\clearpage


\section{Introduction}
\label{se:intro}

A dark sector is a set of fields that include dark matter and low-mass particles that mediate interactions of the dark matter~\cite{Pospelov:2008zw,Pospelov:2008jd, Pospelov:2007mp, Essig:2013lka, Alexander:2016aln, Battaglieri:2017aum}. If these mediators interact with the Standard Model, their signatures may appear in a suite of laboratory based experiments. 
Even if these Standard Model interactions are negligible, the mediators induce long-range potentials between dark matter particles that may be tested astronomically~\cite{Carlson:1992fn, Spergel:1999mh}. This \emph{self-interacting dark matter} framework has been spurred by the observation that it may address potential small-scale structure tensions between simulations of cold dark matter and astronomical observations~\cite{Tulin:2013teo, Tulin:2017ara}.

A single mediator typically produces a Yukawa potential between dark matter particles, $V(r)\sim -e^{-m_\varphi r}/r$, where $m_\varphi$ is the mass of the mediator. 
This long-range behavior can be dramatically altered when the single-mediator exchange picture breaks down, for example when the mediator is represented by a continuum of states. Models of continuum dark sectors have existed for at least a decade in the form of conformal hidden sectors~\cite{Gherghetta:2010cq, vonHarling:2012sz} and closely related work on unparticle hidden sectors~\cite{Strassler:2008bv,Chen:2009ch,Friedland:2009iy,Friedland:2009zg}.
The proposal that such models may lead to novel self-interactions was first identified in Ref.~\cite{Brax:2019koq} for a spin-0 mediator modeled in the holographic description of a warped extra dimension.\,\footnote{In this work we use \emph{continuum} to refer to the discrete set of Kaluza--Klein modes. This could be also referred to as a `discretuum,' as opposed to the `continuous continuum' regime in which the \acro{KK} modes merge \cite{Costantino:2020msc} Because a potential is generated by $t$-channel diagrams, the mediator field carries spacelike four-momentum. This makes it mostly insensitive to whether the spectral distribution is continuous or discrete and no distinction between these  scenarios is necessary.} This paper describes continuum-mediated self-interacting dark matter phenomenology in that benchmark theory. 
The dynamics of  the model generate a long-range potential on the \acro{UV} brane that scales as a non-integer power of separation, 
\begin{align}
   V(r) &\sim \frac{1}{r} \left(\frac{1}{\Lambda r}\right)^\text{non-integer} \ , \label{eq:Vintro}
 \end{align}
where $\Lambda$ is a cutoff scale.

The long-range forces between dark matter particles allow energy exchange in dark matter halos and create a cored density profile compared to standard cold dark matter $N$-body simulations. Observations of small-scale structure anomalies in dwarf spheroidal galaxies are indicative of cored halo profiles and are thus a tantalizing possible signature for dark matter dynamics~\cite{Tulin:2013teo}.
Alternative proposals to address these anomalies include baryonic feedback on the dark matter halo. Future generations of $N$-body simulations may be able to ultimately distinguish between the two scenarios, and it is plausible that nature may even invoke a combination of the two mechanisms. We refer to Ref.~\cite{Bullock:2017xww} for a recent review of the status of these anomalies. A key result of our study is that  continuum-mediated interactions leads to a non-integer velocity dependence on the dark matter self-scattering cross section, a quantity that relates the fundamental particle physics parameters of the dark sector to astronomical observations.
Schematically,
\begin{align}
   \sigma(v) &\sim  v^\text{non-integer} \,. \label{eq:sigmaintro}
 \end{align}

We proceed as follows.
In Section~\ref{se:picture} we motivate a class of conformal models that generate non-integer potentials of the form \eqref{eq:Vintro} and specifically highlight a \acro{5D} dual picture with a mass gap. 
We give a precise definition of the gapped, continuum-mediated self-interacting dark matter model in Section~\ref{se:AdS_SIDM}. 
We discuss experimental constraints beyond self-interactions in Section~\ref{se:SM_constraints}; these constraints can be avoided for the types of parameters needed to address small scale structure puzzles in astronomy.
The long-range potential is derived in Section\,\ref{se:potential} using spectral techniques. We present closed form expressions using  asymptotic limits that we validate numerically. 
In Section~\ref{se:SIDM} we evaluate the figure of merit for astronomical applications, the self scattering transfer cross section. In the so-called Born and classical regimes of dark matter coupling and velocity, we demonstrate novel scaling in the dark matter velocity compared to non-continuum self-interacting models. We confirm the presence of a resonant regime and analyze all regimes numerically. 
Continuum-mediated self-interactions can explain small-scale structure observations even when the slope of its potential differs significantly from a standard Yukawa potential.
In Section~\ref{se:Sommerfeld} we show that Sommerfeld enhancement produces a pattern of resonances that depend on the potential slope and mass gap.
We conclude in Section~\ref{se:conclusion}.
The Appendices include a streamlined review of  \acro{AdS}/\acro{CFT} with a \acro{UV} brane (Appendix~\ref{se:AdS/CFT}), a calculation of the approximate transfer cross section in the non-perturbative classical regime (Appendix~\ref{app:classical}), a proof that there is no Sommerfeld enhancement for a $1/r^2$ potential (Appendix\,\ref{se:r2_sommerfeld}), and a review of the numerical method used to solve for the transfer cross section (Appendix~\ref{app:SIDM}).

\section{Preliminary Observations} 
\label{se:picture}

The simplest assumption for dark matter self-interactions is that dark matter currents, $J_\text{DM}$, interact by exchanging spin-0 or spin-1 mediators at tree-level. 
In momentum space, the matrix elements take schematically the form 
\begin{align}
  \vcenter{
    \hbox{\includegraphics[width=.1\textwidth]{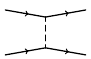}}
    }
    \quad 
    = 
    \quad
  J_\text{DM}(q) \, \frac{1}{q^2-m^2} \, J_\text{DM}(-q) 
  \,. \label{eq:yuk}
\end{align}
The corresponding potential between dark matter currents in position space is Yukawa-like, $V(r)\sim e^{-mr}/{r}$, or Coulomb-like if $m=0$.    
The mediator mass, $m$, cuts off the potential in the infrared and is important for realizing required low-velocity scaling of the dark matter self-scattering cross section for small scale structure anomalies.

The exchange of a single, non-derivatively coupled, weakly-interacting field  in \eqref{eq:yuk} is the simplest dark matter self-interaction.  The resulting $r^{-1}$ potential is the longest ranged potential allowed by the lower bound on the dimension of the exchanged operator set by unitarity, $\Delta \geq 1$. 
However, it is also plausible that the leading self-interaction is shorter range than $1/r$ and thus there are a variety of possibilities that have yet to be thoroughly investigated.
An extreme example is a zero-range interaction, $J_\text{DM}(q)J_\text{DM}(-q)$, which give contact-interactions in position space, $V(r)\sim \delta^{(3)}(r)$. This possibility is too extreme: the contact interactions produce velocity-independent cross sections that are tightly constrained by the upper bound on dark matter self-scattering at high velocities from observations of galaxy cluster collisions like the Bullet Cluster.\footnote{
Other short range possibilities  include tree-level exchange of a pseudoscalar (see {e.g.}~\cite{Fadeev:2018rfl}) and  loop-level mediated processes \cite{Fichet:2017bng, Costantino:2019ixl}, which induce  potentials going as $\propto 1/r^n$ with $n$ integer and $\geq 3$.  }

In this work we explore intermediate possibilities where the self-interaction potential has finite range that is shorter than the Yukawa/Coulomb limit. 
The simplest possibility amounts to a matrix element
\begin{align}
  \vcenter{
    \hbox{\includegraphics[width=.1\textwidth]{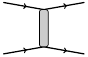}}
    }
    \, 
  &\quad=\quad
  J_\text{DM}(q)
  \,
  \frac{1}{\left(\sqrt{-q^2}\right)^{4-2\Delta}}
  \, 
  J_\text{DM}(-q) 
  \,. \label{eq:CFT}
\end{align}
The parameter $\Delta$ satisfies $\Delta\geq 1$, where $\Delta=1$ recovers the Coulomb case. 
The position-space potential scales as $V(r)\sim r^{-2\Delta+1}$ and becomes steeper near the origin for $\Delta > 1$ such that the interaction has indeed shorter range than the Coulomb case.
The interaction \eqref{eq:CFT} is understood to come from the exchange of a operator of dimension $\Delta$.  Highly non-integer dimensions do not occur in weakly-coupled theories since quantum corrections to the classical scaling dimension are perturbative.
However, if the dark sector has strongly-interacting dynamics, then it is likely that the operators have highly non-integer dimension. 
We focus on a nearly-conformal mediator sector described a conformal field theory (\acro{CFT}); this sector may be a gauge theory with large 't~Hooft coupling.
Currents of elementary dark matter, $J_\text{DM}$, interact with \acro{CFT} operators.
Even though the mediator sector is strongly-interacting, conformal symmetry constrains the \acro{CFT} correlation functions and provides a well-controlled framework for calculations.
The \acro{CFT} two-point function has a continuous spectral representation and so we refer to this scenario as \emph{continuum-mediated self-interacting dark matter}. An analogous description of dark matter--nucleon scattering is used in Ref.~\cite{Katz:2015zba}.

A purely conformal hidden sector does not have a mass gap. This prevents an infrared cutoff that is usually set by the mediator mass. In order to restore the desired exponential damping at long distances, we assume an infrared (\acro{IR}) mass gap in a slightly more evolved model that is most simply described holographically in five dimensional anti-de~Sitter (\acro{AdS}) space. 
In this scenario, a \acro{5D} field $\Phi$ propagates in the bulk and interacts with the brane-localized dark matter currents, $J_\text{DM}$. 

\begin{figure}[H]
\begin{center}
\includegraphics[width=.5\textwidth]{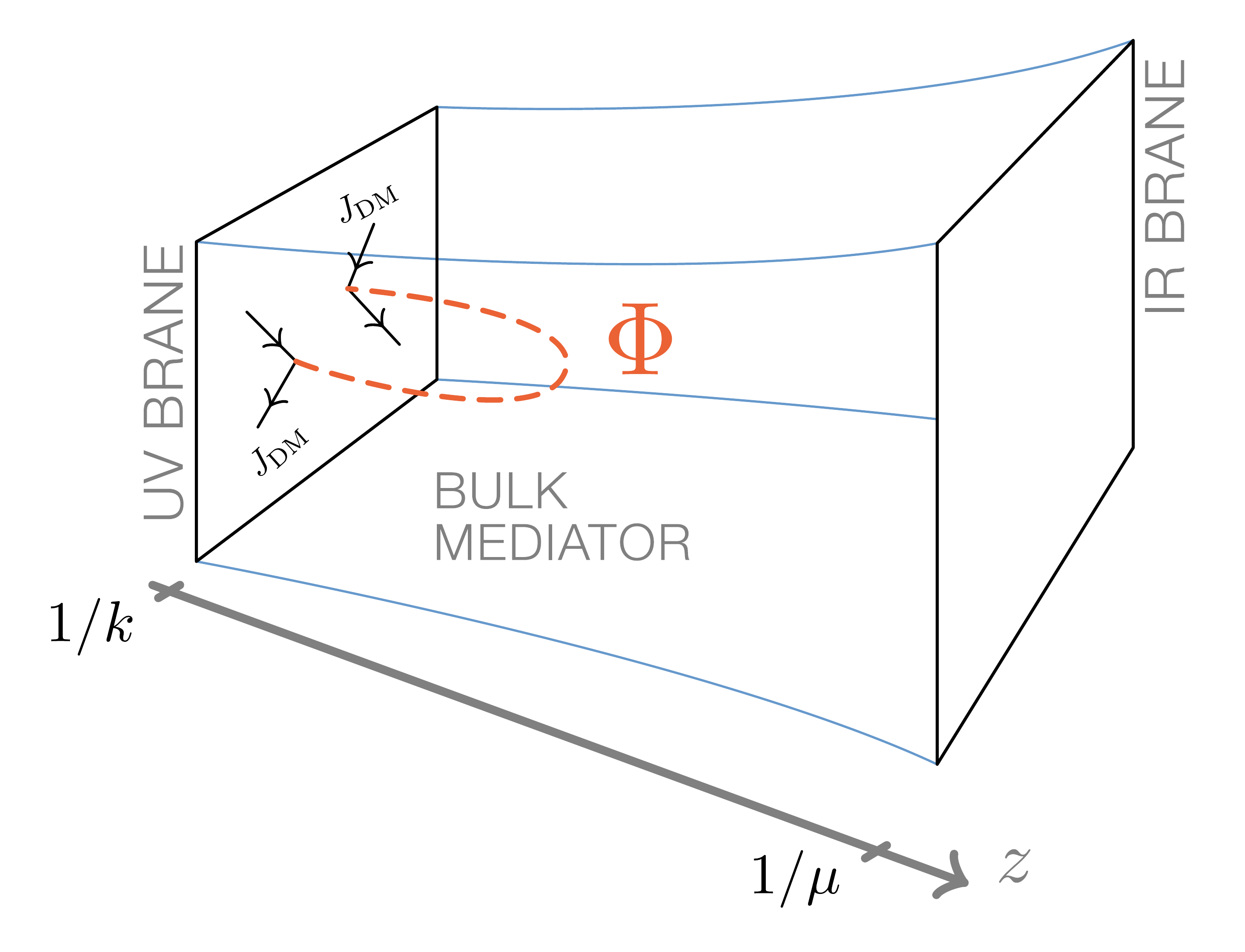}
\end{center}
\caption{Schematic description of the continuum-mediated self-interacting dark matter scenario.}
\label{fig:schematic}
\end{figure}

The \acro{AdS} dual of the ungapped amplitude  \eqref{eq:CFT} is  schematically: 
\begin{align}
  \vcenter{
    \hbox{\includegraphics[width=.1\textwidth]{{Feyn_conf}}}
    }
    \, 
  &\quad=\quad
  \vcenter{
    \hbox{\includegraphics[width=.1\textwidth]{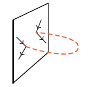}} 
    }
  \, \label{eq:AdS:CFT:bulk}
\end{align} 
see Appendix~\ref{se:AdS/CFT} for relevant details from the \acro{AdS/CFT} correspondence. 
In the \acro{5D} description of continuum-mediated self-interacting dark matter, dark matter itself is a \acro{4D} degree of freedom localized on the \acro{UV} brane near the \acro{AdS} boundary. This is identified with an elementary degree of freedom that probes the \acro{CFT} sector. The mediator continuum is a bulk field coupled to the fields on the boundary.  The mass gap in the \acro{AdS} description is encoded by an infrared (\acro{IR}) brane localized further away from the \acro{AdS} boundary:
\begin{align}
  \vcenter{
    \hbox{\includegraphics[width=.1\textwidth]{{Feyn_conf}}}
    }
    \, 
  &\quad=\quad
  \vcenter{
    \hbox{\includegraphics[width=.15\textwidth]{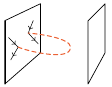}} 
    }
  \,. \label{eq:AdS:CFT:bulk:IRB}
\end{align} 
In the \acro{5D} description, the mass gap follows from the bulk field having two boundary conditions at finite distance. The exact \acro{CFT} limit~\eqref{eq:AdS:CFT:bulk} is recovered when the \acro{IR} brane is decoupled by sending it to spatial infinity. The \acro{5D} model is shown in Figure~\ref{fig:schematic} and is described precisely in the following section.

\section{Continuum-Mediated Self-Interactions from AdS} 
\label{se:AdS_SIDM}

We detail a model in \acro{5D} \acro{AdS} space that realizes the  continuum-mediated self-interacting dark matter scenario; the choices of parameters are discussed in the following section. 
The model is based on the warped dark sector framework~\cite{Brax:2019koq}, which is itself closely related to the Randall--Sundrum~2 model of a warped extra dimension~\cite{Randall:1999vf}.

\subsection{Geometry and Action} \label{se:action}

\paragraph{Geometry.} The metric for the \acro{AdS} spacetime in conformal coordinates is
\begin{align} 
  \label{eq:metric}
  ds^2 &= \left( \frac{1}{ k z} \right)^2 \left(\eta_{\mu\nu}dx^\mu dx^\nu - dz^2\right) 
\end{align}
where $k$ is \acro{AdS} curvature. 
We restrict to a slice of this \acro{AdS} space and place \acro{UV} and \acro{IR} branes at the endpoints,
\begin{align}
  z_\text{UV}
  &\leq 
  z 
  \leq z_\text{IR} 
  &
  z_\text{UV}&=\frac{1}{k}
  &
  z_\text{IR}&=\frac{1}{\mu}
  \label{eq:z:Defs}
  \ .
\end{align}
In our model, the scale $\mu$ characterizes the mass gap of the mediator sector; we take $\mu \ll k$.

We assume that some stabilization mechanism prevents the two branes from falling into one another; though we may remain agnostic about the specific choice as the details are not crucial to our study. For concreteness, one may assume the Goldberger--Wise mechanism~\cite{Goldberger:1999uk}.  We ignore gravitational backreaction effects near the \acro{IR} brane and approximate the metric to be exactly \acro{AdS} over the entire space.  

The action for the theory includes bulk and brane-localized quadratic terms for the \acro{5D} real scalar mediator $\Phi$, \acro{UV} brane-localized quadratic terms for the dark matter $\chi$, and interactions between dark matter and mediator: 
\begin{align}
  S &= 
  \int_{z_\text{UV}}^{z_\text{IR}} \int d^4x \;
  \sqrt{g}\mathcal L_\Phi 
  + \sqrt{\bar{g}}
    \left(
      \mathcal L_\chi
      + \mathcal L_\text{int}
      + \mathcal L_\Phi^\text{UV}
    \right)
    \delta(z-z_\text{UV}) 
  + \sqrt{\bar{g}}
      \mathcal L_\Phi^\text{IR}
    \delta(z-z_\text{IR}) 
    \ ,
\end{align}
where $\bar g$ is  the induced metric on the brane, with $\sqrt{\bar{g}} = (kz)^{-4}$. 
Additional  terms that do not play a role in the self-interaction phenomenology are the \acro{5D} Einstein--Hilbert term, the \acro{4D} Standard Model action localized to the \acro{UV} brane, and possible Standard Model interactions with the mediator.
The dark matter Lagrangian terms encode a \acro{4D} mass $m_\chi$ and Yukawa coupling to the bulk mediator:
\begin{align}
  \mathcal L_\chi &=
  \bar\chi \gamma^\mu \partial_\mu \chi - m_\chi \bar\chi\chi
  &
  \mathcal L_\text{int} &=
  \frac{\lambda}{\sqrt{k}}
  \Phi \bar\chi\chi \ .
  \label{eq:X:lagrangian}
\end{align}
Writing \acro{5D} Lorentz indices $M$, the bulk mediator Lagrangian is
\begin{align}
  \mathcal L_\Phi &=
  \frac{1}{2}
  \left[
    (\partial_M\Phi)(\partial^M\Phi)
    - M_\Phi^2 \Phi^2
  \right] \ ,
\end{align}
where the bulk mass $M_\Phi$ is tied to the dimension $\Delta$ of the operator exchanged between dark matter particles in the \acro{CFT} picture.
The brane-localized Lagrangian terms for the bulk scalar encode mass and kinetic terms:
\begin{align}
  \mathcal L_{\Phi}^{\text{UV}} &=
  \frac{1}{2k}
  \Phi
    B_{\text{UV}}[\partial^2]
  \Phi
  &
  \mathcal L_{\Phi}^{\text{IR}} &=
  \frac{1}{2k}
  \Phi
    B_{\text{IR}}[\partial^2]
  \Phi 
  &
  {B}_{i}[\partial^2] = m^2_{i} +c_i \partial^2+\ldots
  \ .
  \label{eq:boundary:L}
\end{align}
The $B_i[\partial^2]$ are polynomials in the \acro{4D} Laplacian $\partial^2=\partial_\mu\partial^\mu$; the constant term is the brane-localized masses $m_i^2$. Higher order terms are typically small and irrelevant for our study.

We remark that the low-energy effective theory also contains a radion that is identified with the dilaton in the \acro{4D} theory. This mode is light, but localized on the \acro{IR} brane and hence has negligible contributions to the dark matter dynamics on the \acro{UV} brane. We thus do not include it in our analysis as it would produce only a minor shift in the long-range potential.

\subsection{Effective Field Theory Consistency} \label{se:EFT}

\acro{5D} interacting theories are non-renormalizable and are understood to be low-energy effective field theory (\acro{EFT}) valid up to a cutoff, $\Lambda$.
The cutoff is tied to the strongest \acro{5D} interaction---either gravity or another interaction in the theory. 
\acro{5D} na\"ive dimensional analysis (\acro{NDA})~\cite{Manohar:1983md,Georgi:1986kr,Georgi:1992dw,Luty:1997fk,Jenkins:2013sda}, in turn, relates the cutoff to the \acro{AdS} curvature~\cite{Costantino:2020msc},
\begin{align}
  \Lambda \gtrsim \frac{\ell_5}{\ell_4} k \sim \pi k  \ ,
  \label{eq:lambdak}
\end{align}
where the \acro{4D} and \acro{5D} loop factors are $\ell_4 =16\pi^2$ and $\ell_5=24\pi^3$, respectively.

In our dark sector model, the cutoff sets the dark matter--mediator Yukawa coupling $\lambda$. Thus \acro{5D} \acro{NDA} bounds the Yukawa coupling by 
\begin{align}
\lambda
\lesssim \sqrt{\frac{\ell_5 k}{\Lambda}}  
\lesssim 4\pi
\ ,
\label{eq:lambdabound}
\end{align}
where we have used \eqref{eq:lambdak} in the second inequality.

While the \acro{5D} theory is valid below $\Lambda$, the \acro{AdS}/\acro{CFT} dictionary is valid only up to a cutoff scale on the order of $k < \Lambda$. 
From the \acro{4D} perspective, a \acro{CFT} coupled to gravity has a cutoff parametrically smaller than $M_\text{Pl}$ because of the large degrees of freedom of the \acro{CFT}. This cutoff turns out to be $k$, for example by using the species scale conjecture (see \textit{e.g.} \cite{Dvali:2008ec}).

Our \acro{5D} \acro{EFT} contains isolated degrees of freedom localized on a brane. In a realistic theory with gravity, localized \acro{4D} fields are special modes from \acro{5D} bulk fields and are necessarily accompanied by a spectrum of \acro{KK} modes~\cite{Fichet:2019owx}. We assume an appropriate limit where the observable effects of these modes are negligible.

\subsection{Model Parameters} \label{se:params}

For the purposes of studying novel, continuum-mediated dark matter self-interactions, we restrict the parameters presented in the \acro{5D} model in Section~\ref{se:action}. 
The \acro{AdS} curvature, $k$, corresponds to the cutoff of the theory, as described in Section~\ref{se:EFT}.
To ensure that the cutoff of the theory is beyond the experimental reach of the Large Hadron Collider to detect, e.g., Kaluza--Klein gravitons, we set $k$ to be
\begin{align}
  k = 10~\text{TeV} \ .
\end{align}
This sets the position of the \acro{UV} brane $z_\text{UV} = k^{-1}$ and the upper bound on all other dimensionful parameters in the theory. The \acro{AdS} curvature is much smaller than the Planck scale, in the spirit of `little Randall--Sundrum' models~\cite{Davoudiasl:2008hx}. 

The mediator mass, $M_\Phi$ is related to the dimension $\Delta$ of the continuum mediator operator and is conveniently described by the dimensionless parameter $\alpha$,
\begin{align}
  \alpha^2 \equiv 4 + \frac{M_\Phi^2}{k^2} &= \left(2 - \Delta\right)^2 
  \ .
  \label{eq:alpha}
\end{align}
The range of $\alpha$ corresponds to the $\Delta_-$ branch of \acro{AdS}/\acro{CFT} (see details in Appendix~\ref{se:AdS/CFT})  and can be established as follows. Unitarity of \acro{CFT} operators requires $\Delta \geq 1$, implying $\alpha\leq 1$. The Breitenlohner--Freedmann bound for the stability of \acro{AdS}  implies $\alpha^2 \geq 0$~\cite{Breitenlohner:1982bm,Breitenlohner:1982jf},
and we restrict to $\alpha \geq 0$ without loss of generality. We thus obtain $0\leq \alpha \leq 1$ .

Observe that the slope of the resulting long-range potential scales like $V(r)\sim r^{-1}$ for $\alpha =1$ and $V(r)\sim r^{-3}$ for $\alpha=0$. For potentials more singular than $r^{-2}$, solving for the phenomenology becomes computationally intractable and, furthermore, the theory is unlikely to produce the effects relevant for small scale structure anomalies.
We thus further restrict the range to $\alpha \geq 1/2$ to avoid the regime where the long-range potential is steeper than the centrifugal term.

Our theory includes brane-localized masses $m_i^2\Phi(x,z_i)^2$ and kinetic terms $c_i[\partial\Phi(x,z_i)]^2$ for the mediator. It is convenient to parameterize the former into dimensionless variables,
\begin{align}
&b_{\text{IR}} \equiv  \frac{m^2_{\text{IR}}}{k^2}+(2-\alpha) 
&
&b_{\text{UV}} \equiv  \frac{m^2_{\text{UV}}}{k^2}+(2-\alpha) 
\ . 
\label{eq:bUVIR}
\end{align}
The \acro{IR} parameters $b_\text{IR}$ and $c_\text{IR}$ generically have $\mathcal O(1)$ values so we set them all to one. These only have a mild impact on the self-interaction phenomenology. Conversely, we tune $b_\text{UV}=0$ as required to reproduce the \acro{CFT} behavior in \eqref{eq:CFT} since $b_\text{UV}$ corresponds to a double trace deformation in the conformal theory. The \acro{UV} brane kinetic coefficient $c_\text{UV}$ is assumed to be $\mathcal O(1)$, though it is only significant in the limiting case $\alpha=1$. 

With these benchmark values in place, the theory is described by the parameters in Table~\ref{tab:parameter:range}. The \acro{IR} scale $\mu$ defines the mass gap of the theory by setting the scale of the lightest Kaluza--Klein mode and its lower bound is set by dark radiation constraints, described in Section~\ref{se:SM_constraints}.

\begin{table}
  \renewcommand{\arraystretch}{1.3} 
  \centering
  \begin{tabular}{ @{} l @{\qquad} l @{\qquad} l @{} } \toprule 
    Parameter &  Range  &  What sets the range
    \\ \midrule
    Bulk mass
    &  $1/2 \leq \alpha \leq 1$ 
    & Calculability, unitarity 
    \\
    Mass gap
    & $\text{MeV} \lesssim \mu \ll k$ 
    & Early universe
    \\
    Dark matter mass
    & $\mu \lesssim m_\chi \lesssim k$
    & Nonlocal potential, \acro{EFT} validity
    \\
    Yukawa coupling
    & $ \lambda \leq 4\pi$
    & \acro{EFT} perturbativity
    \\ \bottomrule
  \end{tabular}
  \caption{
    Range of parameters in our model. The \acro{AdS} curvature is set to $k = 10~\text{TeV}$; larger values generically suppress self-interaction effects. The dimensionless brane-localized masses and kinetic terms defined in \eqref{eq:boundary:L} and \eqref{eq:bUVIR} are assumed to be $\mathcal{O}(1)$, with the exception of $b_\text{UV}$ which is tuned to zero to reproduce the long-range behavior, \eqref{eq:CFT}. The early universe bound on $\mu$ is described in Section~\ref{se:SM_constraints}.
    \label{tab:parameter:range}
  }
\end{table}

\subsection{Mediator Propagator and Spectrum} \label{se:prop}

It is convenient to work in position space for the $z$-direction but momentum space along \acro{4D} Minkowski slices. The mediator field is decomposed as
\begin{align}
  \Phi_p(z) &= \int d^4x \; e^{i p\cdot x}\;\Phi(x^\mu, z)
  &
  p\cdot x & = p_\mu x^\mu
  \ .
\end{align}
The norm $p=\sqrt{\eta_{\mu\nu} p^\mu p^\nu}$ is real for timelike $p^\mu$ and imaginary for spacelike $p^\mu$. In these coordinates, the free scalar propagator is the two-point Green's function, see e.g.~\cite{Fichet:2019owx},
\begin{align} 
	G_p(z,z')
  &=
 	i \frac{\pi k^3 (zz')^2}{2}
 	\frac{
 		\left[
      \tilde{Y}^\text{UV}_\alpha J_\alpha(pz_<)
      -
      \tilde{J}^\text{UV}_\alpha Y_\alpha(pz_<)
    \right]
    \left[
      \tilde{Y}^\text{IR}_\alpha J_\alpha(pz_>)
      -
      \tilde{J}^\text{IR}_\alpha Y_\alpha(pz_>)
    \right]
 	}{
 		\tilde{J}^\text{UV}_\alpha\tilde{Y}^\text{IR}_\alpha
    -
    \tilde{Y}^\text{UV}_\alpha\tilde{J}^\text{IR}_\alpha
 	}\ , 
 \label{eq:propa}
 \end{align} 
where
$z_{<,>}$ is the lesser/greater of the endpoints $z$ and $z'$. 
The quantities $\tilde{J}^\text{UV,IR}$ are
\begin{align}
  \tilde{J}^\text{UV}_\alpha &= 
  \frac{p}{k} J_{\alpha-1}\!\left(\frac{p}{k}\right) 
  + 
  B_\text{UV}(p^2)\,J_\alpha\!\left(\frac{p}{k}\right)
  &
  \tilde{J}^\text{IR}_\alpha 
  &= 
  \frac{p}{\mu} J_{\alpha-1}\!\left(\frac{p}{\mu}\right) 
  + 
  B_\text{IR}(p^2)\,J_\alpha\!\left(\frac{p}{\mu}\right)
  \ ,
\end{align}
with similar definitions for $\tilde{Y}^\text{UV,IR}$. The boundary functions $B_i(p^2)$ encode brane-localized operators and are defined in \eqref{eq:boundary:L}.  We refer to \eqref{eq:propa} as the canonical representation of the propagator.

The propagator has an infinite series of isolated poles {set by the zeros of $	\tilde{J}^\text{UV}_\alpha\tilde{Y}^\text{IR}_\alpha
    -    \tilde{Y}^\text{UV}_\alpha\tilde{J}^\text{IR}_\alpha$ and  referred to as} Kaluza--Klein (\acro{KK}) modes.  The free propagator can thus equivalently be written as a series
\begin{align}
  G_p(z,z')&=i \sum_n \frac{f_n(z)f_n(z')}{p^2-m_n^2+i\epsilon} \ ,
  \label{eq:propa_KK}
\end{align}
we refer to this particular momentum-space spectral representation as the \acro{KK} representation of the propagator. Depending on the context, either the canonical or \acro{KK} representation may be more convenient.
Assuming that the \acro{UV} brane mass parameter is zero, $b_\text{UV} = 0$, and that the other brane parameters have $\mathcal O(1)$ coefficients, then the \acro{KK} spectrum for $p\gg \mu$ is 
\begin{align}
  m_n
  &\approx\left(n-\frac{\alpha}{2}+\frac{1}{4}\right)\pi\mu
  &
  n &> 0
  \ ,
\end{align}
as can be seen from identifying the poles in the limiting form of the propagator in \eqref{eq:prophighenergy}.
The mass of the lightest mode $m_0$ depends on the brane-localized parameters and is  detailed in Section~\ref{se:asymptotics}.

\subsection{Qualitative Description of 4D Near-Conformal Theory}
\label{se:CFT_qualitative}

The \acro{AdS}/\acro{CFT} correspondence describes the equivalence between a quantum field theory on \acro{AdS}$_{d+1}$ space and a conformal gauge theory with large 't\,Hooft coupling and large-$N$ in flat $d$-dimensional space (for initial works see \cite{Maldacena:1997re,
Gubser:1998bc,
Witten:1998qj,Freedman:1998bj,
Liu:1998ty,
Freedman:1998tz,
DHoker:1999mqo,
DHoker:1999kzh},
for some reviews see
\cite{Aharony:1999ti,Zaffaroni:2000vh,Nastase:2007kj,Kap:lecture}). \acro{AdS} bulk fields correspond to \acro{CFT} operators in a way that is exact (to the best of our knowledge) in the full \acro{AdS} spacetime and in the presence of a \acro{UV} brane.

Fields localized on the \acro{UV} brane are understood to be external fields probing the \acro{CFT}; these are equivalently called \emph{elementary} states in contrast to \acro{CFT} degrees of freedom. In the context of our model, dark matter and Standard Model particles are elementary fields. We require that dark matter couples to a scalar operator of the mediator \acro{CFT} sector; this scalar operator corresponds to the \acro{5D} bulk mediator field $\Phi$. The mediator \acro{CFT} two-point correlation function gives the self-interaction amplitude in \eqref{eq:AdS:CFT:bulk}.

The understanding of the \acro{4D} dual theory is only qualitative in the presence of \acro{IR} brane cutting off large $z$ values. The \acro{IR} brane is interpreted as a spontaneous breaking of the conformal symmetry analogous to confinement in a strongly-interacting gauge theory \cite{ArkaniHamed:2000ds,Creminelli:2001th}. The theory is thus only approximately conformal at scales much larger than $\mu$, however we follow the common colloquiual practice of referring to the \acro{4D} theory as a \acro{CFT}. The scale $\mu = z_\text{IR}^{-1}$ is naturally associated to the mass gap characterizing conformal symmetry breaking, similar to the \acro{QCD} confinement scale.
\acro{KK} modes are identified with composite states that are allowed when conformal invariance is broken. In the simplest realization, the composite states are glueballs of adjoint gauge fields.

Either the \acro{AdS} or \acro{CFT} description of the theory may be more convenient depending on the context. We primarily focus on the \acro{5D} description where the model is concretely defined. The qualitative behavior of this theory is general and captures what is expected for a purely \acro{4D} near-conformal mediator; one may view the \acro{5D} construction as a simple quantitative tool to describe such a theory.

\section{Phenomenological Constraints}
\label{se:SM_constraints}

We briefly comment on implications of our model beyond the dark matter self-interaction phenomenology that is our primary focus.

\subsection{Cosmological Dark Radiation}

Models of near-conformal dark sectors necessarily introduce large numbers of degrees of freedom. Many of these may be relativistic in the early universe and are thus constrained by big bang nucleosynthesis (\acro{BBN}) and the cosmic microwave background (\acro{CMB}). 
There are at least three ways to avoid the tight constraints on the effective number of relativistic degrees of freedom, $N_\text{eff}$:
\begin{enumerate}
\item The theory may have a sufficiently large mass gap, $\mathcal O(\text{MeV})$, so that all states are non-relativistic at the relevant times. In this case there is no dark radiation.

\item The relativistic states decay quickly enough that they do not affect \acro{BBN} or the \acro{CMB}~\cite{Katz:2015zba}.

\item The dark sector may be much colder than the Standard Model so that the density of states is suppressed compared to visible matter. This is a natural possibility and has been studied in the context of gravitational interactions in \acro{AdS} \cite{Hebecker:2001nv,Langlois:2002ke,Langlois:2003zb}. Dark radiation from a bulk scalar will be studied in an upcoming work~\cite{us:DR}.
\end{enumerate}
With these features in mind, we focus on $\mu \gtrsim \mathcal O(\text{MeV})$, but allow for 
 $\mu$ the possibility of lower scales subject to additional model building to accommodate $N_\text{eff}$ limits.

\subsection{Fifth Force}

Bulk graviton exchange leads to deviations from the Newtonian gravitational potential of the form~\cite{Randall:1999vf,Giddings:2000mu}
\begin{align}
  V_\text{N} (r) 
  = 
  -\frac{G_N}{r}
  \left[ 
  	1+ \mathcal{O}\!\left(\frac{1}{k^2r^2}\right)
  \right] \ .
\end{align}
Constraint from fifth force searches set  $k \gtrsim 5\,$meV or $k^{-1} \lesssim 50 \, \mu$m and hence can be ignored; see {e.g.}~\cite{Lee:2020zjt} for a recent measurement, \cite{Brax:2017xho} for a review of $r^{-3}$ constraints.

\subsection{Deviations from the Standard Model}

Standard Model fields are assumed to be localized on the \acro{UV} brane. For the purposes of dark matter self interaction phenomenology, we neglect any direct \acro{UV}-brane interactions between the dark matter and Standard Model and assume that the mediator--Standard Model couplings are negligible. These couplings are phenomenologically relevant, for example in dark matter direct detection experiments~\cite{Katz:2015zba} or in searches for novel forces between Standard Model particles~\cite{Brax:2019koq, Costantino:2019ixl}, but are not directly related to the small scale structure anomalies that are the primary phenomenological focus of this paper.

In principle the brane-localized fields are limits of \acro{5D} fields with heavy \acro{KK} modes~\cite{Fichet:2019owx}. The most significant effects of these modes are deviations in the Standard Model gauge sector: gauge bosons can scatter off \acro{5D} gravitons and the gauge couplings pick up an anomalous logarithmic running above the \acro{IR} scale $\mu$. 
Both of these effects are small enough to be undetected with current data in the limit where $\Lambda$ is sufficiently close to $k$. 
Since we already assume this in \eqref{eq:lambdak}, the model is safe from these effects.

\section{The Continuum-Mediated Potential}
\label{se:potential}

The potential $V$ between two particles is obtained from the $t$-channel scattering amplitude with the external legs taken to the appropriate non-relativistic limit,\footnote{%
In principle $u$-channel diagrams contribute when the scattering particles are identical. This is an $\mathcal O(\text{few})$ effect~\cite[App.~C]{Kahlhoefer:2017umn}. We neglect the $u$-channel contribution for simplicity and ease of direct comparison to Ref.~\cite{Tulin:2013teo}.
}
\begin{align}
i \mathcal{M} 
& \equiv 
-4i m_{\chi}^2 \tilde{V}\left(|\mathbf{q}|\right) 
= 
-4\frac{\lambda^2 }{k}
G_{|\mathbf{q}|}\left(z_\text{UV},z_\text{UV}\right)
\, ,
\end{align}
with $t\approx-|\mathbf{q}|^2$ where $\mathbf{q}$ is the three-dimensional momentum transfer. On the right-hand side we insert the expression from the exchange of a $t$-channel bulk mediator between dark matter currents.
The position-space potential is related by a Fourier transform
\begin{equation}
V(r)=\int \frac{d^3 \mathbf{q}}{\left(2 \pi\right)^3} \tilde{V}\left(|\mathbf{q}|\right) e^{i \mathbf{q}\cdot \mathbf{r}} \,, \label{eq:VFourier}
\end{equation}
with $r=|\mathbf{r}|$. 
Even though our effective theory has a cutoff,
one may integrate \eqref{eq:VFourier} to infinite $|\mathbf{q}|$ under the assumption of a smooth cutoff, as shown in the Appendix~B of Ref.~\cite{Costantino:2019ixl}.

Simply inserting the exact propagator \eqref{eq:propa} is analytically challenging. We proceed by using a spectral representation where the discontinuity of the two-point function is evaluated in the appropriate asymptotic limits from Section~\ref{se:asymptotics}.

\subsection{Spectral Representation}

The spectral representation for the bulk propagator is~\cite{Zwicky:2016lka}
\begin{align}
  G_p(z,z') 
  &=
  \frac{1}{2\pi i}
  \int_0^\infty 
  d\rho\, 
  \frac{ 
    \text{Disc}_\rho\left[G_{\sqrt{\rho}}(z,z')\right] 
  }{
    \rho-p^2
  }
\ ,
\label{eq:spectral:discontinuity}  
\end{align}
where $\text{Disc}_\rho \left[g(\rho)\right]$ is the discontinuity of $g(\rho)$ across the 
branch cut along the real line, $\rho \in \mathbbm{R}^+$:
\begin{align}
  \text{Disc}_\rho[g(\rho)]
  &=\lim_{\epsilon\to 0}\, g(\rho+i\epsilon)-g(\rho-i\epsilon)
  & \epsilon
  &>0
  \ .  
\end{align}
We compute the non-relativistic potential using this spectral representation of the propagator. Performing the $d^3\mathbf{q}$ integral yields a general representation of the long-range potential:
\begin{align}
  V(r)
  &= 
  - \frac{1}{8\pi^2 } 
  \frac{\lambda^2}{k}
  \int_0^\infty d\rho \;
  \text{Disc}_\rho\left[G_{\sqrt{\rho}}(z_\text{UV},z_\text{UV})\right] 
  \;
  \frac{e^{-\sqrt{\rho}r}}{r} 
  \ .
\label{eq:V_spectral_gen}  
\end{align}

\paragraph{Kaluza--Klein representation.}
One may use the \acro{KK} representation of the free propagator \eqref{eq:propa_KK} in the spectral representation of the potential \eqref{eq:V_spectral_gen}; this amounts to identifying the exchange of a \acro{5D} bulk scalar with the sum of $t$-channel diagrams with each \acro{KK} mode:
\begin{align}
  \vcenter{
    \hbox{\includegraphics[width=.15\textwidth]{{CartoonwIR}}} 
    }
    =
    \vcenter{
    \hbox{\includegraphics[width=.1\textwidth]{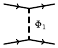}}
    }
    +
    \vcenter{
    \hbox{\includegraphics[width=.1\textwidth]{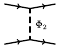}}
    }
    +
    \vcenter{
    \hbox{\includegraphics[width=.1\textwidth]{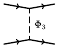}}
    }
    +\cdots
    \label{eq:diagram:5D:sum4D}
\end{align}
The spectral distribution is 
$\text{Disc}_\rho\left[G_{\sqrt{\rho}}(z,z')\right]=\sum_n f_n(z) f_n(z') 2\pi \delta(p^2-m_n^2)$, so that the potential is an infinite sum of Yukawa potentials from each \acro{KK} mode:
\begin{align}
  V(r) &= 
  -\frac{1}{4\pi}
  \frac{\lambda^2}{k}\,
  \sum_n f_n(z_0)^2
  \frac{e^{-m_n r}}{r} 
  \ . \label{eq:V_KK_prop}  
\end{align}
While this \acro{KK} representation of $V$ is exact, it requires knowledge of the entire spectrum of \acro{KK} masses and wavefunctions.

\paragraph{Canonical representation.}
One may alternatively use the canonical representation of the propagator \eqref{eq:propa} in the spectral representation of the potential \eqref{eq:V_spectral_gen}. In this case, one may apply the closed-form asymptotic expressions derived in the following section.
These asymptotic expressions carry the same poles as the \acro{KK} representation. 
The momentum flowing through the propagator is necessarily spacelike in diagrams that contribute to the potential.
Thus we may readily use the asymptotic expressions for large $|p|$ that are valid away from the poles, \,\eqref{eq:Propcont} for $\alpha < 1$ and \eqref{eq:Propalpha1contren} for $\alpha=1$. We numerically validate this approximation in Section~\ref{sec:validity}.

\subsection{Propagator Asymptotics} \label{se:asymptotics}

We present the limits of the bulk propagator $G_p$ for Minkowski momenta $p$ much smaller and larger than the mass gap, $\mu$. We focus on propagation to and from the \acro{UV} brane where the dark matter currents are localized.  These limits illuminate the properties of the theory and yield simplifications for the self-interaction potential.

We treat the $\alpha<1$ and $\alpha=1$ cases separately; the asymptotic behavior of Bessel functions with near integer order have an extra contribution that is neglected for non-integer order.\footnote{This is due to the expression for the Bessel function of the second kind with integer index $\alpha\to n$,
$$Y_n(z) = 
\frac{1}{\pi}
\left.\frac{\partial J_\alpha(z)}{\partial \alpha}\right|_{\alpha=n}
+ \frac{(-)^n}{\pi}
\left.\frac{\partial J_\alpha(z)}{\partial \alpha}\right|_{\alpha=-n} \ .
$$
\label{foot:bessel:alpha:to:1:limit}
}
As a result, one typically cannot obtain the $\alpha=1$ asymptotic behavior as the $\alpha\to 1$ limit of the $\alpha<1$ asymptotic behavior. The $\alpha=1$ case is a meaningful benchmark as it is equivalent to the exchange of a single \acro{4D} mediator.

\subsubsection{Propagator Asymptotics for $0<\alpha<1$}

\paragraph{Small momentum asymptotic, $|p| \ll \mu$.}
For Minkowski momenta much less than the mass gap we find a single \acro{4D} pole:
\begin{align}
    G_p(z_\text{UV},z_\text{UV})
  & = 
  i 
  \frac{
    2 k (1-\alpha)
    \left( 2\alpha+b_{\text{IR}} \right)
  }{
    \alpha
    (2+b_{\text{IR}})
    p^2
    -
    4 \alpha (1-\alpha) b_{\text{IR}} \mu^2 
  } 
   \left( \frac{\mu}{k} \right)^{2-2\alpha} 
   \ .
   \label{eq:proplowenergy}
\end{align}
All other poles are be heavier than $\mathcal O(\mu)$.
For $b_\text{IR}\lesssim \mathcal O(1)$, the light 4D mode mass is 
\begin{align}
    m_0^2 
    & =
    \frac{
      4(1-\alpha)b_{\rm IR}
    }{
      2+b_{\rm IR}
    }
    \mu^2 
    \label{eq:m0alphaneq1}  
    \ . 
\end{align}

\paragraph{Large momentum asymptotic, $|p| \gg \mu$.}
For momenta much larger than the mass gap,
\begin{align}
  G_p(z_\text{UV},z_\text{UV})
  & =
  \frac{i}{2 k}
  \frac{
    \Gamma\left(\alpha\right)
  }{
    \Gamma\left(-\alpha+1\right)
  }
  \left( \frac{4 k^2}{p^2} \right)^{\alpha} 
  S_{\alpha}(p)
  & 
  S_\alpha(p)
  &=
  \frac{
    \sin\!\left(\frac{p}{\mu}-\frac{\pi}{4}(1-2\alpha)\right)
  }{
    \sin\!\left(\frac{p}{\mu}-\frac{\pi}{4}(1+2\alpha)\right)
    }
  \ .
  \label{eq:prophighenergy}
  \end{align}
The tower of \acro{KK} poles are encoded in $S_\alpha(p)$. 
The propagator further simplifies when the momentum has an imaginary part $\text{Im}(p/\mu) \gtrsim 1$:
\begin{align}
  G_p(z_\text{UV},z_\text{UV})
  &=
  \frac{i}{2 k}
  \frac{\Gamma\left(\alpha\right)}{\Gamma\left(-\alpha+1\right)}
  \left(
    \frac{4 k^2}{-p^2}
    \right)^{\alpha} 
  \ ,
  \label{eq:Propcont}
\end{align}
where we have used $S_{\alpha} \approx (-1)^{\alpha}$ in this limit.\footnote{Loops from bulk interactions cause heavy \acro{KK} modes to acquire large widths and give an effective imaginary part to timelike four-momentum in the bulk propagator~\cite{Fichet:2019hkg, Costantino:2020msc,Costantino:2020vdu}.  This physical imaginary part is important for timelike processes but is not  for spacelike processes, hence it is irrelevant for the potential. 
}
This includes the case of spacelike momentum.
In this limit the conformal scaling appears: recalling that $\alpha=2-\Delta$, the propagator reproduces the scaling of the amplitude~\eqref{eq:CFT}.  
Observe that the \acro{UV} brane kinetic term does not appear in this expression. This reflects the fact that none of the modes are localized near the \acro{UV} brane.

\subsubsection{Propagator Asymptotics for $\alpha=1$}

\paragraph{Small momentum asymptotic, $|p| \ll \mu$.} 
For Minkowski momenta much less than the mass gap, we find
\begin{align}
    G_p \left(z_\text{UV},z_\text{UV}\right)
    & =
    \frac{
      (2+b_{\text{IR}})2 i k
    }{
      p^2
      \left[ 
        (2+b_{\text{IR}})
        (2 c_{\rm UV} k+\log(k^2/\mu^2))
        -
        b_{\text{IR}}
      \right]
      - 4b_{\text{IR}}\mu^2
    } 
    \ . 
    \label{eq:propalpha1lowenergy}
\end{align}
This carries a single \acro{4D} pole. The mass of this light mode is 
\begin{align}
    m_0^2
    & =
    \frac{
      4 b_{\text{IR}} \mu^2
    }{
      (2+b_{\text{IR}})
      \left[
        2 c_\text{UV} k
        + \log(k^2/\mu^2)
      \right]
      -b_{\text{IR}}
    } \ .
\label{eq:m0alpha1}
\end{align}
This mass is suppressed by $c_\text{UV}+\log(k/\mu)$, where $c_\text{UV}$ is the coefficient of the \acro{UV} brane-localized kinetic term and $\log(k/\mu)$ describes the bulk volume. 
One may understand \eqref{eq:m0alpha1} as a dressing of the zero mode with an \acro{IR} brane-localized mass.

\paragraph{Large momentum asymptotic, $|p| \gg \mu$.} 
For momenta much larger than the mass gap,
\begin{align}
   G_p \left(z_\text{UV},z_\text{UV}\right)
   &=
   \frac{
    2 i k
  }{
    p^2
    \left[
      2c_\text{UV}
      -
      \pi \cot\left(
        \frac{p}{\mu}
        +
        \frac{\pi}{4}
      \right)
      -
      \log\left( \frac{p^2}{4 k^2} \right)
      -2\gamma
    \right]
  } 
  \, . \label{eq:Propalpha1}
\end{align}
When $\text{Im}(p/\mu)\gtrsim 1$ the cotangent approaches $-i$ and the propagator simplifies,
\begin{align}
  G_p \left(z_\text{UV}, z_\text{UV}\right)
  &=
  \frac{
    2 i k
  }{
    p^2
    \left[
      2c_\text{UV}
      -
      \log\left(-\frac{p^2}{4 k^2}\right)
        -2\gamma
    \right]
    } 
    \, . 
    \label{eq:Propalpha1cont} 
\end{align}
In contrast to the $\alpha < 1$ case \eqref{eq:Propcont}, the \acro{UV} brane kinetic term is not negligible. 
This propagator describes a \acro{4D} mode with a logarithmic running of its wavefunction.
It is similar to the well known case of a bulk gauge field in \acro{AdS}. We can absorb a large logarithm by redefining the brane wavefunction coefficient $c_\text{UV}$ at a physical scale $p_0$:
\begin{align}
  \hat c_\text{UV}
  &=
  c_\text{UV} 
  +
  \left[
    \log\left(k/p_0\right)
    -\gamma 
  \right]
&
  G_p\left(z_\text{UV},z_\text{UV}\right)
  &=
  \frac{
    2 i k
  }{
    p^2
    \left[
      2\hat c_{\rm UV}
      -\log\left(-\frac{p^2}{ p_0^2}\right)
    \right]
  } \, . \label{eq:Propalpha1contren} 
\end{align}
For the astrophysical applications of self-interacting dark matter, the energy transfer ranges over only a few orders of magnitude and the logarithmic running is thus negligible. The $\alpha=1$ case thus reproduces the standard single-mediator self-interacting dark matter model and serves as a useful benchmark.

\subsection{\texorpdfstring{Potential, $\alpha <1$}{Potential, alpha less than 1}}

For bulk masses in the range $0<\alpha<1$ and with generic \acro{IR} brane mass parameter $b_\text{IR} \sim \mathcal O(1)$, the lightest excitations have mass on the order of $\mu$; see \eqref{eq:m0alphaneq1}.
Since there is no light mode to contribute to non-analyticities of $G_p$ for $|p| < \mu$, we may apply the $|p|\gg\mu$ approximation of the propagator to the spectral integral \eqref{eq:V_spectral_gen}. 
The lower limit of the spectral integral is formally the mass of the lightest \acro{KK} mode, 
\begin{align}
  V(r)
  &= 
  - \frac{1}{8\pi^2} 
  \frac{\lambda^2}{k}
  \int_{m_1^2}^\infty 
  d\rho \;
  \text{Disc}_\rho
  \left[
    G_{\sqrt{\rho}}(z_\text{UV},z_\text{UV})
  \right]
  \frac{e^{-\sqrt{\rho}r}}{r} 
  \ .
\label{eq:V_spectral_m1}
\end{align}
However, because $m_1=\mathcal O(\mu)$, by using the $\rho\gg\mu^2$ approximation for the propagator \eqref{eq:Propcont}, we introduce some uncertainty in the lower bound of the spectral integral. We verify the validity of this approximation in Section~\ref{sec:validity}.

The discontinuity across the branch cut along $\rho>0$ is
\begin{align}
  \text{Disc}_{\rho}
  \left[\Delta_{\sqrt{\rho}}(z_0,z_0)\right]
  &=
  \frac{1}{k}
  \left(\frac{4 k^2}{\rho}\right)^{\alpha}
  \frac{\Gamma(\alpha)}{\Gamma(1-\alpha)}\sin(\pi \alpha) 
  \ ,
\end{align}
where we have used \eqref{eq:Propcont}.
This is valid for $\text{Im}(p/\mu) \gtrsim 1$, which we assume because $p$ is spacelike.
Evaluating the integral across the discontinuity using the $\Gamma$ reflection and duplication formulas\footnote{Namely: $\Gamma\left(1-z\right) \Gamma\left(z\right)=\pi/\sin\left(\pi z\right)$ and $\Gamma\left(2 z\right)=\pi^{-1/2}2^{2z-1}\Gamma\left(z\right) \Gamma\left(z+1/2\right)$.} gives the main expression we use in our analysis:
\begin{align}
  V(r)
  &=
  -\frac{\lambda^2}{2 \pi^{3/2}} 
  \frac{\Gamma(3/2-\alpha)}{\Gamma(1-\alpha)} 
  \frac{1}{r}
  \left(\frac{1}{k r}\right)^{2-2\alpha} 
  Q(2-2\alpha, m_1 r)  \ ,
  \label{eq:Vgamma}
\end{align}
where  $Q\left(2-2\alpha,m_1 r\right)$ is the regularized incomplete $\Gamma$ function, 
\begin{align}
  Q\left(p,z\right)=\frac{1}{\Gamma\left(p\right)}\int_{z}^{\infty}dx~x^{p-1}e^{-x} \, .
\end{align}
For $r\gg m_1^{-1}$, the potential is exponentially suppressed at long distances,
\begin{align}
  V(r)
  &\propto 
  -\left(\frac{m_1}{k}\right)^{1-2\alpha}
  \frac{1}{kr^2} e^{-m_1 r}    \ .
  \label{eq:Q:at:large:r}
\end{align}
We see that $Q(2-2\alpha,r)$ takes the place of the $e^{-mr}$ Yukawa factor that encodes the mass gap in the single-mediator scenario. In turn, this mass gap is a key ingredient for cutting off unwanted long-range dark forces.

It is illustrative to check the behavior in the gapless limit $\mu\to 0$. The large, spacelike momentum approximation of the propagator \eqref{eq:Propalpha1cont} is exact in this limit and potential can be evaluated exactly. We recover the gapless limit in \eqref{eq:Vgamma} the gapless limit is recovered by taking $m_1\to 0$, giving 
\begin{align}
  V_\text{gapless}(r)
  &=
  -\frac{\lambda^2}{2 \pi^{3/2}} 
  \frac{\Gamma(3/2-\alpha)}{\Gamma(1-\alpha)} 
  \frac{1}{r} 
  \left(\frac{1}{k r}\right)^{2-2\alpha} \ ,
  \label{eq:Vnogamma}
\end{align}
which matches the  result from \cite{Brax:2019koq}.
The power law behavior obtained matches the proposed scaling in \eqref{eq:CFT} with the \acro{AdS}/\acro{CFT} identification $\Delta=2-\alpha$.

\subsection{\texorpdfstring{Potential, $\alpha=1$}{Potential, alpha equals 1}}
\label{sec:alpha:1:potential}

For bulk mass parameter $\alpha =1$ and with generic \acro{IR} brane mass parameter $b_\text{IR} \sim \mathcal O(1)$, there is a mode lighter than the scale $\mu$. The suppression relative to $\mu$ is the kinetic factor $(c_{\rm UV}+\log(k/\mu))^{1/2}$ in \eqref{eq:m0alpha1}. This is in contrast to the $\alpha <1$ case.
The spectral integral over the discontinuity in $G_{\sqrt{\rho}}$ must thus take into account  this pole in the $\rho\ll \mu^2$ regime in addition non-analyticities in the  $\rho\gg \mu^2$ regime. We separate the potential into two pieces accordingly, $V=V_\text{light} + V_\text{KK}$.

\paragraph{Light mode contribution.}
The light mode contributes a simple Yukawa potential:
\begin{align}
  V_\text{light}
  &=   
  -\frac{\lambda^2}{4 \pi k}
  f_0(z_\text{UV})^2
  \frac{e^{-m_0 r}}{r} 
  \label{eq:V_light}
\end{align}
where the profile evaluated on the \acro{UV} brane is
\begin{align}
  f_0(z_\text{UV})^2 
  &=
  \frac{
    (2+b_{\text{IR}})2  k
  }{
    (2+b_{\text{IR}})
    \left[
      2 c_{\rm UV} +\log\!\left(\frac{k^2}{\mu^2}\right)
    \right]
    -b_{\text{IR}}
  } 
  \approx
  \frac{k}{ \hat c_\text{UV} +\log\!\left(\frac{p_0}{\mu}\right)+\gamma}
  \ .
  \label{eq:f0}
\end{align}
as can be derived from the pole of the small momentum transfer limit of the propagator \eqref{eq:propalpha1lowenergy}. 
On the right-hand side we
use the assumption that $b_\text{IR}\sim\mathcal O(1)$,
apply the $\mu\ll k$ limit, insert the renormalized brane kinetic term coefficient  $\hat c_\text{UV}$ defined at the scale $p_0$ from \eqref{eq:Propalpha1contren}.

\paragraph{KK mode contribution.}
The \acro{KK} mode contribution uses the $|p|\gg \mu$ asymptotic of the bulk $\alpha=1$ propagator \eqref{eq:Propalpha1contren} applied to the large-momentum spectral integral, \eqref{eq:V_spectral_m1}. 
To obtain an analytically tractable expression we take the limit $\hat c_\text{UV} \gg \log(\rho / p_0^2)$ over the range $\rho\in[m_1^2,r^{-2}]$; the upper bound comes from the $\exp\!\left(-\sqrt{\rho} r\right)$ factor in the spectral integral. The resulting propagator is
\begin{align}
      G_{p} \left(z_\text{UV},z_\text{UV}\right)
      &=
      \frac{i k}{p^2 \hat c_\text{UV} }
      \left[
        1
        +\frac{\log(-p^2 / p_0^2)}{2\hat c_\text{UV} }
        + \mathcal O\!\left(
          \frac{1}{\hat c_\text{UV}^2}
        \right) 
      \right]
      \ .
      \label{eq:Propalpha1exp} 
\end{align}
The discontinuity in the spectral intergal is
\begin{align}
  \text{Disc}_{\rho}
  \left[G_{\sqrt{\rho}}(z_\text{UV},z_\text{UV})\right]
  &= 
  \frac{ 2\pi  k}{ \hat c_{\rm UV}} \delta(\rho) 
  + 
  \frac{k}{\hat c_{\rm UV}^2 } \frac{\pi}{\rho}   
  + \mathcal O\!
  \left(
  \frac{1}{\hat c_{\rm UV}^3 }
  \right)\,.
 \label{eq:Discalpha1} 
\end{align}
The singular $\delta(\rho)$ term is outside the range of integration and does not contribute. The leading contribution comes from the  $\mathcal O\!\left(\hat c_\text{UV}^{-2}\right)$ term and evaluates to
\begin{align}
  V_\text{KK} (r)
  &= 
  - \frac{1}{4\pi r} 
  \frac{\lambda^2}{\hat c_\text{UV}^2}
  \Gamma(0,m_1 r)\,.  
\end{align}

\paragraph{The $\alpha=1$ Potential and Limits.}
Since we have used the $\hat c_\text{UV} \gg \log(\rho/p_0^2)$ limit in the \acro{KK} potential, we may apply the same approximation to the light mode contribution. This produces the full $\alpha=1$ potential
\begin{align}
  V(r) 
  &= 
  -\frac{\lambda^2}{4 \pi r} 
  \left[ 
    \frac{1}{\hat c_\text{UV} } 
    \left(
      1
      -\frac{\log(p_0/\mu)+\gamma}{\hat c_{\rm UV} }
    \right) 
    e^{-m_0 r} 
    + 
    \frac{\Gamma(0,m_1 r) }{\hat c_\text{UV}^2}  
  \right]
  + \mathcal O\!\left(\frac{1}{\hat c_\text{UV}^{3} }\right)  
  \ .
\end{align}
At long distances, $r\gg m_1^{-1}$, 
\begin{align}
  \frac{\Gamma(0,m_1 r) }{\hat c_\text{UV}^2}  
  \quad\to \quad
  \frac{1}{\hat c_\text{UV}^2 }
  \frac{e^{-m_1 r}}{m_1 r} 
  \ .
\end{align}
%
One can explicitly see the exponential suppression from both the light mode and \acro{KK} mode mass gaps. In the short distance $r\ll m_1^{-1}$ limit, the incomplete $\Gamma$~function is $\Gamma(0,x)\approx-(\log\,x+\gamma)$ and one has  $e^{-m_0 r}\sim 1$. Since $m_0 r\ll1$,  we obtain 
\begin{align}
  V(r)
  &=
  -\frac{\lambda^2 }{\hat c_\text{UV}}
  \frac{1}{4\pi r}
  \left[
    1
    -\frac{1}{\hat c_{\rm UV}}
    \log\!\left( \frac{r}{r_0} \right)
  \right] 
  + \mathcal O\!\left(\frac{1}{c_\text{UV}^3 }\right) 
  \ ,
  \label{eq:Valpha1mu0}
\end{align}
where we introduce the scale $r_0$ 
\begin{align}
  \log r_0 &= \log p_0 +2\gamma +\log\left(\frac{m_1}{\mu}\right)
\end{align}
to absorb  $\mathcal O(1)$ coefficients. The explicit $\mu$ dependence vanishes because the $\log\mu$ from the light mode and the $\log m_1=\log\mu +\mathcal O(1)$ from the \acro{KK} modes cancel. 

While \eqref{eq:Valpha1mu0} could be understood as the $\mu\to 0$ limit of the $\alpha=1$ potential, the $\hat c_\text{UV} \gg \log(\rho/p_0^2)$ assumption we used to evaluate the spatial potential formally does not hold in this limit. Instead the full $\log^n r$ series would need to be resummed. Nevertheless, we verify that the Fourier transform of the propagator \eqref{eq:Propalpha1exp} matches the potential \eqref{eq:Valpha1mu0}. Interestingly, in this limit the  contribution from the light mode is replaced by the $\delta(\rho)$ contribution in the discontinuity across the propagator, \eqref{eq:Discalpha1}, which is otherwise cut off at finite $\mu$.  Details of this calculation are presented in Appendix~\ref{se:Valpha1}.

The expressions in this section show that the \acro{KK} mode contribution tends to be  small with respect to the light mode for both large and small $r$.  This logarithmic correction is negligible in our self-interacting dark matter calculations and thus the $\alpha=1$ case matches the standard single \acro{4D} mediator scenario. It can thus be used as a benchmark comparing to $\alpha\neq 1$ phenomenology. 

\subsection{Validation of Potential}
\label{sec:validity}

\begin{figure}
\centering
  \begin{subfigure}[b]{0.49\textwidth}
    \centering
    \includegraphics[width=\textwidth]{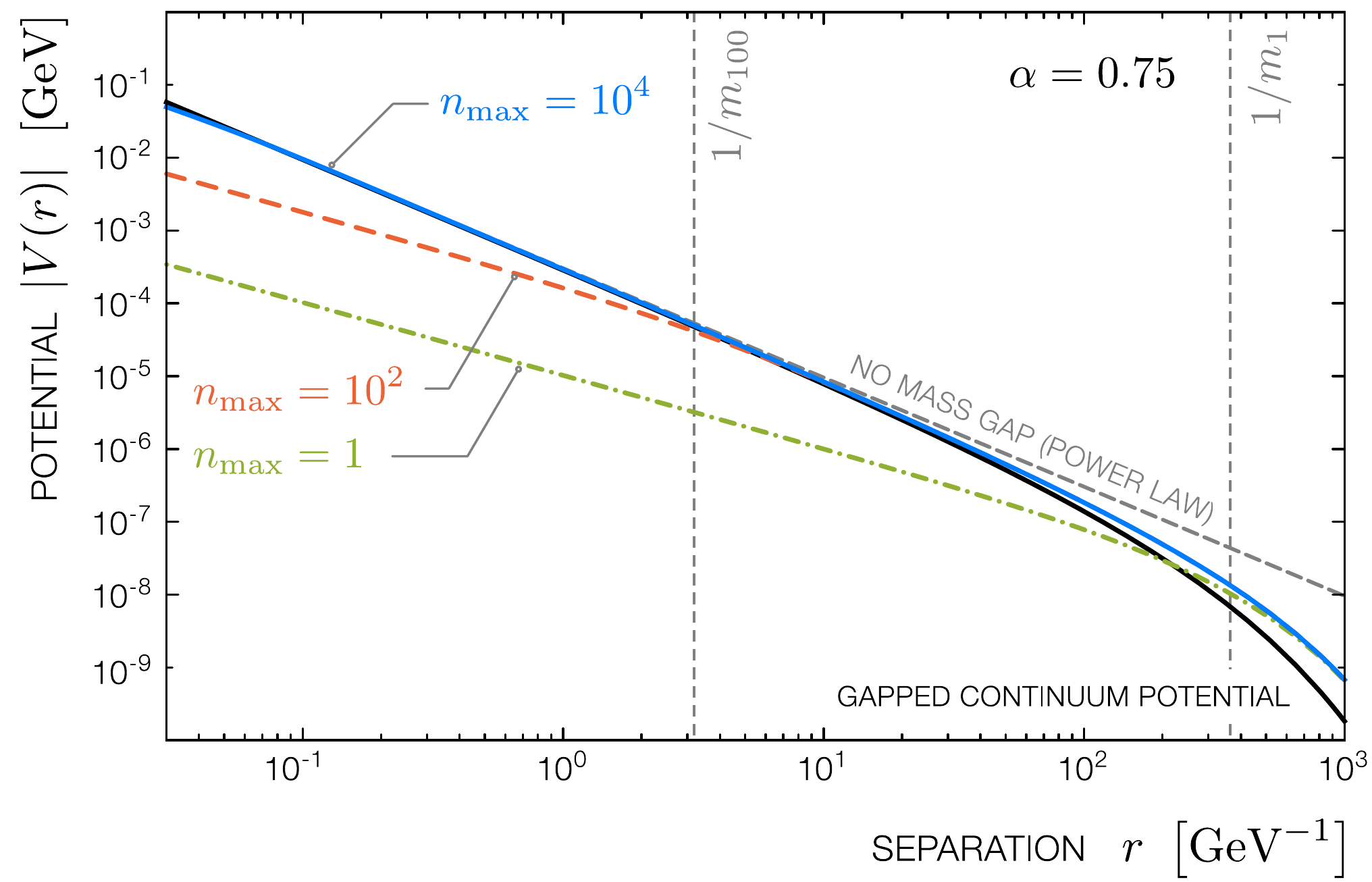}
  \end{subfigure}
  \hfill
  \begin{subfigure}[b]{0.49\textwidth}  
    \centering 
    \includegraphics[width=\textwidth]{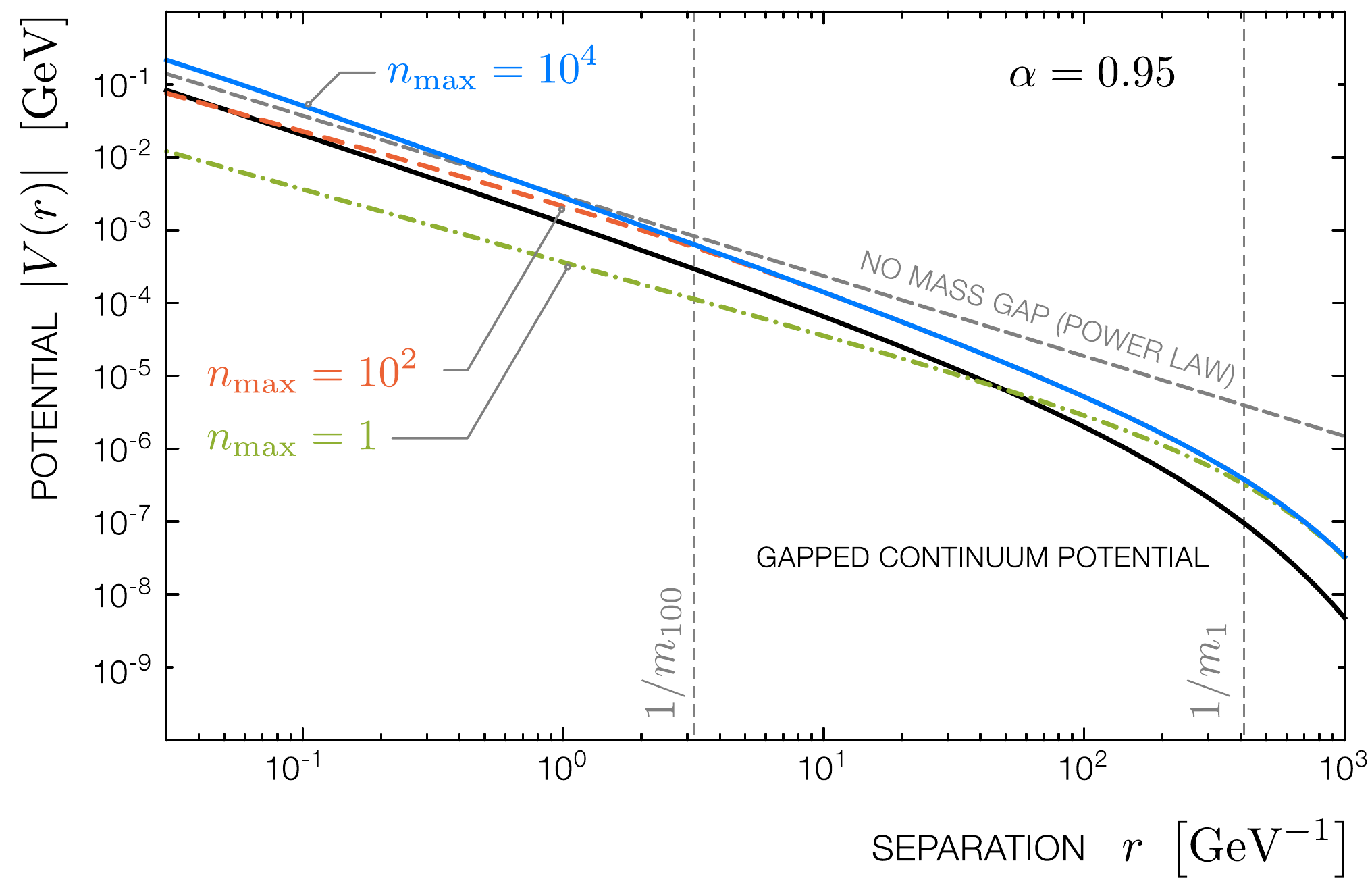}
  \end{subfigure}
\caption{%
Absolute potential $|V(r)|$ plotted to validate the continuum-mediated potential with a mass gap~(black) against a sum over $n_\text{max}$ Kaluza--Klein modes~(colored). The potential with $n_\text{max}$ \acro{KK} modes is valid for separations larger than $r\gtrsim m_{n_\text{max}}^{-1}$.
The disagreement at long separations between the blue and black lines represents our numerical error and does not change the quantitative behavior of integrals over the potential. 
Also shown: the non-integer power law limit (dashed gray) that is realized in the gap-less limit $m_1\to0$.
}
\label{fig:fractionalvskksum}
\end{figure}

\noindent

\noindent
In this study we use the asymptotic approximation of the gapped continuum-mediated potential~\eqref{eq:Vgamma}. In order to quantify its validity, we compare our approximation to an explicit sum over Kaluza--Klein mediated Yukawa potentials~\eqref{eq:V_KK_prop}. This is a meaningful check since a sum over $n_\text{max}$ \acro{KK} modes is a valid approximation to the full sum on scales longer than the inverse mass of the heaviest mode, $r\gtrsim m^{-1}_{n_\text{max}}$. We thus test for agreement of the gapped continuum-mediated potential with the sum over a large number of \acro{KK} in the regime where the latter is valid. 

We present our validation in Figure~\ref{fig:fractionalvskksum}. The key comparison is between sum over $n_\text{max}=10^4$ \acro{KK} modes (blue) and the continuum-mediated potential (black). For values of $\alpha \lesssim 0.95$, the sum over $n_\text{max}$ \acro{KK} modes agrees with the continuum potential in the regime where the finite \acro{KK} sum is valid, $r\gtrsim m^{-1}_{n_\text{max}}$. However, at distances longer than the inverse mass gap, $r\gtrsim m_1^{-1}$, the curves diverge slightly while maintaining the same qualitative gapped behavior.
This discrepancy is caused by the $|p|\gg\mu$ limit assumed in the derivation of the continuum-mediated potential \eqref{eq:Vgamma}. This discrepancy grows when $\alpha\approx 1$; see Footnote~\ref{foot:bessel:alpha:to:1:limit}.
Practically, we restrict the continuum-mediated potential for $\alpha \lesssim 0.95$. 
In this range, the large-$\alpha$ discrepancy does not change the qualitative behavior of the continuum-mediated potential, nor the quantitative behavior of integrals of this potential. 
For larger values of $\alpha$, the potential reproduces the well-known case of a single \acro{4D} mediator, as described in Section~\ref{sec:alpha:1:potential}.

Figure~\ref{fig:fractionalvskksum} also demonstrates how  a sum of Yukawa potentials can reproduce a potential that goes like a non-integer power of the separation, \eqref{eq:Vintro}. The lightest \acro{KK} mass sets a long-range length scale, $m_1^{-1}$. In the regime $m_{n_\text{max}}^{-1} \lesssim r \ll m_1$, the sum over Yukawa potentials from $n_\text{max}$ \acro{KK} modes produces a total potential that matches the power law of~\eqref{eq:Vnogamma}. 

\section{Astrophysical Phenomenology}
\label{se:SIDM}

We apply our continuum-mediated model to the phenomenology of self-interacting dark matter for small-scale structure. 
The quantity that connects particle physics parameters to astronomy is the transfer cross section. We demonstrate the dependence of this cross section on our model parameters and provide representative fits.

\subsection{Review of Self-Interacting Dark Matter Cross Sections}
\label{eq:review:SIDM}

\begin{figure}
\centering
\includegraphics[width=.8\textwidth]{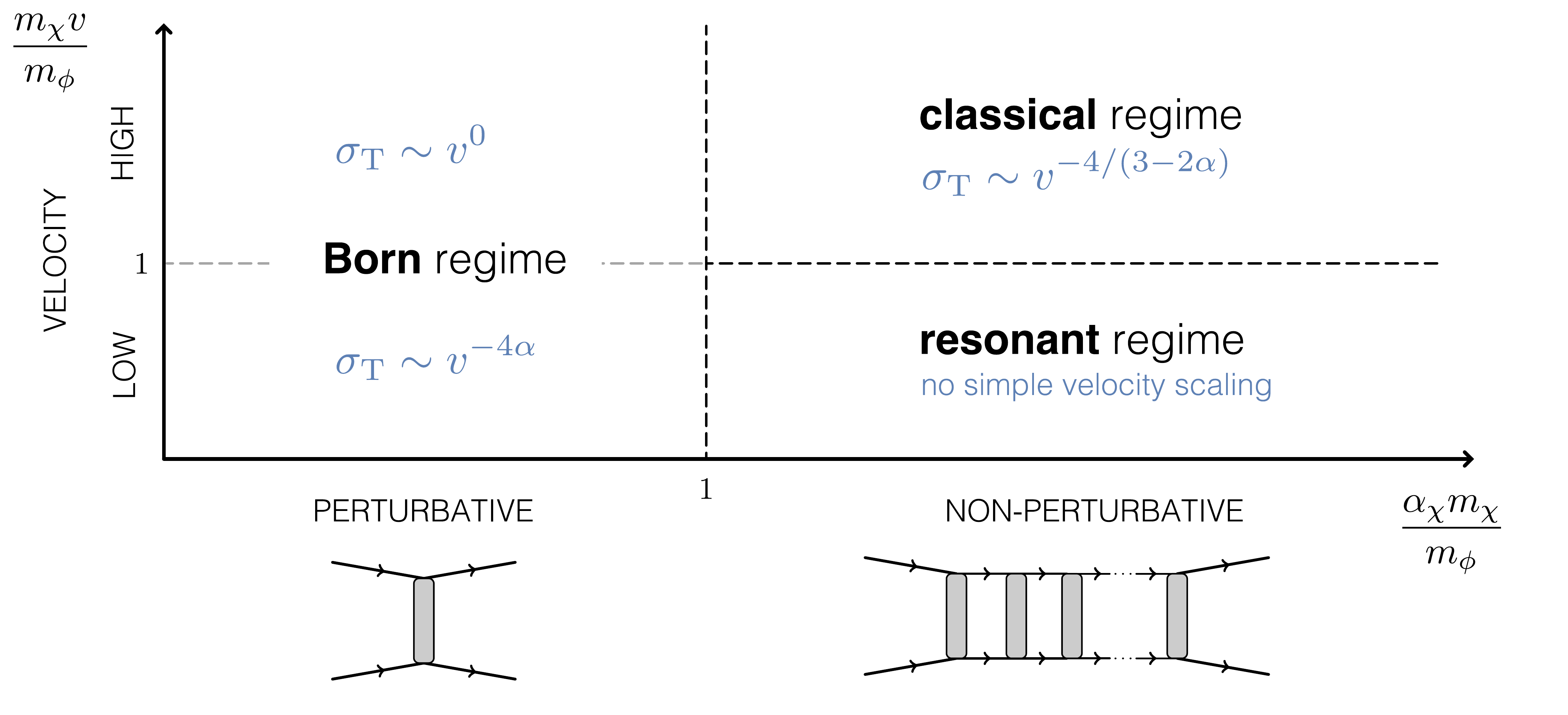}
\caption{Regimes of self-interacting dark matter. The horizontal axis measures whether the ladder of mediator exchanges can be approximated by a single mediator exchange. The vertical axis is a measure of the velocity. The figures of merit are scaled by the ratio of the dark matter mass to the mediator mass (or mass gap) so that the regimes are limits relative to unity. The perturbative regime is described by the Born approximation over the range of all velocities, whereas the non-perturbative regime is separated into a classical regime at high velocities and a resonant regime at low velocities. 
Blue: asymptotic velocity scaling of the transfer cross section $\sigma_\text{T}$ in the continuum-mediated scenario. No simple scaling exists in the resonant regime. 
The standard case of a single \acro{4D} mediator corresponds to $\alpha= 1$.
}
\label{fig:regimes}
\end{figure}

We summarize key results of self-interacting dark matter phenomenology; see Ref.~\cite{Tulin:2017ara} for a detailed review.
Long-range dark matter self-interactions affect halo density profiles by thermalizing the inner halo and reducing the central density.
The effect of dark matter self-interactions on halos depends on the scattering rate, $\sigma v (\rho_\chi/m_\chi)$. Since the dark matter density $\rho_\chi$ and the relative velocity $v$ are known for the relevant astrophysical systems, the figure of merit is the ratio of the cross section to the dark matter mass, $\sigma/m_\chi$. Dwarf spheroidal galaxies have low relative velocities ($v\sim 10~\text{km/s}$) and exhibit small-scale structure anomalies that could be explained by sufficient self-interactions~\cite{Kaplinghat:2015aga,Tulin:2013teo, Dave:2000ar}. On the other hand, galaxy clusters have large relative velocities ($v\sim 1500~\text{km/s}$) and typically set {upper} bounds on these interactions:
\begin{align}
  \left(\frac{\sigma}{m_\chi}\right)_\text{dwarf} 
  &\sim 1~\frac{\text{cm}^2}{\text{g}} 
  &
  \left(\frac{\sigma}{m_\chi}\right)_\text{cluster} 
  &\lesssim 0.1~\frac{\text{cm}^2}{\text{g}} \ .
\end{align}
The small-scale target and large-scale upper limit are simultaneously satisfied in self-interacting dark matter models due to the velocity dependence of the cross section.
In fact, a more relevant quantity for fitting to astronomical observations is the \emph{transfer cross section}, which is weighted by the amount of transverse momentum transferred between dark matter particles: 
\begin{align}
\sigma_\text{T}= \int d\Omega \frac{d\sigma}{d \Omega} \left(1-\cos \theta\right) \ .
\end{align}
This accounts for the fact that back-to-back scattering does not change the distribution of energy between halo dark matter particles.\footnote{A more symmetric treatment is to use the viscosity cross section, $\sigma_\text{V} = \int d\Omega \sin^2\theta d\sigma/d\Omega$\ . In order to map to the standard self-interacting dark matter literature, we use $\sigma_\text{T}$ which differs from $\sigma_\text{V}$ by at most an $\mathcal O(1)$ factor~\cite{Tulin:2017ara}.}
The transfer cross section is the figure of merit for determining the effect of self-interactions on the dark matter halo profile. The behavior is classified according to regimes along two axes: perturbativity and relative velocity, see Figure~\ref{fig:regimes}.

\paragraph{Perturbativity.} 
The horizontal axis of Figure~\ref{fig:regimes} distinguishes whether the transfer cross section is accurately described by the exchange of a single mediator (perturbative) or otherwise requires a sum over ladder diagrams (non-perturbative).  In the former case, one may use the Born approximation.
For a \acro{4D} dark sector with a single mediator of mass $m_\phi$ and corresponding potential $V\sim \alpha_\chi e^{-m_\phi r}/r$, these regimes correspond to
\begin{align}
  \text{Born:}\quad \frac{\alpha_\chi m_\chi}{m_\phi} &\ll 1
  &
  \text{non-perturbative:}\quad 
  \frac{\alpha_\chi m_\chi}{m_\phi} &\gg 1 \ .
  \label{eq:SIDM:Born}
\end{align} 
The weighted coupling, $\alpha_\chi m_\chi/m_\phi$, measures whether the Hamiltonian eigenstates are distorted from the non-interacting case~\cite[(7.2.13)]{Sakurai:1167961}. The sum over ladder diagrams in the non-perturbative regime reproduces the distortions of the asymptotic states relative to the non-interacting eigenstates. 

\paragraph{Velocity.}  The horizontal axis of Figure~\ref{fig:regimes} distinguishes whether the dark matter relative velocity (kinetic energy) is large enough to ignore the effect of the mediator mass. When the theory is perturbative, the Born approximation may be applied across the entire range of velocities. On the other hand the velocity separates the non-perturbative case into two regimes according to whether the de~Broglie wavelength (inverse momentum) $(m_\chi v)^{-1}$ is comparable to the screening length (inverse mediator mass), $m_\phi^{-1}$:
\begin{align}
  \text{resonant:}\quad \frac{m_\chi v}{m_\phi} &\ll 1   
  &
  \text{classical:}\quad \frac{m_\chi v}{m_\phi} &\gg 1
  \ .
  \label{eq:SIDM:Resonant}
\end{align} 
The \emph{classical regime} is the case where the zeroth-order \acro{WKB} approximation is valid; this corresponds to the $\hbar\to 0$ limit. For a \acro{4D} dark sector with a single mediator of mass $m_\phi$, the classical regime is the case where the mediator mass is negligible and the theory reproduces the case of Rutherford/Coulomb scattering. In contrast, in the \emph{resonant regime} the Yukawa factor deforms the potential away from the Coulomb limit enough to support quasi-bound states. In this regime, one must numerically solve the Schr\"odinger equation in a partial wave expansion to determine the transfer cross section~\cite{Tulin:2013teo}.

Figure~\ref{fig:regimes} shows that  $v<\alpha_\chi$ is a necessary condition for the existence of resonances over some range of $v$. Conversely, $v>\alpha_\chi$ is a sufficient condition for having no resonance for any value of $v$.

\subsection{Analytical Behavior of a Continuum Mediator}

The transfer cross section from a continuum-mediated potential can be mapped onto the self-interacting dark matter regimes described above and pictured in Figure~\ref{fig:regimes}.

\paragraph{Effective coupling.} The condition for perturbativity depends on the dark fine structure constant, which is $\alpha_\chi = g_\chi^2/4\pi$ for a single \acro{4D} mediator. We can identify an effective fine structure constant $\alpha_\chi^\text{eff}$ for our continuum mediator.
For bulk mass parameters $1/2 < \alpha < 1$, 
\begin{align}
    \alpha_{\chi}^{\text{eff}}
    =
    \frac{\lambda^2 m_1}{4\pi k} 
    \sum_n \frac{f_n^2(z_{\text{UV}})}{m_n}
    \approx
    \frac{\lambda^2}{4\pi}
    \left[\frac{4}{2\alpha-1}\frac{1}{\Gamma(1-\alpha)^2}\right]
    \left(\frac{m_1}{2k}\right)^{2-2\alpha}
    \ .
    \label{eq:Born:alphaeff}
\end{align}
This follows from applying the Born approximation condition~\eqref{eq:SIDM:Born} to the sum of Kaluza--Klein potentials  \eqref{eq:V_KK_prop}. On the right-hand side we use the spectral representation~\eqref{eq:spectral:discontinuity} 
to evaluate the sum.
 This calculation is detailed in Appendix~\ref{sec:BornValidity}, where we also discuss the limiting cases where the bulk masses satisfy $\alpha = 1/2$ and $\alpha=1$. We note that the factor of  $(m_1/k)^{2-2\alpha}\sim(\mu/k)^{2-2\alpha}$ in \eqref{eq:Born:alphaeff} 
 suppresses the effective coupling compared to a na\"ive estimate $\lambda^2/4\pi$.

\paragraph{Transfer cross section regimes.} The self-interaction regimes in Figure~\ref{fig:regimes} are mapped to the continuum-mediated scenario by identifying the mediator mass with the lightest \acro{KK} mode mass (the mass gap), $m_\phi \to m_1$.  We find that the effective coupling $\alpha_\chi^\text{eff}$ replaces $\alpha_\chi$ in the demarcation of the perturbative (Born) and non-perturbative regimes,
\begin{align}
  \text{Born:}\quad 
  \frac{\alpha_\chi^\text{eff} m_\chi}{m_1} &\ll 1
  &
  \text{non-perturbative:}\quad 
  \frac{\alpha_\chi^\text{eff} m_\chi}{m_1} &\gg 1 \ .
  \label{eq:SIDM:Born:eff}
\intertext{
We can likewise divide the non-perturbative regime into the classical and resonant regimes:
}
  \text{Classical:}\quad \frac{m_\chi v}{m_1} &\gg 1
  &
  \text{Resonant:}\quad \frac{m_\chi v}{m_1} &\ll 1 
  \ .
  \label{eq:CMSIDM:Resonant}
\end{align}
Unlike the case of a Yukawa potential, there are no analytic expressions for the transfer cross section in the entire non-perturbative classical regime.\footnote{See Ref.~\cite{Chiron:2016lzn} for a discussion of scattering in the  limit of no mass gap.}  We show the scaling of the transfer cross section for the classical regime in the small mass gap/high velocity limit and give a closed form result in the low velocity regime below, see Appendix~\ref{app:classical}.

\paragraph{Continuum-mediated Born regime.}

\begin{figure}
\centering
\begin{subfigure}[t]{0.49\textwidth}
  \centering
    \includegraphics[width=\textwidth]{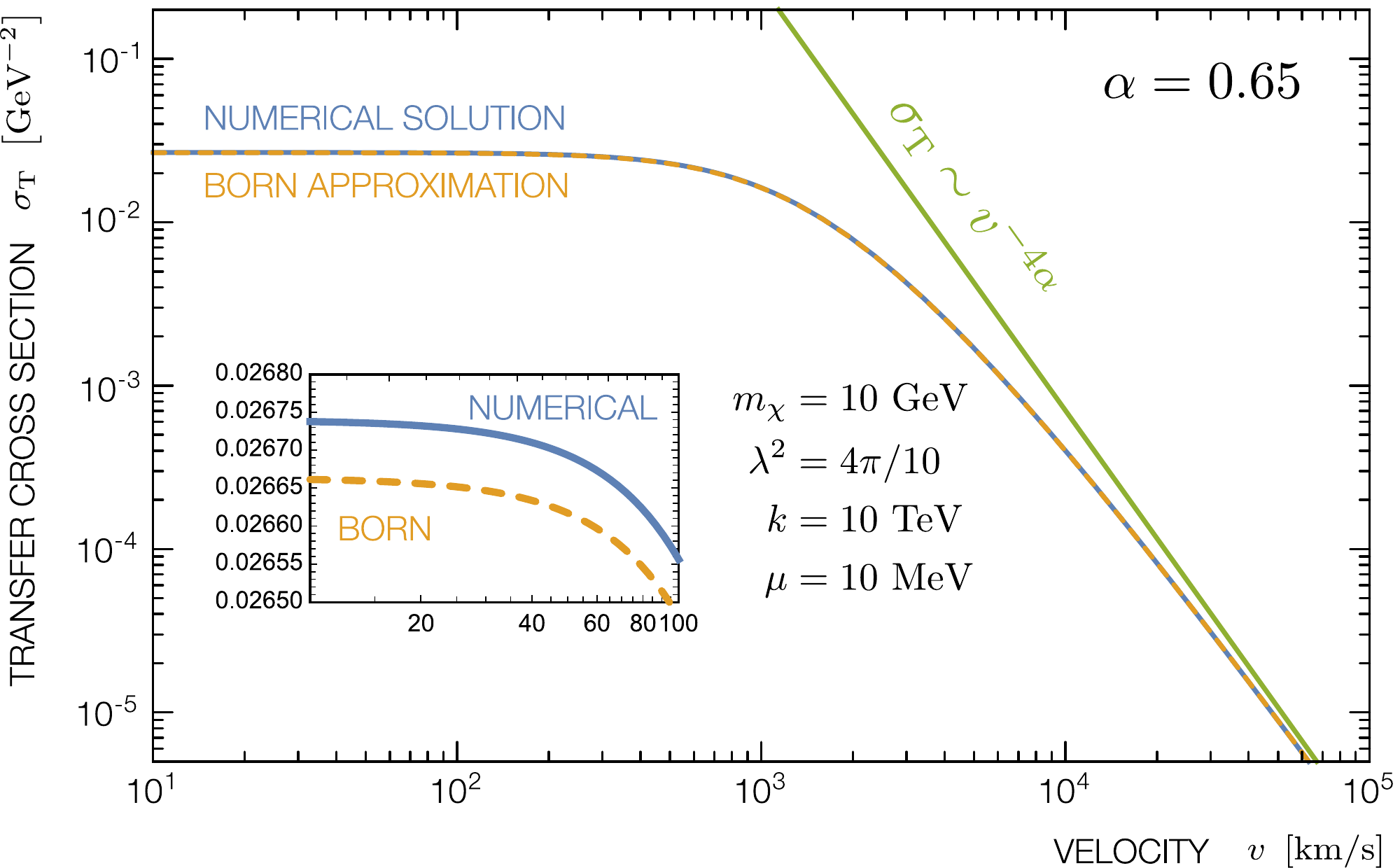}
\end{subfigure}
\hfill
\begin{subfigure}[t]{0.49\textwidth}  
  \centering 
    \includegraphics[width=\textwidth]{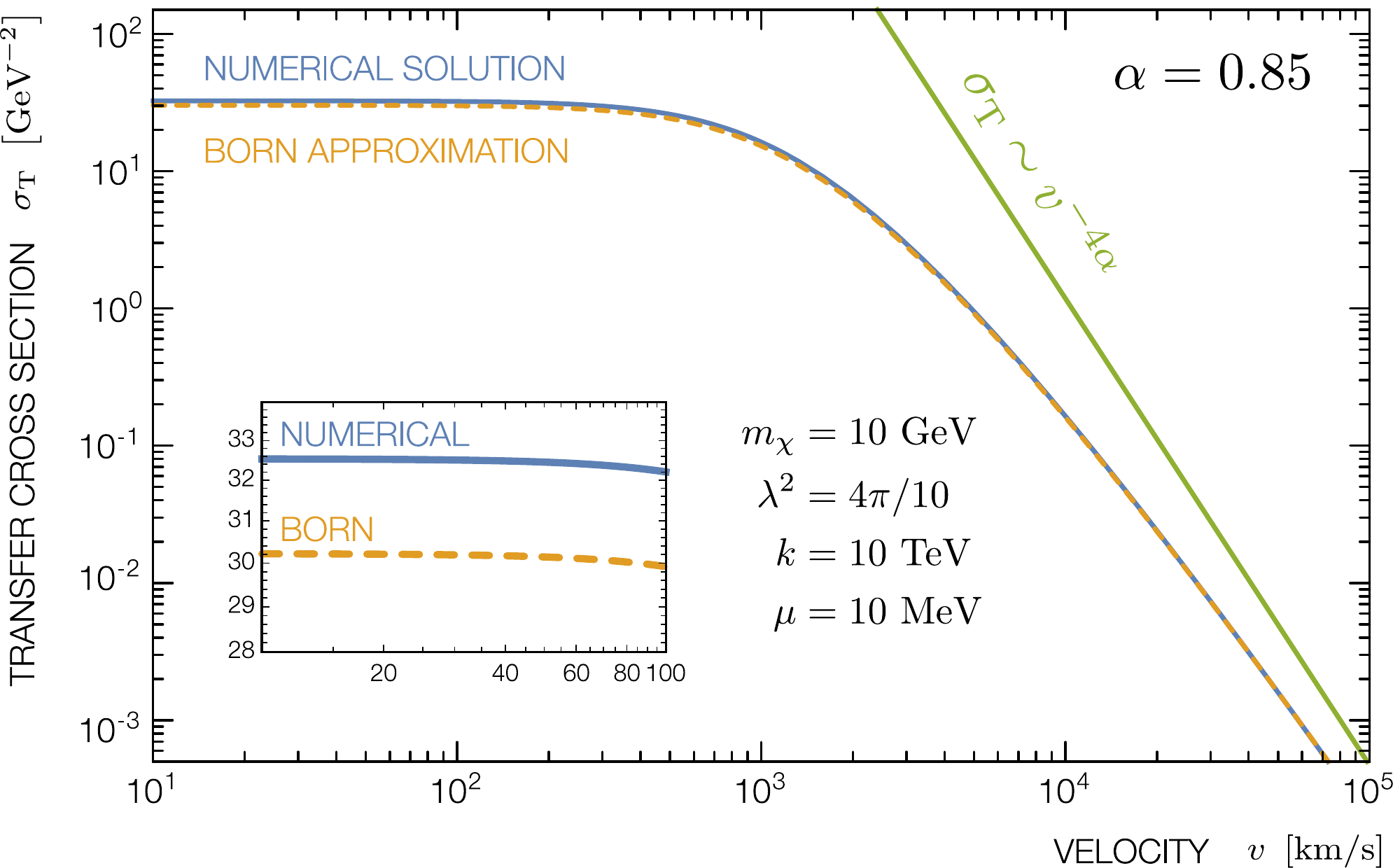}
\end{subfigure}
\caption{Velocity dependence of the transfer cross section in the Born regime. Comparison between the Born approximation and (blue/solid) and the numerical result from a sum of partial waves (orange/dashed). The results asymptotically scale like $v^{-4\alpha}$ at large velocity (green).}
\label{fig:Born_v_scaling_v2}
\end{figure}

In the Born regime, the transfer cross section computed perturbatively from the $1/2 < \alpha < 1$ continuum-mediated potential \eqref{eq:Vgamma} is 
\begin{align}
    \left(\frac{d\sigma}{d\Omega}\right)^{\text{Born}}
    &=
    \frac{\left(\alpha_{\chi}^{\text{eff}}\right)^2 m_\chi^2}{4m_1^4}\left(2\alpha-1\right)^2 
    \left._2 F_1\right. \left(1,\alpha;1+\alpha;-|\vec{q}|^2/m_1^2\right)^2
    \ ,
    \label{eq:CMSIDM:sigmaBorn}
\end{align}
where 
$_2 F_1 \left(1,\alpha;1+\alpha;-|\vec{q}|^2/m_1^2\right)$
is the hypergeometric function that encodes the mass gap. The transferred three-momentum, $\vec{q}$, satisfies 
$|\vec{q}|^2= \tfrac{1}{2}m_{\chi}^2 v^2(1-\cos \theta)$ 
where $\theta$ is the scattering angle in the center of mass frame. We compute the angular integral numerically. 

We may examine \eqref{eq:CMSIDM:sigmaBorn} in the limits of large and small transferred three-momentum. For a transfer momentum much larger than the mass gap, $|\vec{q}| \gg m_1$, the transfer cross section is
\begin{align}
\sigma_\text{T}^{\text{Born}}
&\approx
\frac{\lambda^4 m_{\chi}^2}{16\pi k^4(1-\alpha)}
\left[\frac{\Gamma(\alpha)}{\Gamma(1-\alpha)}\right]^2
\left(\frac{2k}{m_{\chi} v}\right)^{4\alpha} 
&
|\vec{q}| &\gg m_1
\ .
\label{eq:CMSIDM:sigmaBorn:largeq}
\\
\intertext{This matches the result from the gapless potential, \eqref{eq:Vnogamma}. In the opposite limit, $|\vec{q}| \ll m_1$, the transfer cross section approaches a constant:
}
    \sigma_\text{T}^{\text{Born}}
    &\approx 
    \frac{\lambda^4 m_\chi^2}{64\pi^2 \alpha^2 k^4 \Gamma(1-\alpha)^4} 
    \left(\frac{2 k}{m_1}\right)^{4 \alpha} 
    &
    |\vec{q}| &\gg m_1
    \ .
    \label{eq:CMSIDM:sigmaBorn:smallq}
\end{align}
Figure~\ref{fig:Born_v_scaling_v2} compares these asymptotic behaviors to a numerical solution.

Early astrophysical simulations of self-interacting dark matter assumed a constant $\sigma_\text{T}$ and found that the cross sections required to address small-scale structure anomalies were inconsistent with bounds from the upper limits set by galaxy cluster collisions. One of the key insights of Ref.~\cite{Tulin:2013teo} was that suppression of the cross section at transfer momenta relative to a light mediator would alleviate this tension. In the continuum-mediated scenario, we see that the bulk mass parameter $\alpha$ controls the velocity-scaling in the high-velocity Born limit. This parametric control is not possible for the exchange of a single mediator.

\paragraph{Continuum-mediated classical regime.}
Unlike in the Born regime, in the classical regime closed form results for the transfer cross section do not follow from straightforward calculation.
While in the case of a Yukawa potential closed form expressions can be determined for the entire non-perturbative classical regime, see~{e.g.}~Ref.~\cite[eqn.~(7)]{Tulin:2013teo}, analytic expressions for the continuum mediated transfer cross section are harder to come by. In the limit of a small mass gap/large velocity,  one can determine its velocity dependence. In the opposite low velocity limit, one finds a closed form expression. The calculations are detailed in Appendix~\ref{app:classical}.

One can write the transfer cross section in this regime as an integral over the impact parameter $\rho$. It is convenient to introduce the dimensionless quantities  $\xi = \rho/\rho_0$, where $\rho_0$ is a characteristic length scale defined from the potential~\eqref{eq:Vgamma},
\begin{align}
    &\sigma_\text{T}^{\text{classical}}=2\pi \rho_0^2 \int_0^\infty \left[1-\cos\theta(\xi, m_1 \rho_0)\right]\xi d\xi
    &
    &\rho_0&\equiv \left[\frac{\lambda^2}{2 \pi^{3/2} m_\chi v^2 k^{2-2\alpha} }\frac{\Gamma(3/2-\alpha)}{\Gamma(1-\alpha)}\right]^{\tfrac{1}{3-2\alpha}}
    \ .
    \label{eq:CMSIDM:sigmaclassical}
\end{align}
When $m_1 \rho_0 \ll1$, corresponding to the small mass gap/high velocity limit, the scattering angle $\theta$ is a function of the ratio $\xi$ only~\cite{Chiron:2016lzn}. In this case, the transfer cross section depends on a non-integer power of the relative velocity, $-4/(3-2\alpha)$. A finite mass gap induces corrections to this scaling.

While this scaling holds in the small mass gap/high velocity limit of the classical regime, an approximate closed form solution for the transfer cross section can be computed for lower velocities. Following the methodology of Ref.~\cite{Khrapak:2003kjw}, we calculate the transfer cross section in terms of the parameter
\begin{align}
    \beta=\frac{2\alpha_{\chi}^\text{eff}m_1}{v^2 m_\chi}(2\alpha-1)
    \ .
    \label{eq:beta}
\end{align}
In the limit $\beta \gg 1$, the transfer cross section is found to approximately be
\begin{align}
     \sigma_\text{T}^{\text{classical}}\approx
    \frac{\pi}{m_1^2}\left[1+\log\left(\frac{\beta}{\log\beta}\right)-\frac{(2\alpha-1)}{\log\beta}+\frac{\left(2\alpha-\frac{3}{2}\right)}{\log\left(\frac{\beta}{\log\beta}\right)}\right]^2 
    \label{eq:classical:approx}
 \, .
\end{align}
See Appendix\,\ref{app:classical:lowv} for details. 
Our analytical result is shown to be in good agreement with the numerical solution to the Schr\"{o}dinger equation, see Figure~\ref{fig:sigmavsmuregions}.

\paragraph{Summary of Velocity Scaling}
We summarize the velocity scaling in the different regimes:
\begin{align}
    \sigma_\text{T} \sim
    \begin{cases}
    v^{0} & \text{Born (low velocity)} \\
    v^{-4\alpha} & \text{Born (high velocity)} \\
    v^{-4/(3-2\alpha)} & \text{Classical} \\
    \text{no simple scaling} & \text{Resonant}
    \end{cases}
    \ 
    \label{eq:scaling}
\end{align}
The dependence on the bulk mass parameter $\alpha$ is a key difference from the standard \acro{4D}, single mediator case. The \acro{4D} scenario corresponds to  $\alpha =1$.

\subsection{Numerical Methodology and Results}
\label{sec:numerical}

To make quantitative statements about the transfer cross section that extend to the classical and resonant regimes, we numerically solve the Schr\"{o}dinger equation using a partial wave expansion, 
\begin{align}
  \sigma_\text{T}
  &= 
  \frac{4\pi}{\left(m_\chi v/2\right)^2} 
  \sum_\ell (\ell+1) 
  \sin^2\left(\delta_{\ell+1}-\delta_\ell\right) 
  \label{eq:partial:waves}
  \ ,
\end{align}
where $\delta_\ell$ is the scattering phase shifts partial wave $\ell$. 
We follow the methodology of Ref.~\cite{Tulin:2013teo} with a more relaxed numerical algorithm described in Appendix \ref{app:SIDM}.

For bulk mass parameters $\alpha \leq 1/2$, the potential dominates over the repulsive centrifugal barrier for $r \to 0$. In this case one must place a short distance cutoff on $r$ that encodes data from the \acro{UV} completion. Practically, the partial wave expansion converges poorly and becomes numerically intractable for potentials more singular than $r^{-2}$. As such, we restrict the bulk mass parameter to the range $1/2 < \alpha < 1$, where the upper limit is the theoretical upper limit established in Section~\ref{se:params}.

\paragraph{Realization of the transfer cross section regimes.}

The scattering rate density relevant for thermalizing the cores of dark matter halos is the transfer cross section times the dark matter number density,  $\sigma_\text{T} n_\chi \sim \sigma_{T} \rho_\chi /m_\chi$. The dark matter density $\rho_\chi$ is a measured input, so a useful figure of merit is the ratio $\sigma_\text{T}/m_\chi$, for which the typical value required for small-scale structure is $\sigma_\text{T}/m_\chi \sim \mathcal O(1)$.

\begin{figure}[ht]
\centering
\begin{subfigure}[b]{0.49\textwidth}
  \centering
  \includegraphics[width=\textwidth]{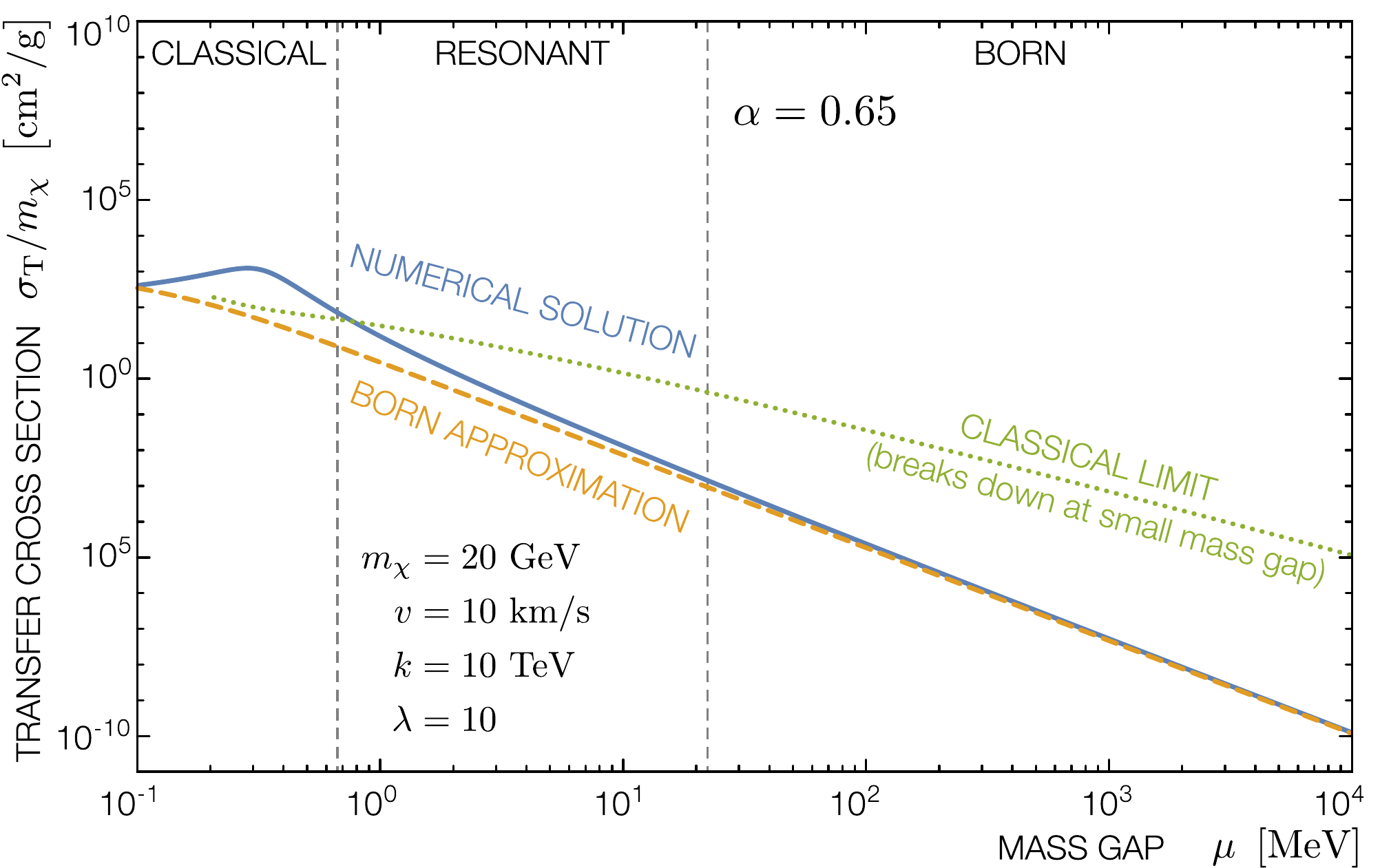}
\end{subfigure}
\hfill
\begin{subfigure}[b]{0.49\textwidth}  
  \centering 
    \includegraphics[width=\textwidth]{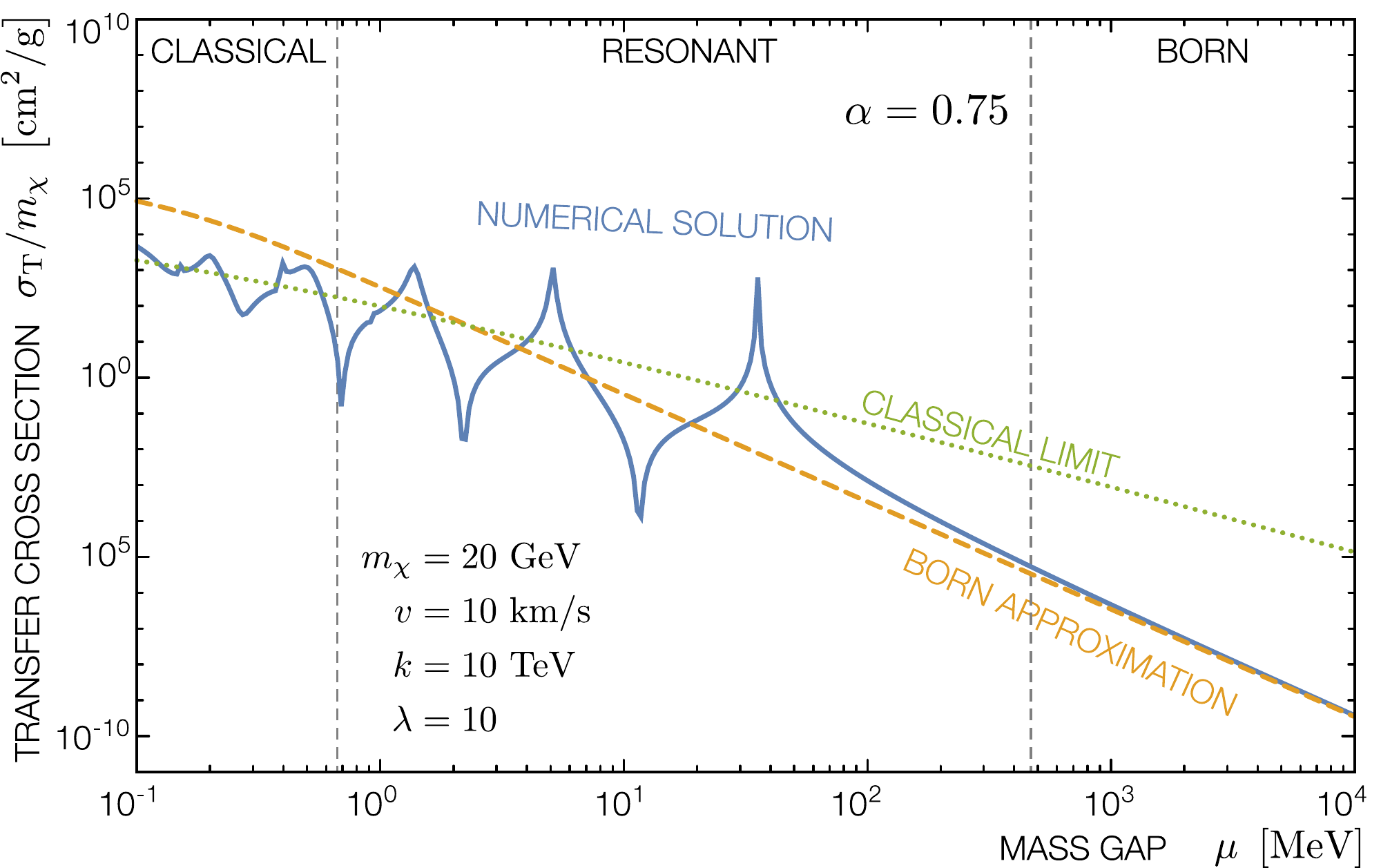}
\end{subfigure}
\vskip\baselineskip
\begin{subfigure}[b]{0.49\textwidth}   
    \centering 
    \includegraphics[width=\textwidth]{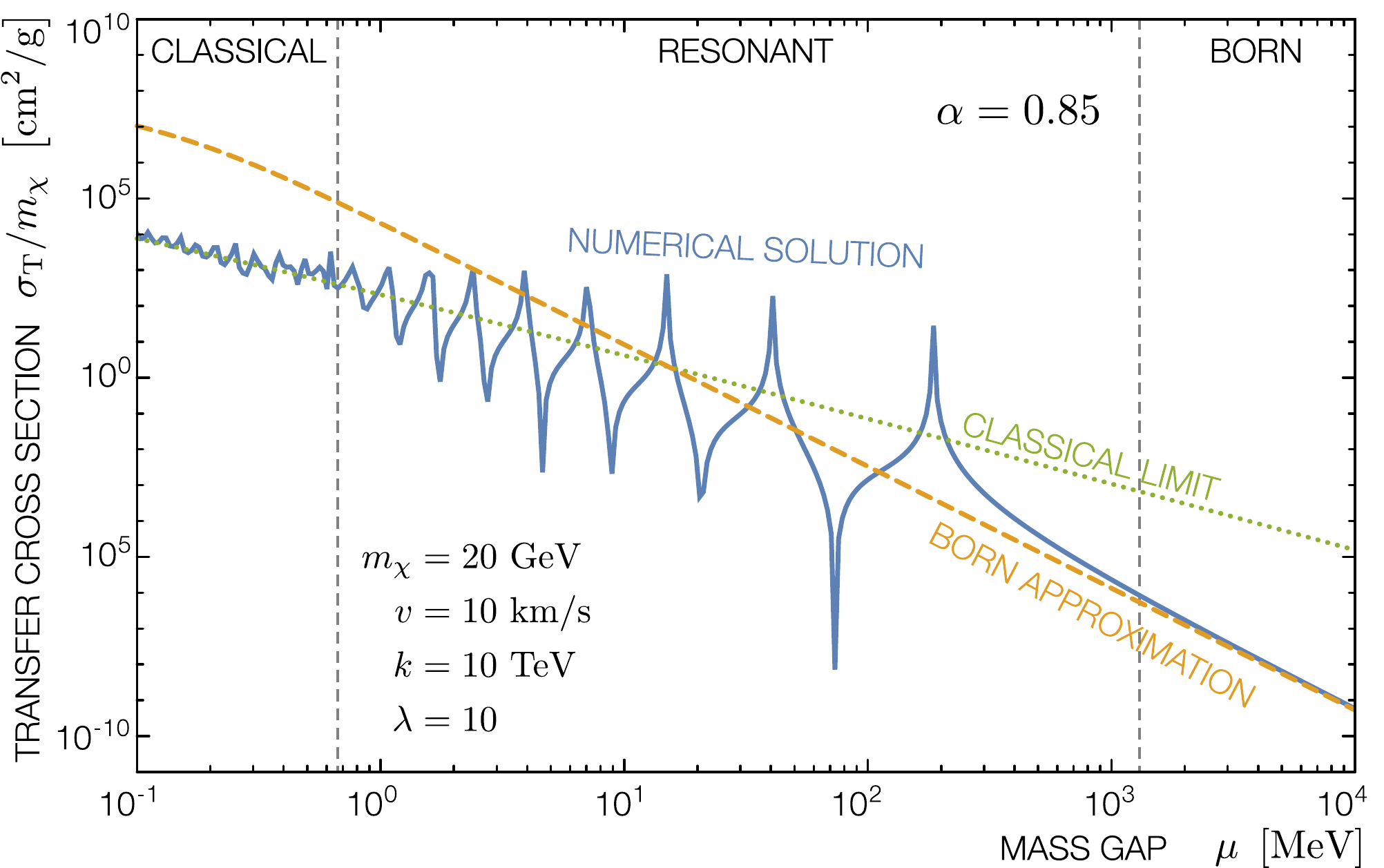}
\end{subfigure}
\hfill
\begin{subfigure}[b]{0.49\textwidth}   
    \centering 
    \includegraphics[width=\textwidth]{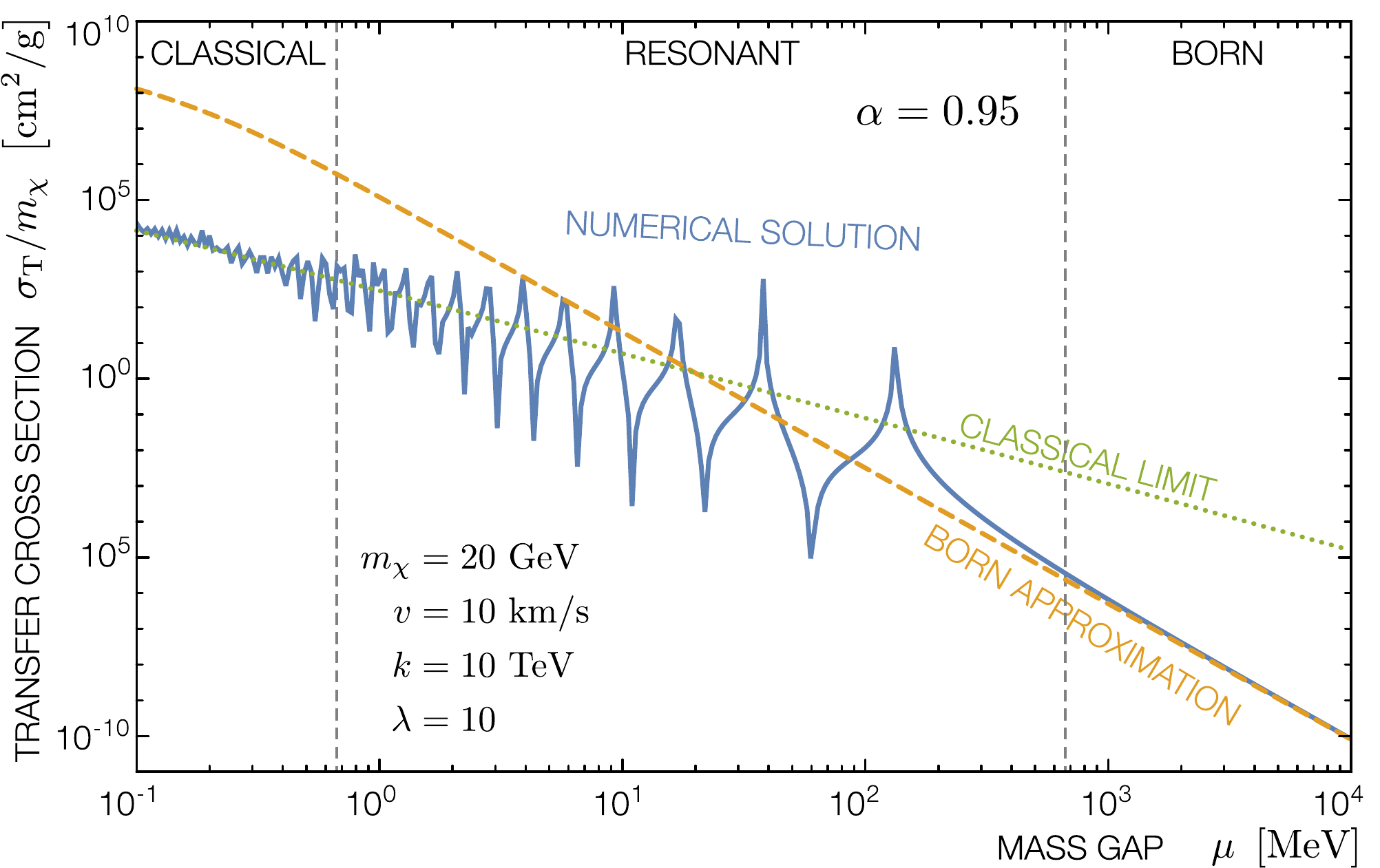}
\end{subfigure}
\caption{Comparison of the numerically calculated transfer cross section to the analytic approximations introduced in Figure~\ref{fig:regimes}. The general behavior displays distinct regimes, similar to that of a single mediator, see e.g.~Ref.~\cite[Fig.~2]{Tulin:2013teo}.  
The blue line is the numerical solution.
Orange (dashed)/green (dotted) lines correspond to analytic Born/classical approximations valid in their respective regimes; \eqref{eq:CMSIDM:sigmaBorn} and \eqref{eq:classical:approx} .
}
\label{fig:sigmavsmuregions}
\end{figure}

To demonstrate the self-interacting dark matter regimes discussed in this section,  Figure~\ref{fig:sigmavsmuregions} scans  the ratio $\sigma_\text{T}/m_\chi$ over the mass gap $\mu \sim m_1$ for different values of the bulk mass parameter $\alpha$. These one-dimensional plots are slices of the transfer cross section over the two-parameter space of regimes in Figure~\ref{fig:regimes}. 
For each of these plots, large values of $\mu$ correspond to the low-velocity Born regime. Figure~\ref{fig:sigmavsmuregions} confirms the agreement with the Born approximation in this limit. As one decreases $\mu$, one moves upward and to the right in Figure~\ref{fig:regimes}, crosses the resonant regime with pronounced peaks in the cross section, and finally enters the classical regime. 
%
Figure~\ref{fig:sigmavsmuregions} confirms that our approximate analytical results in the classical regime agree with the numerical solution to the Schr\"{o}dinger equation. For smaller values of  $\alpha$, our approximation for the transfer cross section in the classical regime breaks down as expected. 
%

\begin{figure}
\centering
\includegraphics[width=\textwidth]{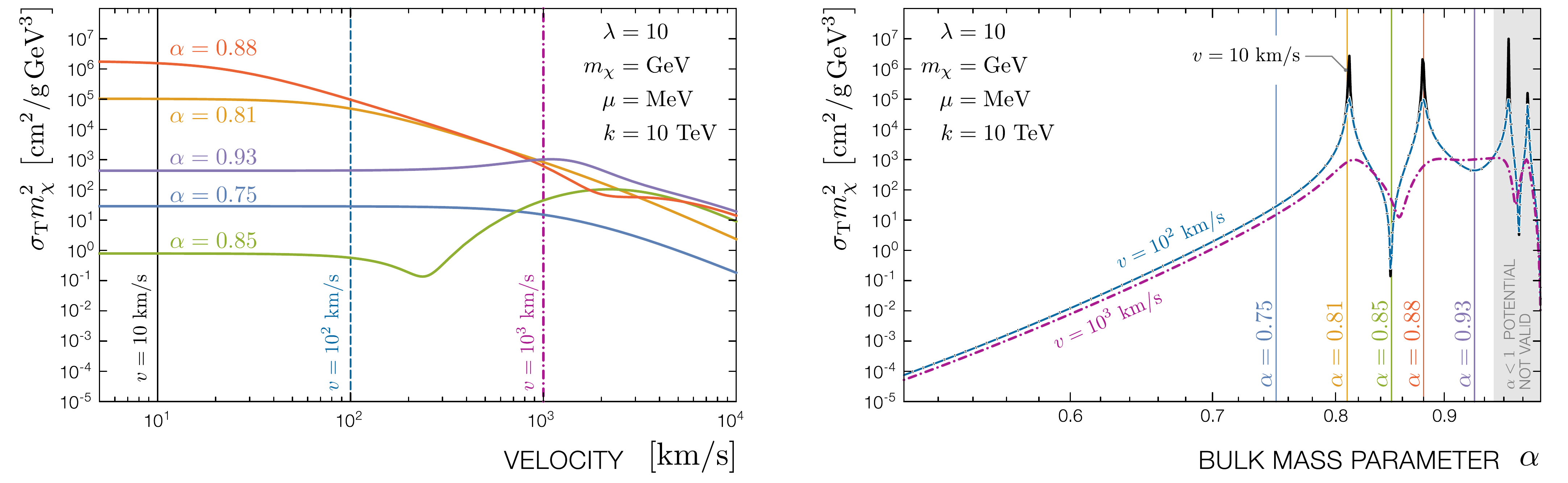}
\caption{Transfer cross section as a function of relative dark matter velocity $v$ (left) and bulk mass parameter $\alpha$ (right). 
The plots demonstrate the presence of resonances and anti-resonances.
Vertical markers identify parameters used in the opposite plot.
}
\label{fig:sigmaalphadependence}
\end{figure}

\paragraph{Resonances and the bulk mass parameter.}
The resonance structure of transfer cross section can be very sensitive to the bulk mass parameter $\alpha$. This parameter has no analog in \acro{4D} self-interacting dark matter models with a single mediator and represents a new model degree of freedom to affect phenomenology. The bulk mass feeds into both the overall effective coupling $\alpha_\chi^\text{eff}$ \eqref{eq:Born:alphaeff} and the slope of the potential at short distances \eqref{eq:Vgamma}.
We demonstrate the $\alpha$-sensitivity of the transfer cross section  with a set of benchmark parameters in Figure~\ref{fig:sigmaalphadependence}. The two plots scan over both $\alpha$ and the relative velocity $v$ to highlight the interplay in the resonance structure.

We remark that Figure~\ref{fig:sigmaalphadependence} plots $\sigma_\text{T} m_\chi^2$ to make it straightforward to use scaling relations to connect results to different parameters. 
The partial wave expansion \eqref{eq:partial:waves} makes it clear that $\sigma_\text{T} \sim m_\chi^{-2}$. The additional $m_\chi$ dependence of the phase shifts $\delta_\ell$ depend only on the ratios $m_\chi/m_1$ and $m_\chi/k$; see Appendix \ref{app:SIDM:method}. Thus the plots are unchanged by the following rescaling of parameters by $\eta$:
\begin{align}
  m_\chi &\to \eta m_\chi
  &
  k &\to  \eta k
  &
  \mu &\to  \eta \mu \ .
\end{align}
This extends the scaling arguments in Ref.~\cite{Tulin:2013teo} to the case of a continuum mediator.

\subsection{Comparison to Astrophysical Data}
\begin{figure}[t]
\includegraphics[width=\textwidth]{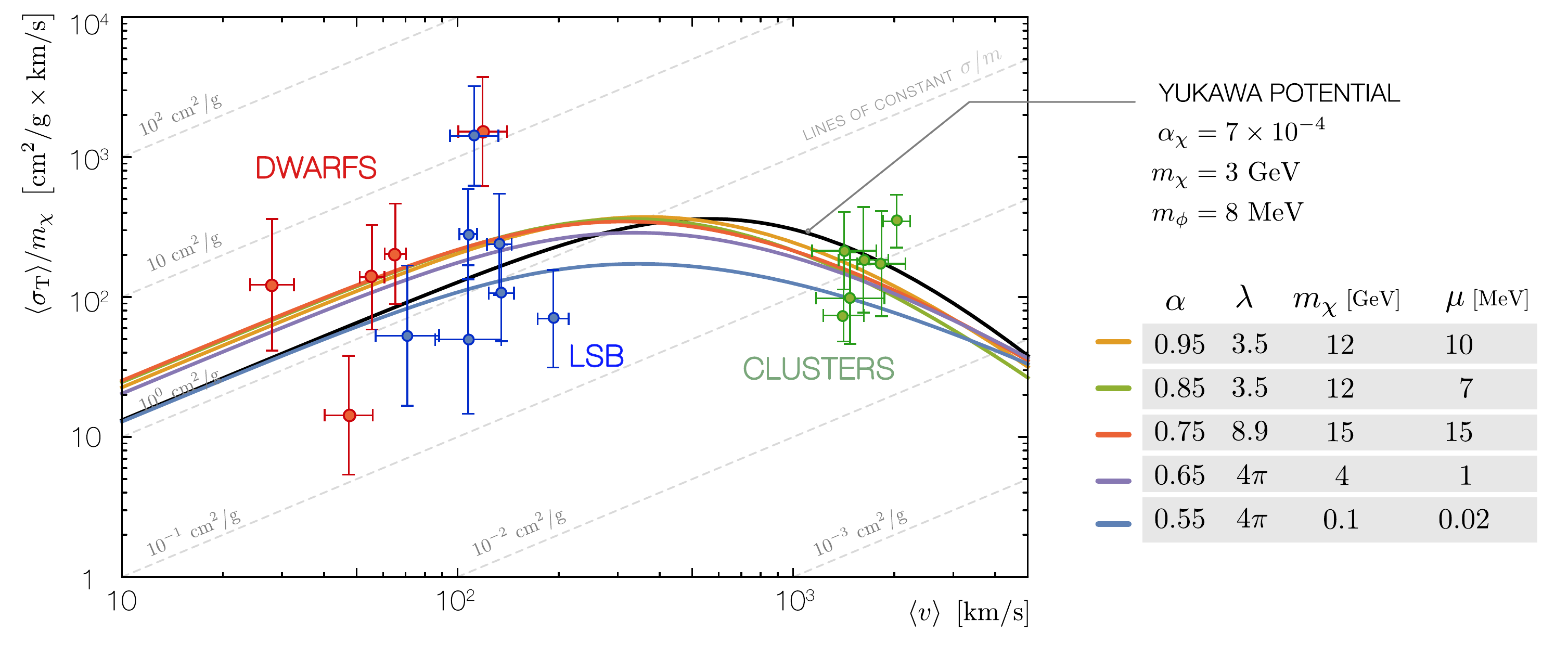}
\caption{Velocity dependence of the thermally averaged transfer cross section. The parameters are chosen to be reasonably fit astronomical  data. A benchmark \acro{4D} self-interacting dark matter model with a scalar mediator is shown for comparison. 
The data points for velocities $v\sim 30-200~\text{km}/\text{s}$ are determined from the observed rotation curves of dwarf (red) and low-surface brightness galaxies (blue) respectively; points for velocities $v\gtrsim 10^3~\text{km}/\text{s}$ correspond to galaxy clusters (green) and are determined from stellar line-of-sight velocity dispersion data~\cite{Kaplinghat:2015aga}.}
\label{fig:sigmavastrodata}
\end{figure}

The scattering rate, $\sigma_\text{T} v (\rho_\chi/m_\chi)$, determines the energy transfer in dark matter halos. 
Figure~\ref{fig:sigmavastrodata} plots the figure of merit $\sigma_\text{T} v/m_\chi$ for a set of benchmark parameters compared to the astronomical data points presented in Ref.~\cite{Tulin:2013teo}. 
The plot includes a Yukawa potential to represent the \acro{4D} single mediator case.
These benchmarks correspond to a range of bulk mass parameters $\alpha$. The other parameters are set to give fits of comparable $\chi^2$ to the Yukawa potential. We remark that this is not a scan to minimize $\chi^2$ and is only meant to demonstrate the range of parameter possibilities that can fit the data.
The ultimate cause for the dark matter halo density profile observations may partially (or wholly) include contributions from baryonic feedback, see Ref.~\cite{Bullock:2017xww} for a recent status report. Thus one may conservatively interpret the data in Figure~\ref{fig:sigmavastrodata} as upper limits on the transfer cross section for a viable model. 

The mass hierarchy between the dark matter and lightest \acro{KK} mass is comparable to that of the benchmark \acro{4D} self-interacting dark matter theory, $m_\chi/m_\phi \sim \mathcal O(10^3)$.
While $\lambda$ can vary over a few orders of magnitude, the effective coupling $\alpha_{\chi}^{\text{eff}}$ remains approximately constant for the benchmarks in Figure~\ref{fig:sigmavastrodata}. 
In the extreme case $\alpha=0.55$, the effective coupling  $\alpha_\chi^\text{eff}$ is small compared to the other benchmarks.  This is compensated by a small dark matter mass.
This interplay between $\alpha$ (\textit{i.e.} the bulk mass) and the dark matter--mediator coupling $\lambda$ may be used, for example, to maintain the fit to data in Figure~\ref{fig:sigmavastrodata} while adjusting a mediator--Standard Model coupling $\lambda_\text{SM}$ to realize other phenomenology.

We remark that while we restrict to the range of bulk mass parameters $1/2 < \alpha < 1$ for theoretical reasons, we also observe that the model phenomenology gives a mild preference for values away from the lower limit. For small values of $\alpha \approx 0.55$, reproducing the desired $\langle\sigma_\text{T}v\rangle/m_\chi$ behavior requires sub-GeV dark matter and a Kaluza--Klein scale of $\mathcal O(\text{10}~\text{keV})$, which may cause tension with cosmological constraints~\cite{Cyburt:2015mya}. On the other hand, for large values of $\alpha\to 1$, one must take care to use the appropriate limiting form of the bulk propagator, as discussed in Section~\ref{se:asymptotics}. Since the $\alpha=1$ case essentially describes a single \acro{4D} mediator, this limit approaches that of ordinary self-interacting dark matter models.

\begin{figure}[t]
\centering
\begin{subfigure}[t]{0.49\textwidth}
  \centering
    \includegraphics[width=\textwidth]{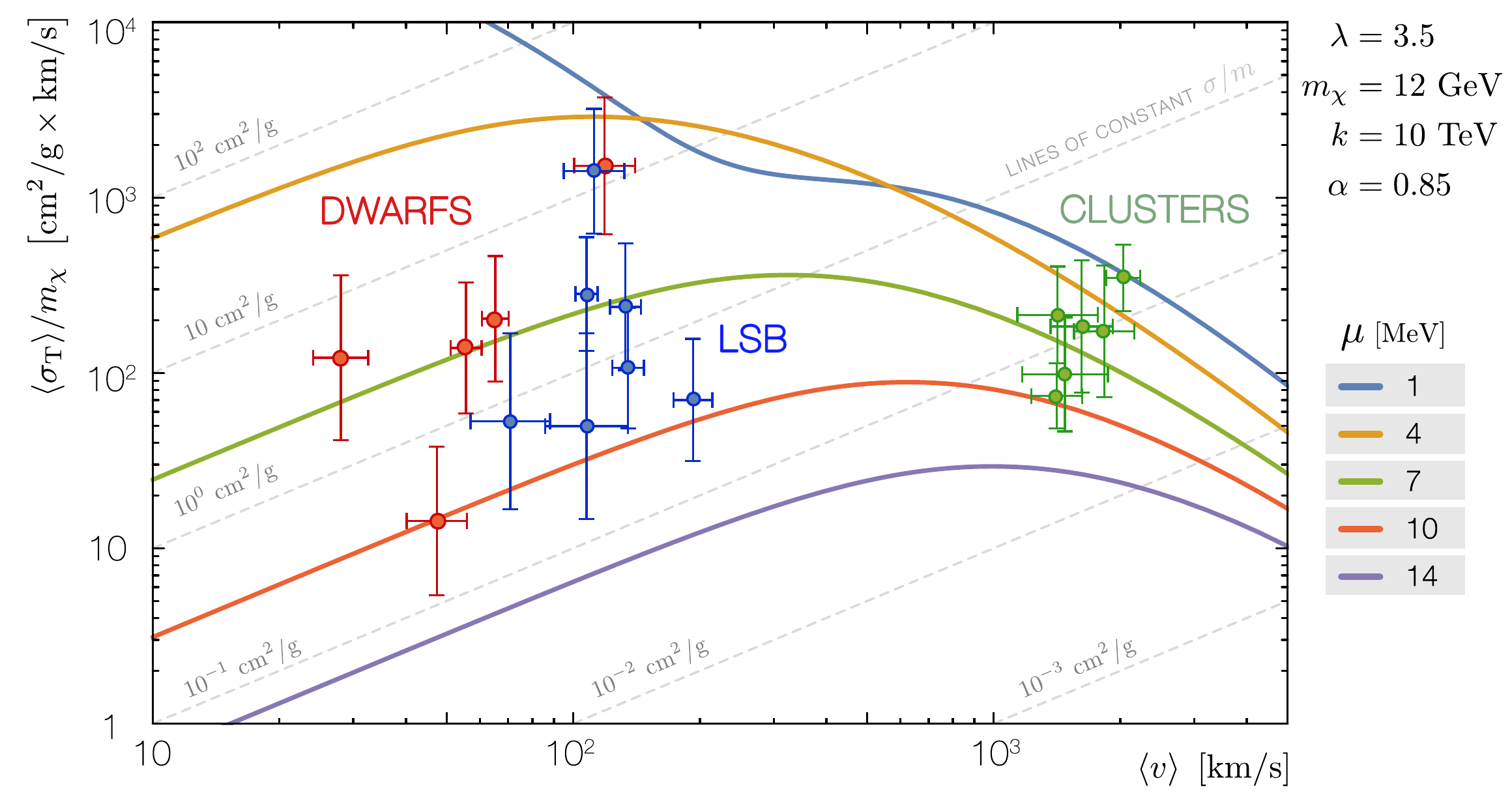}
\end{subfigure}
\hfill
\begin{subfigure}[t]{0.49\textwidth}  
  \centering 
    \includegraphics[width=\textwidth]{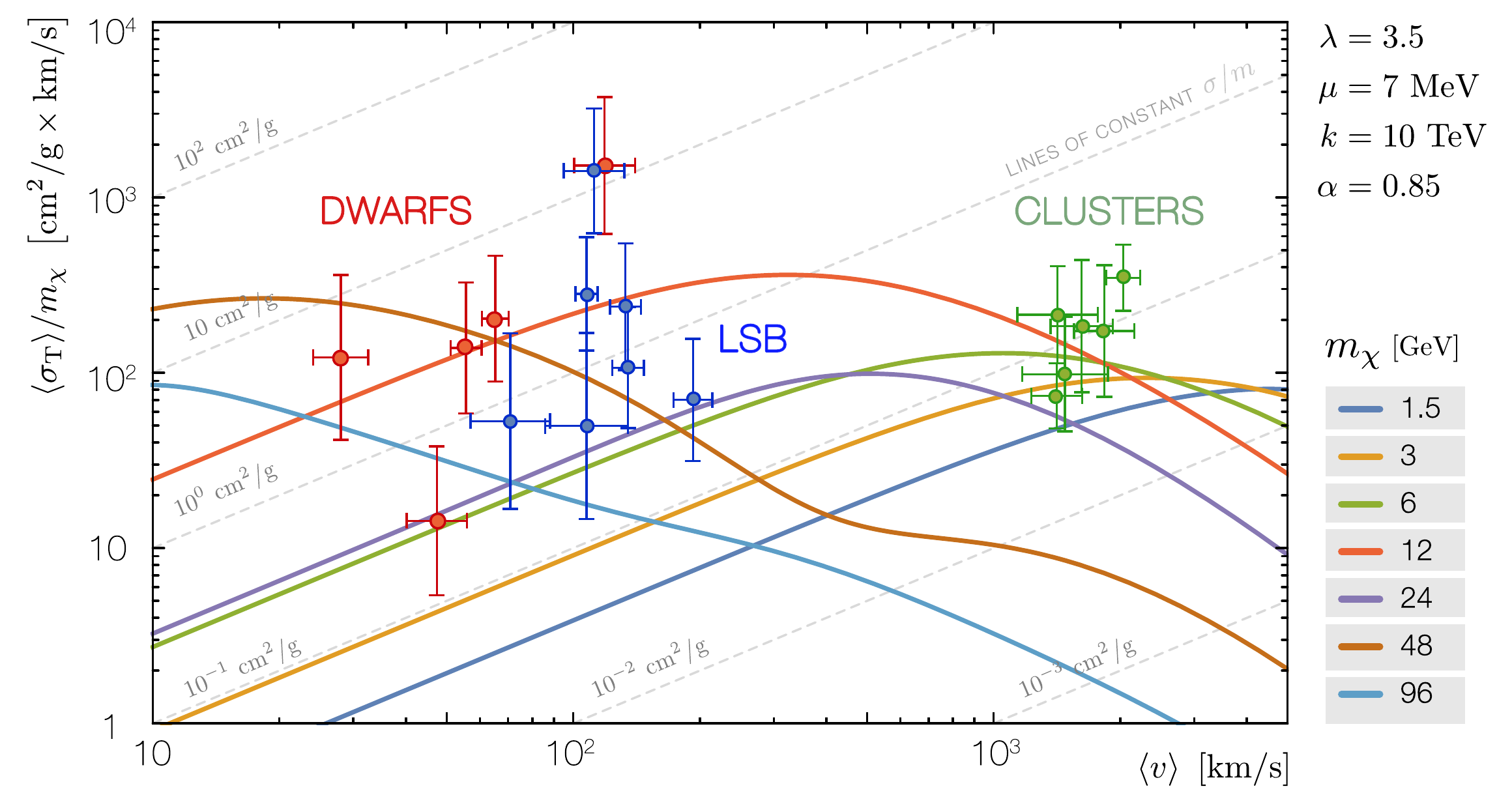}
\end{subfigure}
\caption{
Velocity dependence of the thermally averaged transfer cross section, analogous to Figure~\ref{fig:sigmavastrodata}, for a range of $\mu$ and $m_\chi$ choices to demonstrate the behavior with respect to these parameters.
}
\label{fig:mx_mu_scan}
\end{figure}

Beyond simply describing the model parameters that reproduce astrophysical data, it is also illustrative to plot a range of model parameters to see how they distort the $\langle \sigma_\text{T} v\rangle/m_\chi$ behavior from the ideal case. Figure~\ref{fig:mx_mu_scan} presents such a scan over the mass gap $\mu$ and $m_\chi$ with other parameters fixed. 

Varying the mass gap $\mu$ primarily affects the behavior at low velocities (low momentum transfer), though it leads to an overall rescaling because it is a multiplicative factor in the effective coupling $\alpha_\chi^\text{eff}$ \eqref{eq:Born:alphaeff}. Thus for a set of parameters that fit the cluster data well, one can tune the mass gap to help fit the low-velocity data.

The dependence on the dark matter mass $m_\chi$, on the other hand, is highly nontrivial. One can see this because the phase shifts in \eqref{eq:partial:waves} depend on the dimensionless combinations $m_\chi/m_1$ and $m_\chi/k$, as described in Appendix~\ref{app:SIDM:method}. Varying $m_\chi$ thus affects two independent quantities in the numerical solution of the partial waves. 

\subsection{Comment on Annihilation and Relic Abundance}
The purpose of this study is to demonstrate the distinctive self-interaction phenomenology of our model and we have remained agnostic about whether or not dark matter is a relic from thermal freeze out.
Thus we have not restricted  the dark matter mass $m_\chi$ and bulk coupling $\lambda$ to fit that of a thermal relic, even though such a restriction would itself be an interesting benchmark. Indeed, one of the constraints on typical \acro{4D} self-interacting dark matter models is that the required self-interactions for small scale structure are generally too large for dark matter to be a thermal relic in the simplest cosmological scenarios.
Recent work has shown that in the presence of bulk self-interactions, the high \acro{KK}-number states of the \acro{5D} scalar are not valid asymptotic states due to the breakdown of the narrow width approximation~\cite{Costantino:2020msc}. As a result, the production of \acro{KK} modes is heavily suppressed by phase space. This can lead to a tantalizing mechanism to suppress the annihilation rate: by increasing the bulk scalar self-interaction---a new parameter in the theory---one may control the total number of effectively allowed final states. We leave this topic for future work.

\section{Continuum-Mediated Sommerfeld Enhancement}

\label{se:Sommerfeld}

\begin{figure}[t]
    \centering
    \includegraphics[width=0.7\textwidth]{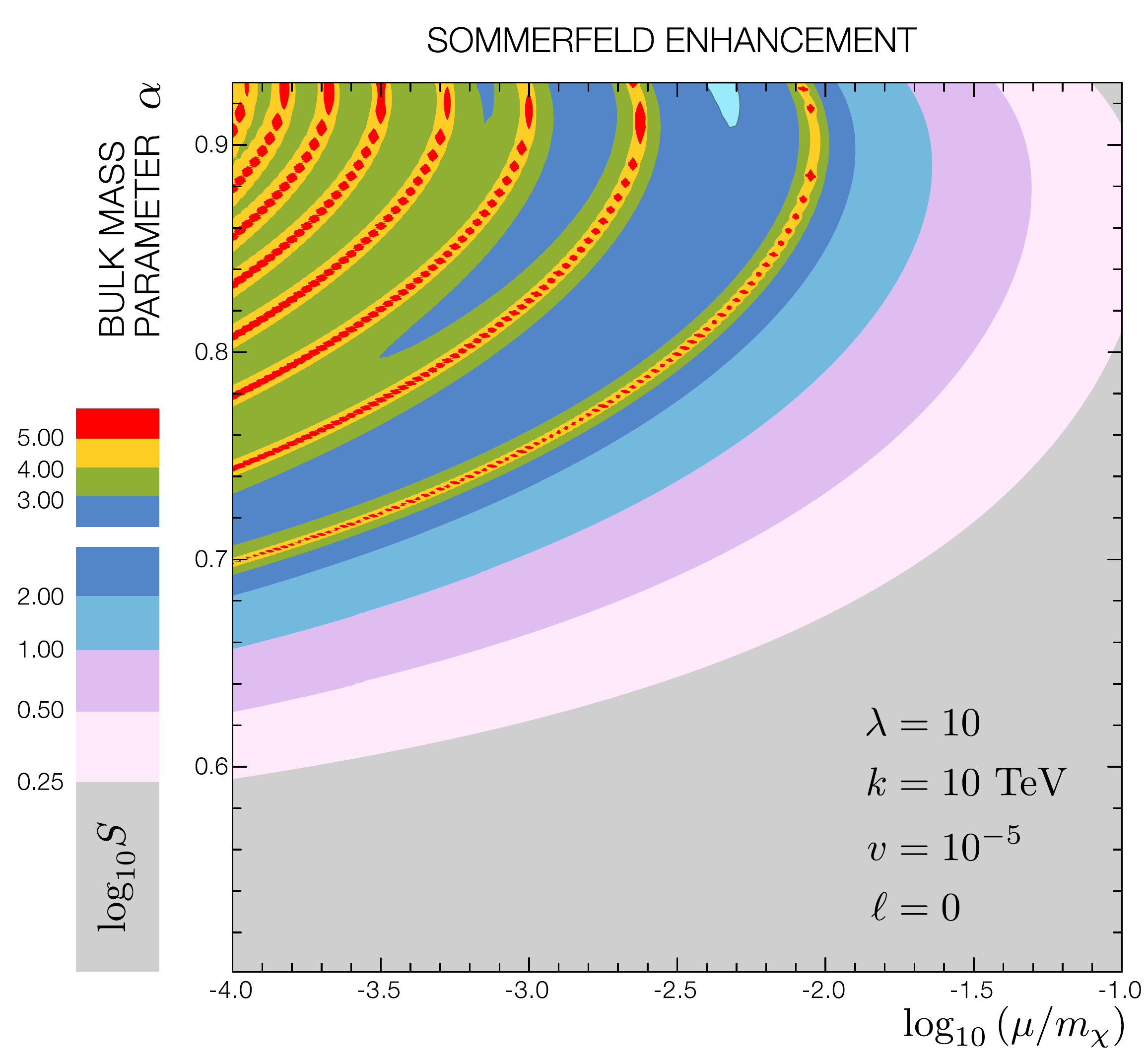}
    \caption{Sommerfeld enhancement of the $\ell=0$ partial wave for a range of $\alpha$ and the ratio $\mu/m_\chi$. 
     Our approximation of the potential breaks down near $\alpha=1$ and hence this region is removed. }
    \label{fig:S_contour}
\end{figure}

\begin{figure}
\centering
\begin{subfigure}[t]{0.475\textwidth}
  \centering
  \includegraphics[width=\textwidth]{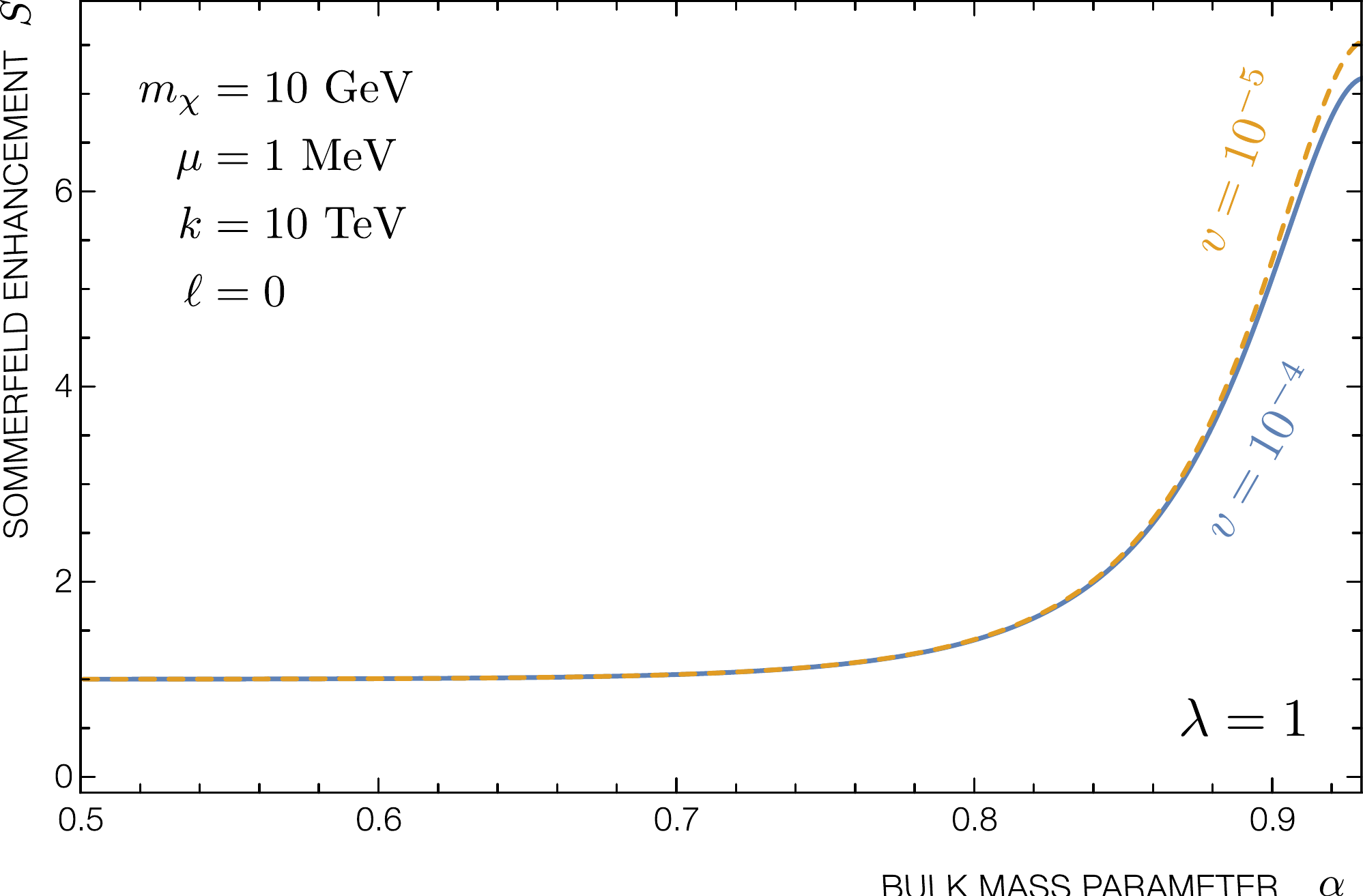}
\end{subfigure}
\hfill
\begin{subfigure}[t]{0.475\textwidth}  
  \centering 
    \includegraphics[width=\textwidth]{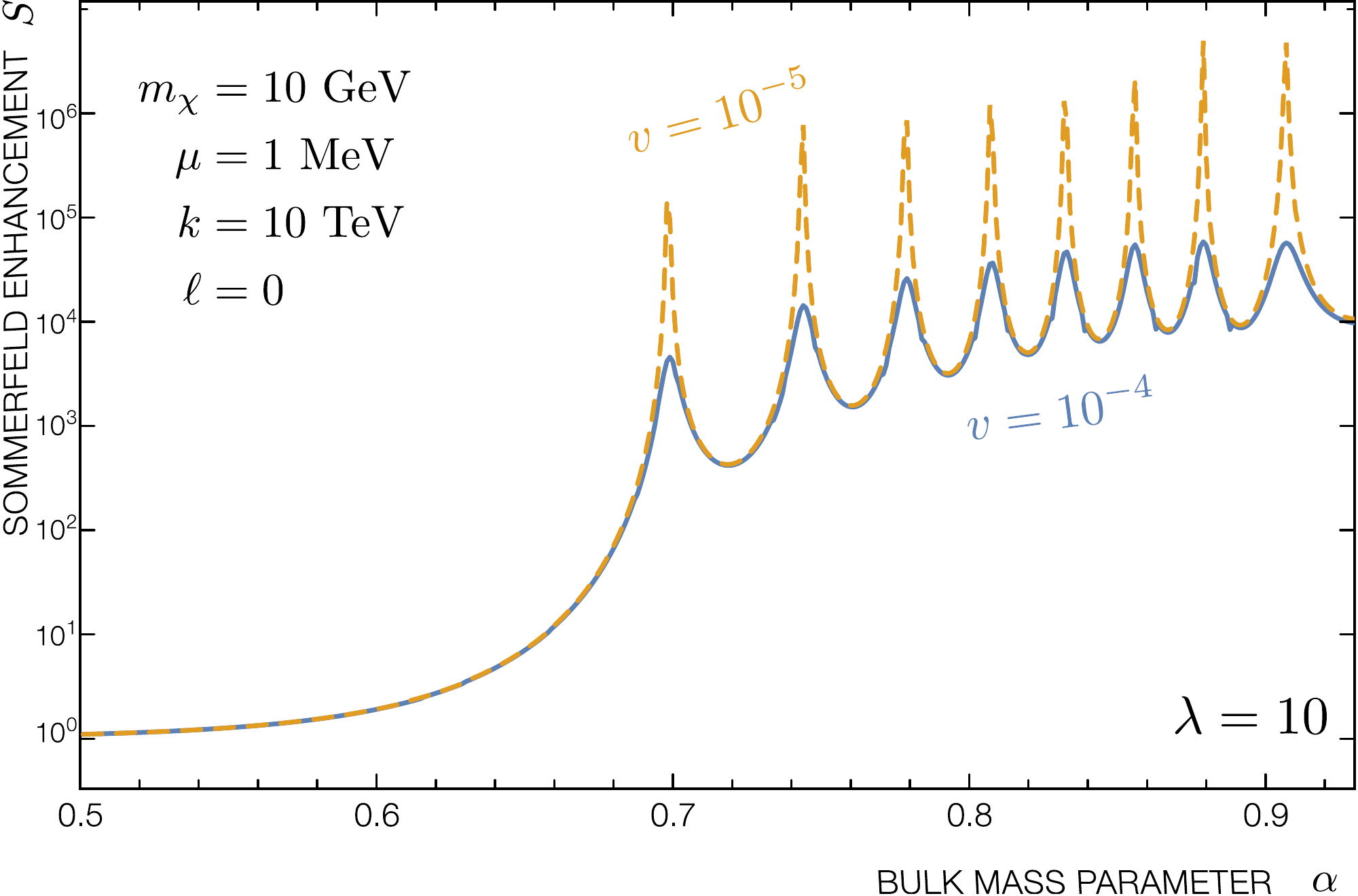}
\end{subfigure}
\caption{Sommerfeld enhancement of the $\ell=0$ partial wave as a function of $\alpha$.
}
\label{fig:somm_a_lambda}
\end{figure}

The same dynamics that generate dark matter self-interactions also lead to Sommerfeld enhancements.
Sommerfeld enhancements encode the effect of the long-range force on a short-distance process (annihilation) and so depend on the solution to the two-body Schr\"odinger equation at the origin, $\Psi(0)$~\cite{Hisano:2004ds,ArkaniHamed:2008qn,Lattanzi:2008qa,Iengo:2009ni,Iengo:2009xf,Cassel:2009wt, Hannestad:2010zt}. In contrast, the dark matter self-interactions that are the main focus of this paper are intrinsically long-ranged.
Diagrammatically both processes involve a ladder of exchanged force mediators between the dark matter initial states. When the potential has a mass gap, the potential supports resonances at large enough coupling.

We investigate the Sommerfeld effect in our continuum-mediated model. 
The continuum-mediated potentials we consider are shorter-ranged than the $1/r$ factor in Yukawa potentials. Since the Sommerfeld enhancement is a long range effect, one may expect that the continuum-mediated Sommerfeld effect is suppressed as compared to the Coulomb case. However, the possibility of resonances may compensate for this and a detailed quantitative analysis is required.

Analytical results for Sommerfeld enhancement are only available for Coulomb potentials. More generally, one must use numerical methods to solve for the enhancement from more general potentials, see e.g.~Ref.~\cite{Bellazzini:2013foa}. This method is valid for potentials that scale like $r^{-1}$ to $r^{-2}$, corresponding to bulk masses $1/2\leq\alpha\leq 1$ in our continuum-mediated model. Potentials that are strictly steeper than $r^{-2}$ require a separate treatment because the potential term dominates the centrifugal term at small distances. These potentials require a short-distance cutoff as expected from a low-energy effective theory.%

We numerically explore the Sommerfeld enhancement over the range of bulk masses  $1/2\leq \alpha <1 $ range for a continuum mediator with a mass gap. This is described by the same potential used for self-interactions, \eqref{eq:Vgamma}. To avoid the breakdown of asymptotic approximations described in Section~\ref{se:potential}, we restrict to $\alpha \lesssim 0.9$. Figure~\ref{fig:S_contour} shows the Sommerfeld enhancement as a function of $\alpha$ and $\mu/m_\chi$ for a benchmark coupling $\lambda=10$.  The key result is that Sommerfeld enhancement occur even when $\alpha < 1$ where the potential is shorter-ranged than a Yukawa potential. The enhancement increases for smaller mass gap relative to the dark matter mass, $\mu/m_\chi$. For example for $\mu/m_\chi=10^{-4}$ we find $S\sim 10$ for  $\alpha\sim 0.66$.  

Resonances appear in a nontrivial pattern in the $\alpha$--$\mu$ plane. With the assumptions in Figure~\ref{fig:S_contour}, the theory exhibits resonant behavior occurring for bulk masses as low as $\alpha\sim 0.7$. These resonances are expected to vanish at lower coupling; this is shown in Figure~\ref{fig:somm_a_lambda} where the Sommerfeld enhancement is plotted for constant $\mu$. The large coupling case $\lambda=10$ exhibits resonances, while a smaller coupling $\lambda=1$ does not. In this case the Sommerfeld effect is found to quickly decrease with $\alpha$.

The Sommerfeld enhancement decreases quickly with $\alpha$ and eventually vanishes near $\alpha=1/2$, corresponding to a $r^{-2}$ potential. 
We remark that the enhancement for an ungapped $V(r)\propto r^{-2}$ potential can be solved exactly. In this case, the centrifugal term has the same scaling as the potential so that 
the $\ell=0$ solution is singular and dependent on the \acro{EFT} cutoff. To the best of our knowledge, Sommerfeld enhancement for this case has not been discussed in the literature. We present details of this calculation in Appendix~\ref{se:r2_sommerfeld}. We find $S=1$ whenever the dark matter mass is much smaller than the EFT cutoff. 
In other words, the $1/r^2$ potential is too short-ranged to induce any Sommerfeld enhancement, confirming the numerical result in Figure~\ref{fig:S_contour}.

\section{Conclusion}
\label{se:conclusion}

We propose a model where dark matter self-interacts through a continuum of \acro{4D} mediators. This generalizes work on self-interacting dark matter that has otherwise focused on the case of a single massive mediator producing a Yukawa potential. 
A continuum mediator may arise in a strongly-coupled gauge sector. We assume that this mediator sector is nearly conformal so that its features are dictated by symmetry. 
Applications of the self-interacting dark matter paradigm to small-scale structure anomalies require a mass gap to cut off the potential at long distances. 
A natural choice to realize this mass gap is to assume that the strongly-coupled sector has a large number of colors so that the theory is described holographically by a brane-localized dark matter interacting with a bulk field in a slice of \acro{5D} \acro{AdS} space.

We present a concrete realization where the \acro{5D} continuum mediator is a scalar. We address aspects of effective field theory and constraints from experiments and cosmology. The key parameter that characterizes the hallmark features of our model is $\alpha$, which encodes the scalar field's bulk mass and maps onto the conformal dimension $\Delta$ of the dual scalar operator. 

We evaluate the non-relativistic potential induced by a continuum mediator with a mass gap using the spectral representation and asymptotic expressions for the \acro{5D} propagator. We obtain simple closed-form expressions for the $\alpha < 1$ and $\alpha = 1$ cases and validate them numerically. The $\alpha=1$ case corresponds to a Yukawa potential. At long distances, the potential scales like a non-integer power, $V\sim r^{2\alpha-3}$. We focus on the range $1/2 < \alpha \leq 1$ where calculations are tractable and the potential satisfies constraints from \acro{CFT} unitarity.

The astronomical phenomenology of dark matter self-interactions depends on the transfer cross section, $\sigma_\text{T}$. We calculate this quantity in the continuum-mediated scenario and demarcate three types of qualitative behavior---the Born, resonant, and classical regimes. These regimes are qualitatively similar to those of a single \acro{4D} mediator, but in the continuum-mediated model the regimes depend on  $\alpha$ in addition to to the strength of the dark matter coupling and mass gap.

The velocity-dependence of the transfer cross section allows a self-interacting dark matter model to explain small-scale structure anomalies while avoiding cluster-scale constraints.
In contrast to the single \acro{4D} mediator, the transfer cross section in the continuum-mediated model exhibits non-integer velocity scaling. For example, in the perturbative Born regime, $\sigma_\text{T}\sim v^{-4\alpha}$ for large velocities. In the non-perturbative classical regime, $\sigma_\text{T}\sim v^{-4/(3-2\alpha)}$ in the small mass gap limit. In contrast, a Yukawa potential in both of these regimes has a transfer cross section scaling of $\sigma_\text{T}\sim v^{-4}$.

We present benchmark fits of the transfer cross section to astrophysical data. 
In the extreme case of bulk mass parameters $\alpha \sim 0.55$, fits typically require sub-GeV dark matter and sub-MeV mass gaps, which may be cosmologically challenging. Larger values of $\alpha$ permit higher mass scales so long as the ratio of the dark matter mass to the mass gap is $m_\chi/\mu \sim 10^{3}$. Larger bulk masses cause \acro{KK} mode profiles to localize away from the \acro{UV} brane that contains dark matter. Thus larger bulk masses typically require larger dark matter--mediator couplings between the dark matter and mediator.

Our model necessarily leads to continuum-mediated Sommerfeld enhancement. We demonstrate the pattern of resonances that occur in the $(\mu, \alpha)$ plane. 
The enhancements vanish as $\alpha \to 1/2$, consistent with our analytical results for a $1/r^2$ potential.

We conclude that models of dark matter with continuum mediators introduce novel power-law scalings in self-interaction effects. The bulk mass parameter, $\alpha$, has no analog in standard \acro{4D} self-interacting dark matter models and is a new way to control the phenomenology. Since the bulk mass controls the localization of the mediator, it naturally plays a role in possible effective couplings to the Standard Model. These observations open new possibilities for dark matter phenomenology.

 \section*{Acknowledgments}

We thank Gerardo Alvarez, Lexi Costantino, and Hai-Bo Yu for useful comments and discussions. 
\textsc{p.t.} thanks the Aspen Center for Physics (NSF grant \#1066293) for its hospitality during a period where part of this work was initiated. \textsc{p.t.} is supported by the DOE grant \textsc{de-sc}/0008541.
\textsc{p.t.}~and \textsc{i.c.}~thank the Physics 40B (Winter 2021, Section 001) students of \acro{UC} Riverside for their patience with grade postings while this manuscript was being completed.

\appendix

\section{
 AdS/CFT with UV brane
}
\label{se:AdS/CFT}

The \acro{AdS}/\acro{CFT} correspondence states that boundary correlators of quantum field theory in \acro{AdS}$_{d+1}$ spacetime are equivalent to correlators of a conformal field theory in $d$-dimensional spacetime~\cite{Aharony:1999ti,Zaffaroni:2000vh,Nastase:2007kj,Kap:lecture}.   
For a given bulk field in \acro{AdS}, the corresponding \acro{CFT} operator arises through the asymptotic behavior of the field near the \acro{AdS} boundary. 
In this appendix we revisit and streamline  the two branches of the correspondence in the presence of a \acro{UV} brane.

\subsection{The Two Branches}

A scalar bulk field $\Phi$ in \acro{AdS}$_{5}$ corresponds to a scalar operator $\cal O$ of a \acro{CFT}.  The conformal dimension of $\cal O$ is denoted $\Delta$. An analysis of the boundary asymptotics shows that the relation between \acro{AdS} bulk mass and $\Delta$ is given by~\cite{Maldacena:1997re,
Gubser:1998bc,
Witten:1998qj,Freedman:1998bj,
Liu:1998ty,
Freedman:1998tz,
DHoker:1999mqo,
DHoker:1999kzh}
\begin{align}
\Delta (\Delta+d)k^2=M^2_\Phi
\ , \label{eq:AdS_matching}
\end{align}
or equivalently \eqref{eq:alpha}. We recall that $M^2_\Phi =(\alpha^2-4)k^2$ and $\alpha\geq 0$ by convention.  The two roots of \eqref{eq:AdS_matching} are 
\begin{align}
\Delta_\pm=2\pm\alpha \,.
\end{align}
These two roots indicate that the correspondence has two branches; for a given \acro{AdS} field there can be {two} \acro{CFT} duals. Unitarity of the operator implies $\Delta\geq 1$. It follows that the $\Delta_+$ branch exists for $\alpha\in \mathbb{R}_+$, but the $\Delta_-$ branch exists only for $0\leq\alpha\leq 1$~\cite{Klebanov:1999tb}. 

The correspondence is formulated as follows. 
We define the value of the bulk field on the  \acro{AdS} boundary $\Phi_0\equiv\Phi(X^M\rightarrow {\rm boundary})$.  
Starting from the \acro{AdS} partition function, one integrates over the bulk degrees of freedom while holding $\Phi_0$ constant. This defines the boundary effective action
\begin{align}
\int_{\Phi_0} {\cal D} \Phi e^{i S_{\rm AdS}[\Phi]}=   e^{i \Gamma_{\rm AdS}[\Phi_0] }\,. 
\end{align}
The two branches of the correspondence are then formulated as follows.

In the $\Delta_+$ branch, the dual  \acro{CFT} is defined by the correspondence
\begin{align}
\Gamma_{\rm AdS}[\Phi_0] \equiv W_{\rm CFT}[\Phi_0]
\label{eq:DeltaP}
\end{align}
with $W_{\rm CFT}[J]$   the generating functional of connected correlators of a \acro{CFT} where  $J$ is the source of the operator $\cal O$ (with $[{\cal O}]=\Delta_+$), 
\begin{align}
Z_{\rm CFT}[J]=
\int {\cal D} \phi_{\rm CFT} e^{i S_{\rm CFT}[\phi_{\rm CFT}] + \int d^4x{\cal O} J }=e^{iW_{\rm CFT}[J]} \label{eq:def_W} \,.
\end{align}
In this branch we can observe that the $\Phi_0$  variable corresponds to the source of the $\cal O$ operator.

In the $\Delta_-$ branch the dual  \acro{CFT} is defined by the correspondence
\begin{align}
\Gamma_{\rm AdS}[{\cal O}] \equiv \Sigma_{\rm CFT}[{\cal O}]
\label{eq:DeltaM}
\end{align}
with $\Sigma_{\rm CFT}[{\cal O}_{\rm cl}]$  the Legendre transform of $W_{\rm CFT}[J]$, 
\begin{align}
\Sigma_{\rm CFT}[{\cal O}]=W_{\rm CFT}[J]-\int dx^\mu {\cal O} J \,. \label{eq:Sigma_CFT}
\end{align}
$\Sigma$ is constructed similarly to an effective action. Its argument is understood to be an expectation value, {e.g.}~${\cal O}_{\rm cl}$, this is left implicit here. 
In the $\Delta_-$ branch we can observe that $\Phi_0$ corresponds  to the expectation value of the ${\cal O}$ operator itself.

\subsection{The Two Branches with a UV brane}

One can truncate \acro{AdS} with a \acro{UV} brane and identify $\Phi_0 = \Phi(X^M\rightarrow {\rm UV \, brane})$. The above \acro{AdS}/\acro{CFT} relations from full \acro{AdS} remain structurally the same, however fields on a brane away from the boundary can be dynamical, hence the \acro{UV} brane has a localized action $S_{\rm UV}$. In particular the $\Phi_0$ variable is in general dynamical instead of being static as in the full \acro{AdS} case. Thus $\Phi_0$ is now a \acro{4D} field, external to the \acro{CFT}.

The \acro{AdS} partition function is
\begin{align}
\int {\cal D}\Phi_0 e^{i S_{\rm UV}[\Phi_0]} \int_{\Phi_0} {\cal D} \Phi e^{i S_{\rm AdS}[\Phi]}= \int {\cal D}\Phi_0  e^{i S_{\rm UV}[\Phi_0]+i\Gamma_{\rm AdS}[\Phi_0] }\,. 
\end{align}
To formulate the \acro{4D} theory in terms of a generating functional of connected correlators, one would have to introduce new static sources coupled to $\Phi_0$ and $\cal O$. Instead we can Legendre transform and describe the theory directly in terms of an effective action $\Gamma_{\rm 4D}$. We introduce $\Gamma_{\rm UV}$, the effective action generated by $S_{\rm UV}$.

Consider the $\Delta_+$ branch. The \acro{4D} theory is identified as in \eqref{eq:DeltaP}, appending $S_{\rm UV}$ on both sides. The $W_{\rm CFT}$ is substituted by its Legendre transform using Eq.\,\eqref{eq:Sigma_CFT}, where the $J$ source is localized on the \acro{UV} brane  and  can now be dynamical.  It follows that the effective action of the \acro{4D} theory is  given by
\begin{align}
\Gamma_{\rm UV}[\Phi_0] + \Gamma_{\rm AdS}[\Phi_0] \equiv \Gamma_{\rm UV}[\Phi_0] + \Sigma_{\rm CFT}[{\cal O}] + \int dx^\mu {\cal O} \Phi_0 = \Gamma_{\rm 4D}[\Phi_0,{\cal O}] \,.
\label{eq:DeltaPUV}
\end{align}
To illustrate the \acro{4D} theory defined by \eqref{eq:DeltaPUV}, consider a dynamical \acro{UV} brane-localized current $J_{\rm UV}$ coupled to $\Phi$ as $S_{\rm UV}=\int d^4xJ_{\rm UV}\Phi_0$, and evaluate the $\langle J_{\rm UV}J_{\rm UV}\rangle$ correlator. One finds that the $J_{\rm UV}$ currents exchange a propagator of $\Phi_0$, which is itself dressed by the two-point function of $\cal O$. 

In the $\Delta_-$ branch the effective action of the \acro{4D} theory is identified as
\begin{align}
\Gamma_{\rm UV}[{\cal O}] + \Gamma_{\rm AdS}[{\cal O}] \equiv \Gamma_{\rm UV}[{\cal O}] + \Sigma_{\rm CFT}[{\cal O}]  = \Gamma_{\rm 4D}[{\cal O}] \,. \label{eq:DeltaMUV}
\end{align}
Consider again the  $\langle J_{\rm UV}J_{\rm UV}\rangle$ correlator from the $S_{\rm UV}=\int d^4xJ_{\rm UV}\Phi_0$ interaction.  What we obtain is that the $J_{\rm UV}$ currents exchange a two-point correlator of $\cal O$. 
This $\Delta_-$ branch of the duality is the one used for our model. Identifying the $J_{\rm UV}$ current  as $J_{\rm DM}$, the $\langle J J \rangle$ correlator discussed here  describes formally the relation given in~\eqref{eq:AdS:CFT:bulk}.

\section{\texorpdfstring{Derivation of Gapless $\alpha=1$ Potential}{Derivation of Gapless alpha=1 Potential}}
\label{se:Valpha1}

In this appendix we show how to evaluate the Fourier transform of~\eqref{eq:Propalpha1exp}. The first term is a simple pole at the origin and thus gives a Coulomb potential. The next-to-leading term goes as $\log( q)/q^2$. To evaluate its Fourier transform we use 
\begin{align}
\frac{\log q^2}{q^{2n}}=-\partial_\alpha \frac{1}{q^{2\alpha}}\bigg|_{\alpha\to n}\,
\end{align}
with $n=1$. The Fourier transform of $q^{-2\alpha}$ is
\begin{align}
\frac{1}{(2\pi)^3} \int d^3{\bf q}  e^{i{\bf q r}} \frac{1}{q^{2\alpha}}= 
\frac{1}{(2\pi)^3}\frac{1}{\Gamma(\alpha)} \int d^3{\bf q} e^{i{\bf q r}} \int \frac{dt}{t}t^{\alpha-1} e^{-tq^2}= \frac{1}{(4\pi)^{3/2}} \frac{\Gamma(3/2-\alpha)}{\Gamma(\alpha)} \left(\frac{4}{r^2}\right)^{3/2-\alpha}\,.
\end{align}
We then evaluate the $\alpha$ derivative and set $\alpha=1$, which gives 
\begin{align}
\frac{1}{(4\pi)^{3/2}} \partial_\alpha \left( \frac{\Gamma(3/2-\alpha)}{\Gamma(\alpha) \left(\frac{4}{r^2}\right)^{3/2-\alpha} }\right)_{\alpha\to 1} = \frac{1}{2\pi} (\gamma+\log r)\,.
\end{align}
Combining these identities gives~\eqref{eq:Valpha1mu0}.

\section{Validity of the Born Approximation}
\label{sec:BornValidity}
In order to determine the validity of the Born approximation, consider the wave function for a dark matter particle scattering off of a potential $V(\Vec{x})$,
\begin{align}
    \psi(\Vec{x})\sim e^{i \Vec{p}\cdot \Vec{x}}-m_\chi\int d^3 x' \frac{e^{i p|\Vec{x}-\Vec{x}'|}}{4\pi|\Vec{x}-\Vec{x}'|}V(\Vec{x}')e^{i\Vec{p}\cdot\Vec{x}'}
    \, ,
\end{align}
where $|\Vec{p}|=m_\chi v/2$ and $\Vec{p}\cdot\Vec{x}'=p r'\cos\theta$. Near the origin $|\Vec{x}-\Vec{x}'|\approx r'$, thus the condition for when the Born approximation is valid is
\begin{align}
\left|\frac{m_\chi}{4\pi}\int d^3 x' \frac{e^{i p r'}}{r'}V(\Vec{x}')e^{i \Vec{p}\cdot \Vec{x}'}\right| \ll 1
\, .
\label{eq:Born:validitygeneral}
\end{align}
For a Yukawa potential $V(r)=\alpha_\chi e^{-m_\phi r}/r$ this condition is simply $\alpha_\chi m_\chi/m_1 \ll 1$. 
At low energies we can replace the exponentials by 1. For a central potential in spherical coordinates the angular integral is trivial.
Evaluating \eqref{eq:Born:validitygeneral} for a Yukawa potential gives the condition $\alpha_\chi m_\chi/m_\phi \ll 1$. This bound can be equivalently determined by considering the typical momentum flowing through a ladder diagram is of order $\alpha_\chi m_\chi$~\cite{gross1999relativistic, Petraki:2016cnz}.
We evaluate \eqref{eq:Born:validitygeneral} for \eqref{eq:V_KK_prop} and arrive at the result
\begin{align}
    \frac{\lambda^2}{4\pi k}\sum_n\frac{f_n^2(z_{\text{UV}})}{m_n} \ll 1
    \, .
    \label{eq:app:Borncondition}
\end{align}
In order to make the connection to the Yukawa case more explicit, we define the effective coupling
\begin{align}
    \alpha_{\chi}^\text{eff}=\frac{\lambda^2 m_1}{4\pi k}\sum_n\frac{f_n^2(z_{\text{UV}})}{m_n}
    \ ,
    \label{eq:app:alphaeff}
\end{align}
such that the condition for when then Born approximation is valid becomes
\begin{align}
    \frac{\alpha_{\chi}^\text{eff}m_\chi}{m_1} \ll 1
    \, ,
\end{align}
analogous to the Yukawa case.

Recalling that the bulk profiles depend on the bulk mass parameter $\alpha$, we note that the sum over \acro{KK} modes in $\alpha_{\chi,\text{eff}}$ diverges for $\alpha \leq 1/2$. This is consistent with the Schr\"{o}dinger equation in which, near the origin, the continuum mediated potential \eqref{eq:Vgamma} dominates over the centrifugal barrier for $\alpha \leq 1/2$.
In order to achieve finite results in the case when $\alpha \leq 1/2$, we introduce a smooth cutoff to \eqref{eq:app:Borncondition} such that
\begin{align}
    \alpha_{\chi}^\text{eff}=\frac{\lambda^2 m_1}{4\pi k}\sum_n\frac{f_n^2(z_{\text{UV}})}{m_n}
    &\quad\quad\quad \longrightarrow&
    &\alpha_{\chi}^\text{eff}\left(\Lambda\right)=\frac{\lambda^2 m_1}{4\pi k}\sum_n\frac{f_n^2(z_{\text{UV}})}{m_n} e^{- m_n/\Lambda}
    \ ,
\end{align}
where $\Lambda^{-1}$ is the short distance cutoff. 
We evaluate the \acro{KK} sum using the spectral representation of the propagator ~\eqref{eq:spectral:discontinuity} and using the large-momentum asymptotics~\eqref{eq:Propcont}. 
We arrive at the result
\begin{align}
 \alpha_{\chi}^\text{eff} =\frac{\lambda^2}{4 \pi \Gamma(1-\alpha)^2}\frac{m_1}{\Lambda}\left(\frac{2k}{\Lambda}\right)^{2\alpha-2}\Gamma\left(1-2\alpha,\frac{m_1}{\Lambda}\right)
 \, .
 \label{eq:app:CMSIDM:alphaeffcut}
\end{align}
When $\alpha>1/2$, the limit $\Lambda\to\infty$ is finite and cutoff independent,
\begin{align}
    \left.\alpha_{\chi}^\text{eff}\right|_{\alpha>1/2}= \frac{\lambda^2}{4\pi}\left[\frac{4}{2\alpha-1}\frac{1}{\Gamma(1-\alpha)^2}\right]\left(\frac{m_1}{2k}\right)^{2-2\alpha}
    \, .
    \label{eq:app:CMSIDM:alphaeff}
\end{align}
This result is identical to evaluating \eqref{eq:Born:validitygeneral}  for the continuum mediated potential \eqref{eq:Vgamma}.

For the special case of a bulk mass parameter $\alpha=1/2$, we find that the effective coupling for the Born approximation is
\begin{align}
    \alpha_{\chi}^\text{eff}
    &=
    \frac{\lambda^2 m_1}{8 \pi^2 k}\log\left(\frac{\Lambda e^{-\gamma}}{m_1}\right)
    \ .
    \label{eq:alphaeff1/2}
\end{align}
The other limit, $\alpha \to 1$ requires special care. Because the asymptotic expansions of the canonical propagator used for the $\alpha < 1$ result break down in this limit, one cannot simply take $\alpha\to 1$ in  \eqref{eq:app:CMSIDM:alphaeff}. Instead, in the case where $\alpha=1$, scattering is governed by the Yukawa potential \eqref{eq:V_light} and we can directly apply \eqref{eq:SIDM:Born} so that
\begin{align}
\alpha_{\chi}^\text{eff}=\frac{\lambda^2 m_1}{4\pi k}f_0^2(z_{\text{UV}})
\label{eq:CMSIDM:alphaeffalpha1}
\end{align}
where $f_0(z_{\text{UV}})$ is given by \eqref{eq:f0}.

The accuracy of the Born approximation improves at higher energies. This can also be shown from \eqref{eq:Born:validitygeneral} by computing the angular integral  for a general central potential,

\section{Classical Transfer Cross Section}
\label{app:classical}

We calculate the transfer cross section in the classical regime and observe its velocity dependence. The angle by which a particle in a central potential is deflected is $\theta(\rho)=|\pi-2\varphi(\rho)|$ where~\cite{Landau1976Mechanics}
\begin{align}
    \varphi(\rho)=\rho \int_{r_{\text{min}}}^\infty \frac{dr}{r^2 \sqrt{1-\rho^2/r^2-4 V(r)/m_\chi v^2}}  \label{eq:app:phiclassical}
\end{align}
and $\rho$ is the impact parameter. The lower limit of integration $r_{\text{min}}$ is the largest root of the denominator of \eqref{eq:app:phiclassical}. In contrast to the quantum case, the classical cross section is typically given in terms of the impact parameter rather than the angular variables. In the classical limit, integration over the deflection angle can be troublesome since the solution to \eqref{eq:app:phiclassical} for $\varphi(\rho)$ and thus $\theta(\rho)$ takes values greater than $\pi$ for cases other than a $1/r$ potential. On the other hand the impact parameter always ranges between zero and infinity.

The differential cross section is $d\sigma=2\pi \rho \, d\rho$. The transfer cross section is thus
\begin{align}
    \sigma_\text{T}^{\text{classical}}=2\pi\int_0^{\infty}\left[1-\cos\theta(\rho)\right]\rho d\rho
    \, . \label{eq:app:sigmaTclassical}
\end{align}
To connect to the deflection angle, we note that
\begin{align}
  \left(  \frac{d\sigma}{d\Omega}\right)^{\text{classical}}=\frac{\rho(\chi)}{\sin\theta}\left|\frac{d\rho}{d\theta}\right|
\end{align}
where in these variables $d\Omega=2\pi \,  d\cos\theta$. We present calculations for the velocity scaling in the small mass gap/high velocity limit and an analytical result in the low velocity region of the non-perturbative classical regime.

\subsection{Velocity Scaling in the  Small Mass Gap/High Velocity Limit} \label{app:classical:highv}
For the sake of this calculation we assume the gapless limit where the potential is  \eqref{eq:Vnogamma}
\begin{align}
    V(r)=-\frac{\lambda^2}{2 \pi^{3/2}}\frac{\Gamma(3/2-\alpha)}{\Gamma(1-\alpha)}\frac{1}{r(k r)^{2-2\alpha}}
    \, .
\end{align}
This approximation also accounts for the high velocity limit where the particle momentum is much greater than the mass gap. Given our potential we can define a characteristic length scale
\begin{align}
    \rho_0\equiv \left[\frac{\lambda^2}{2 \pi^{3/2} m_\chi v^2 k^{2-2\alpha} }\frac{\Gamma(3/2-\alpha)}{\Gamma(1-\alpha)}\right]^{\tfrac{1}{3-2\alpha}}
    \label{eq:app:rho0}
\end{align}
so that after making the change of variables $r=\rho/x$, \eqref{eq:app:phiclassical} becomes~\cite{Chiron:2016lzn}
\begin{align}
    \varphi(\rho)=\int_0^{x_{\text{max}}}\frac{dx}{\sqrt{1-x^2+2(\rho_0 x/\rho)^{3-2\alpha}}}
    \label{eq:app:phiclassical2}
\end{align}
where the limit of integration $x_{\text{max}}$ is  the smallest positive root of the denominator. Observe that $\varphi$ (and by extension $\chi$) and as $x_{\text{max}}$ are functions of the dimensionless combination $\rho/\rho_0$ and not $\rho$ independently. Making the change of variables $\rho=\rho_0 \xi$, the transfer cross section is 
\begin{align}
    \sigma_\text{T}^{\text{classical}}=2\pi \rho_0^2 \int_0^\infty \left[1-\cos\chi(\xi)\right]\xi d\xi
    \, . \label{eq:app:sigmaTclassical2}
\end{align}
Because $\chi$ is a function of $\xi$ and $\alpha$ only, the integral \eqref{eq:app:sigmaTclassical2} only depends on the bulk mass parameter. We can thus conclude from \eqref{eq:app:rho0} that the velocity dependence of the transfer cross section in the classical regime is $\sigma_\text{T}^{\text{classical}}\sim v^{-4/(3-2\alpha)}$.

The presence of the mass gap spoils the velocity dependence derived in \eqref{eq:app:sigmaTclassical2}. For the gapped potential \eqref{eq:Vgamma}, after changing variables, the deflection angle depends on the quantity $m_1 \rho_0$ as well. Thus a small but nonzero $m_1$ induces corrections to \eqref{eq:app:sigmaTclassical2}.

\subsection{Low Velocity Classical Regime}
\label{app:classical:lowv}

We present a closed form result for the transfer cross section in the low velocity region of the non-perturbative classical regime. Following the method of Ref.~\cite{Khrapak:2003kjw}, the transfer cross section is a function of a single unique parameter,
\begin{align}
    \beta=\frac{2\alpha_{\chi}^\text{eff}m_1}{m_\chi v^2}(2\alpha-1)
    \ .
    \label{eq:betaclassical}
\end{align}
The transfer cross section is  $\sigma_\text{T}\approx\pi \rho_*^2$ where $\rho_*$ is found by solving the set of equations
\begin{align}
    &\tilde{V}_{\text{eff}}(r_{\text{max}},\rho_*)=1
    &
    &\left.\frac{d\tilde{V}_{\text{eff}}(r,\rho_*)}{dr}\right|_{r=r_{\text{max}}}=0
    \ 
\end{align}
where $\tilde{V}_{\text{eff}}$ is the effective potential
\begin{align}
    \tilde{V}_{\text{eff}}(r,\rho)=\frac{\rho^2}{r^2}+\frac{4}{m_\chi v^2}V(r)
    \ .
    \label{eq:Veff}
\end{align}
These conditions correspond to the maximum of the effective potential $\tilde{V}_{\text{eff}}$.

We find for $\beta \gg 1$ that the transfer cross section is approximately
\begin{align}
    \sigma_\text{T}^{\text{classical}}\approx
    \frac{\pi}{m_1^2}\left[1+\log\left(\frac{\beta}{\log\beta}\right)-(2\alpha-1)\log^{-1}\beta+\left(2\alpha-\frac{3}{2}\right)\log^{-1}\left(\frac{\beta}{\log\beta}\right)\right]^2
    \ .
    \label{eq:sigma:cl:smallv}
\end{align}
The accuracy of \eqref{eq:sigma:cl:smallv} is confirmed in Figure~\ref{fig:sigmavsmuregions}.

\section{\texorpdfstring{Sommerfeld Enhancement from a $1/r^2$ Potential}{Sommerfeld Enhancement from an inverse square Potential}}
\label{se:r2_sommerfeld}

The Sommerfeld effect amounts to the  enhancement of the particle wavefunction at the point where the local annihilation process happens. It comes from the dressing from ladder diagrams generated by a  potential $V(r)$. The dressed wavefunction  is determined by directly solving the Schr\"{o}dinger equation. The Sommerfeld enhancement factor is defined as $\sigma =S(p) \sigma_0$ with $\sigma_0$ the undressed cross section. 
The method to evaluate the Sommerfeld effect is well known, here we follow \cite{ArkaniHamed:2008qn} (see also \cite{Iengo:2009ni}).

The Schr\"{o}dinger equation is 
\begin{align}
-\frac{1}{2M}\Delta \Psi(r) + V(r) \Psi(r) = \frac{p^2}{2M} \Psi(r)\,.
\end{align}
In any solution of the Schr\"{o}dinger equation with rotational invariance around $z$, the solutions can be expanded as
\begin{align}
\Psi=\sum a_l P_\ell(\cos \theta) R_\ell(r) \,.
\end{align}
The radial wavefunction satisfies 
\begin{align}
-\frac{1}{2M r^2}\frac{d}{dr}\left(r^2\frac{dR_{\ell}}{dr}\right) +\left( V(r) +\frac{\ell(\ell+1)}{2 M r^2} \right) R_\ell(r) = \frac{p^2}{2M} R_\ell(r)\,. \label{eq:app:radialSchrodinger}
\end{align}

In the standard approach one uses the fact that angular momentum with $\ell>0$ gives $R_\ell\sim r^\ell$ at small $r$, which implies that the $\ell>0$ contributions to the wavefunction vanish at the origin. Hence  one can focus on the $\ell=0$ angular momentum.

For our continuum-mediated potential $V(r)\propto r^{2\alpha-3}$, the vanishing of $\ell>0$  remains true for any $\alpha\geq 1/2$. For $\alpha>1/2$, the $\ell=0$ mode gives $R_\ell\sim $\,constant at small $r$. But for $\alpha=1/2$, which is the $V(r)\propto 1/r^2$ potential of our interest, the $\ell=0$ component diverges at small $r$. This feature is not an inconsistency. We work in a low-energy \acro{EFT} so the $r$ coordinate cannot be zero, it is rather cut  at a small value corresponding to the \acro{UV} cutoff, $r=r_0$. In our \acro{AdS}  model the cutoff is at $r_0\sim k^{-1}$. 
Of course, the subsequent results may be cutoff dependent, but this is not a conceptual problem, this simply reflects that an \acro{EFT} prediction can  depend on the unknown \acro{UV} physics. 

Here we parametrize the $\alpha=1/2$ potential as 
\begin{align}
V(r)= -\frac{\kappa}{2r^2}\,.
\end{align}
The matching to the physical couplings from the \acro{AdS}  model is $\kappa=\frac{\lambda^2}{\pi^2 k} $.

Introducing $\chi_\ell(r)=r R_\ell(r)$ the Schr\"{o}dinger equation becomes
\begin{align}
-\frac{1}{2M}\partial^2_r \chi_0(r) + V(r)   \chi_0(r) = \frac{p^2}{2M} \chi_0(r)\,.
\end{align}
From this equation, various equivalent methods lead to the Sommerfeld factor, which differs by the boundary conditions chosen for $\chi$ \cite{ArkaniHamed:2008qn}. We use the following. $\chi_\ell$ is chosen to satisfy $\partial_r \chi_\ell=ip \chi_\ell $ at $r=\infty$. Using this solution, the Sommerfeld factor is
\begin{align}
S=\left|\frac{\chi_0(r=\infty)}{\chi_0(r=r_0)}\right|\,.
\end{align} 
Notice that since we are in an \acro{EFT} with have replaced the $r=0$ by $r=r_0$.

The solution satisfying the condition at $r=\infty$ is found to be
\begin{align} 
\chi_0(r)_\ell \propto \sqrt{r } H^{(1)}_{\eta}(pr) \,,\quad\quad  \eta = \sqrt{\frac{1}{4}-M \kappa}\,. 
\end{align}

The dimensionful $\kappa$ coupling is of order of the inverse cutoff of the \acro{EFT}. The \acro{EFT} would break at $M\kappa \sim 1$,  we are rather interested in $M\kappa\ll 1$, {i.e.} the dark matter mass is much lower than the cutoff $k$. 

Expanding in the small parameter $M\kappa$ we find 
\begin{align}
\chi_0(r_0)\propto i+{\cal O}(p r_0)\,.
\end{align}
 We have that $p r_0$ is necessarily $\ll 1$ since $r_0$ is the inverse cutoff $k^{-1}$, and because the non-relativistic approximation  requires $p<M$ and the \acro{EFT} validity requires $M<k$.

It follows that within the range of validity of the \acro{EFT}, we can simply take  $\eta\approx 1/2$. The Hankel simplifies to $H^{(1)}(z)\propto z^{-1/2} e^{i z} $, thus $\chi_0(r) \propto e^{ip r}$ for any $pr$. The Sommerfeld factor is then  exactly $S=1$ for any $p$.

\section{Self-Interacting Dark Matter Numerical Method} \label{app:SIDM}
\label{app:SIDM:method}

We summarize the methodology for determining the dark matter self-interaction cross section. We closely follow the procedure in Ref.~\cite{Tulin:2013teo} however we employ a slightly more relaxed algorithm.
The relevant quantity is the transfer cross section,
\begin{equation}
\sigma_\text{T}=\int d\Omega \left(1-\cos\theta\right)\frac{d \sigma}{d \Omega} \ , \label{eq:transfercrosssection}
\end{equation}
which characterizes interaction cross section weighted by momentum transfer. This regulates the $\cos\theta \to 1$ divergence where dark matter scattering does not affect halo shapes. There is no known analytical expression for the transfer cross section that is valid for the entire parameter space. A large region of the parameter space corresponds to the resonant regime where both quantum mechanical and non-perturbative effects become important, as such a numerical solution to the non-relativistic Schr\"odinger equation is necessary.

We employ a partial wave analysis. The transfer cross section is related to the $\ell^{\text{th}}$ partial wave phase shift, $\delta_\ell$, by
\begin{equation}
\sigma_\text{T}=\frac{4 \pi}{p^2}\sum\limits_{\ell=0}^{\infty}(\ell+1)\sin^2\left(\delta_{\ell+1}-\delta_{\ell}\right) \label{eq:transfercrosssectionsum} \ .
\end{equation}
The $\delta_{\ell}$ are, in turn, obtained by solving the radial Schr\"{o}dinger equation \eqref{eq:app:radialSchrodinger} taking $M=m_\chi/2$ and $p=m_X v/2$ where $v$ is the relative velocity of the two-particle dark matter system. $\delta_l$ is found by comparing with the asymptotic solution for $R_{\ell}$:
\begin{equation}
\lim\limits_{r\to \infty}R_{\ell}(r) \propto \cos \delta_{\ell} j_{\ell}(pr)-\sin \delta_{\ell} n_{\ell}(pr) \label{eq:asympsol}\ ,
\end{equation}
where $j_{\ell}$ ($n_\ell$) is the spherical Bessel (Neumann) function of the $\ell^{\text{th}}$ order.
We again define the function $\chi_{\ell} \equiv r R_{\ell}$ along with the dimensionless variables
\begin{align}
x &\equiv \alpha_X m_X r & a&=\frac{v}{2\alpha_X} & b&=\frac{\alpha_X m_X}{m_{1}} & c&=\frac{\alpha_X m_X}{k} \ ,
\end{align}
  so that we can rewrite \eqref{eq:app:radialSchrodinger} as~\cite{Buckley:2009in}
\begin{equation}
\left[\frac{d^2}{dx^2}+a^2-\frac{\ell (\ell+1)}{x^2}\pm \frac{2}{\pi^{1/2}}\frac{1}{x}\left(\frac{c}{x}\right)^{2-2\alpha}\frac{\Gamma(3/2-\alpha)}{\Gamma(1-\alpha)}Q\left(2-2\alpha,x/b\right)\right]\chi_{\ell}(x)=0\ . \label{eq:dimlessSchroeq}
\end{equation}
Near the origin for $\alpha>1/2$, the angular momentum term dominates over the potential. This implies that $\chi_{\ell}\propto x^{\ell+1}$ close to $x=0$. When $\alpha\leq 1/2$ and the potential becomes singular, our method breaks down and we cannot determine an initial condition. We choose a normalization for the wavefunctions such that $\chi_{\ell}(x_0)=1$ and $\chi'_{\ell}(x_0)=\left(\ell+1\right)/x_0$ where $x_0$ is a point close to the origin chosen to satisfy $x_0 \ll b$ and $x_0 \ll \left(\ell+1\right)/a$. We take $x_0$ as the lower limit for the range in which we numerically solve the Schr\"odinger equation. Similarly, to define the upper limit, we pick a point $x_m$ satisfying the condition
\begin{align}
a^2 \gg \frac{2}{\pi^{1/2}}\frac{1}{x}\left(\frac{c}{x}\right)^{2-2\alpha}\frac{\Gamma(3/2-\alpha)}{\Gamma(1-\alpha)}Q\left(2-2\alpha,x/b\right) \, .
\end{align}
When $x_m$ satisfies this condition, the potential term is negligible compared to the kinetic term and the solution approaches
\begin{equation}
\chi_{\ell}(x)\propto x e^{i \delta_{\ell}}\left(\cos\delta_{\ell}j_{\ell}(a x)-\sin\delta_{\ell} n_{\ell}(a x)\right).
\end{equation}
The phase shift is then
\begin{align}
\tan \delta_{\ell}&=\frac{a x_m j'_{\ell}(a x_m)-\beta_{\ell}j_{\ell}(a x_m)}{a x_m n'_{\ell}(a x_m)-\beta_{\ell}n_{\ell}(a x_m)}
& \text{where}&&
\beta_{\ell}&=\frac{x_m \chi'_{\ell}(x_m)}{\chi_{\ell}(x_m)}-1 \ .
\label{eq:tandelta}
\end{align}
For an initial guess of the range $(x_0,x_m)$ and the maximum number of partial waves required for convergence, $\ell_\text{max}$, we calculate $\delta_{\ell}$ from \eqref{eq:tandelta}. In Ref.~\cite{Tulin:2013teo} $x_m$ and $x_0$ are increased and decreased respectively, recalculating $\delta_{\ell}$ until the differences of successive iterations converge to be within 1\%. This condition can be quite cumbersome numerically and is \textit{not strictly required} unless one wishes to do a fine grained scan over the parameter space. Instead, we take the value of $\delta_{\ell}$ given by our initial guess. This method is sufficient to reproduce the benchmark results in Ref.~\cite{Kaplinghat:2015aga}.

We then sum \eqref{eq:transfercrosssectionsum} from $\ell= 0$ to $\ell=\ell_{\text{max}}$ to obtain an estimate for $\sigma_\text{T}$.
Next we increment $\ell_{\text{max}}\to \ell_\text{max}+1$ and repeat the procedure until successive values of $\sigma_\text{T}$ converge to be within 1\% and $\delta_{\ell_{\text{max}}}<0.01$.
Ref.~\cite{Tulin:2013teo} iterates $\ell_{\text{max}}$ until $\sigma_\text{T}$ converged and $\delta_{\ell_{\text{max}}}<0.01$ ten consecutive times. We have found that the ``StiffenessSwitching'' method from the \texttt{NDSolveUtilities} package in \emph{Mathematica} to be particularly useful.

We employ this method to calculate the Sommerfeld enhancements as well. The enhancement factor is~\cite{Iengo:2009ni,Bellazzini:2013foa}
\begin{align}
    S=\left[\frac{(2\ell+1)!!}{C}\right]^2
\end{align}
where $C^2$ is 
\begin{align}
    C^2=\left(\chi^2_\ell(x)-\chi^2_\ell(x-\pi/2a)\right)_{x\to\infty}
    \ .
\end{align}

\bibliographystyle{utcaps}  
\bibliography{FSIDM}

\providecommand{\href}[2]{#2}\begingroup\raggedright\begin{thebibliography}{10}

\bibitem{Pospelov:2008zw}
M.~Pospelov, ``{Secluded U(1) Below the Weak Scale},''
  \href{http://dx.doi.org/10.1103/PhysRevD.80.095002}{{\em Phys. Rev.}
  {\bfseries D80} (2009) 095002},
\href{http://arxiv.org/abs/0811.1030}{{\ttfamily arXiv:0811.1030 [hep-ph]}}.

\bibitem{Pospelov:2008jd}
M.~Pospelov and A.~Ritz, ``{Astrophysical Signatures of Secluded Dark
  Matter},'' \href{http://dx.doi.org/10.1016/j.physletb.2008.12.012}{{\em Phys.
  Lett.} {\bfseries B671} (2009) 391--397},
\href{http://arxiv.org/abs/0810.1502}{{\ttfamily arXiv:0810.1502 [hep-ph]}}.

\bibitem{Pospelov:2007mp}
M.~Pospelov, A.~Ritz, and M.~B. Voloshin, ``{Secluded WIMP Dark Matter},''
  \href{http://dx.doi.org/10.1016/j.physletb.2008.02.052}{{\em Phys. Lett.}
  {\bfseries B662} (2008) 53--61},
\href{http://arxiv.org/abs/0711.4866}{{\ttfamily arXiv:0711.4866 [hep-ph]}}.

\bibitem{Essig:2013lka}
R.~Essig {\em et~al.}, ``{Working Group Report: New Light Weakly Coupled
  Particles},'' in {\em {Proceedings, 2013 Community Summer Study on the Future
  of U.S. Particle Physics: Snowmass on the Mississippi (Cs$S^2$013):
  Minneapolis, Mn, Usa, July 29-August 6, 2013}}.
\newblock 2013.
\newblock \href{http://arxiv.org/abs/1311.0029}{{\ttfamily arXiv:1311.0029
  [hep-ph]}}.
\newblock
\url{http://www.slac.stanford.edu/econf/C1307292/docs/IntensityFrontier/NewLight-17.pdf}.
\newblock

\bibitem{Alexander:2016aln}
J.~Alexander {\em et~al.}, ``{Dark Sectors 2016 Workshop: Community Report},''
\newblock 2016.
\newblock \href{http://arxiv.org/abs/1608.08632}{{\ttfamily arXiv:1608.08632
  [hep-ph]}}.
\newblock
\url{http://lss.fnal.gov/archive/2016/conf/fermilab-conf-16-421.pdf}.
\newblock

\bibitem{Battaglieri:2017aum}
M.~Battaglieri {\em et~al.}, ``{Us Cosmic Visions: New Ideas in Dark Matter
  2017: Community Report},'' in {\em {U.S. Cosmic Visions: New Ideas in Dark
  Matter College Park, Md, Usa, March 23-25, 2017}}.
\newblock 2017.
\newblock \href{http://arxiv.org/abs/1707.04591}{{\ttfamily arXiv:1707.04591
  [hep-ph]}}.
\newblock
\url{http://lss.fnal.gov/archive/2017/conf/fermilab-conf-17-282-ae-ppd-t.pdf}.
\newblock

\bibitem{Carlson:1992fn}
E.~D. Carlson, M.~E. Machacek, and L.~J. Hall, ``{Self-Interacting Dark
  Matter},'' \href{http://dx.doi.org/10.1086/171833}{{\em Astrophys. J.}
  {\bfseries 398} (1992) 43--52}.

\bibitem{Spergel:1999mh}
D.~N. Spergel and P.~J. Steinhardt, ``{Observational Evidence for
  Selfinteracting Cold Dark Matter},''
  \href{http://dx.doi.org/10.1103/PhysRevLett.84.3760}{{\em Phys. Rev. Lett.}
  {\bfseries 84} (2000) 3760--3763},
  \href{http://arxiv.org/abs/astro-ph/9909386}{{\ttfamily
  arXiv:astro-ph/9909386}}.

\bibitem{Tulin:2013teo}
S.~Tulin, H.-B. Yu, and K.~M. Zurek, ``{Beyond Collisionless Dark Matter:
  Particle Physics Dynamics for Dark Matter Halo Structure},''
  \href{http://dx.doi.org/10.1103/PhysRevD.87.115007}{{\em Phys. Rev. D}
  {\bfseries 87} no.~11, (2013) 115007},
  \href{http://arxiv.org/abs/1302.3898}{{\ttfamily arXiv:1302.3898 [hep-ph]}}.

\bibitem{Tulin:2017ara}
S.~Tulin and H.-B. Yu, ``{Dark Matter Self-Interactions and Small Scale
  Structure},'' \href{http://dx.doi.org/10.1016/j.physrep.2017.11.004}{{\em
  Phys. Rept.} {\bfseries 730} (2018) 1--57},
\href{http://arxiv.org/abs/1705.02358}{{\ttfamily arXiv:1705.02358 [hep-ph]}}.

\bibitem{Gherghetta:2010cq}
T.~Gherghetta and B.~von Harling, ``{A Warped Model of Dark Matter},''
  \href{http://dx.doi.org/10.1007/JHEP04(2010)039}{{\em JHEP} {\bfseries 04}
  (2010) 039},
\href{http://arxiv.org/abs/1002.2967}{{\ttfamily arXiv:1002.2967 [hep-ph]}}.

\bibitem{vonHarling:2012sz}
B.~von Harling and K.~L. McDonald, ``{Secluded Dark Matter Coupled to a Hidden
  CFT},'' \href{http://dx.doi.org/10.1007/JHEP08(2012)048}{{\em JHEP}
  {\bfseries 08} (2012) 048},
\href{http://arxiv.org/abs/1203.6646}{{\ttfamily arXiv:1203.6646 [hep-ph]}}.

\bibitem{Strassler:2008bv}
M.~J. Strassler, ``{Why Unparticle Models with Mass Gaps are Examples of Hidden
  Valleys},'' \href{http://arxiv.org/abs/0801.0629}{{\ttfamily arXiv:0801.0629
  [hep-ph]}}.

\bibitem{Chen:2009ch}
C.-H. Chen and C.~Kim, ``{Sommerfeld Enhancement from Unparticle Exchange for
  Dark Matter Annihilation},''
  \href{http://dx.doi.org/10.1016/j.physletb.2010.03.054}{{\em Phys. Lett. B}
  {\bfseries 687} (2010) 232--235},
  \href{http://arxiv.org/abs/0909.1878}{{\ttfamily arXiv:0909.1878 [hep-ph]}}.

\bibitem{Friedland:2009iy}
A.~Friedland, M.~Giannotti, and M.~Graesser, ``{On the R$S^2$ Realization of
  Unparticles},'' \href{http://dx.doi.org/10.1016/j.physletb.2009.06.012}{{\em
  Phys. Lett. B} {\bfseries 678} (2009) 149--155},
  \href{http://arxiv.org/abs/0902.3676}{{\ttfamily arXiv:0902.3676 [hep-th]}}.

\bibitem{Friedland:2009zg}
A.~Friedland, M.~Giannotti, and M.~L. Graesser, ``{Vector Bosons in the
  Randall-Sundrum 2 and Lykken-Randall Models and Unparticles},''
  \href{http://dx.doi.org/10.1088/1126-6708/2009/09/033}{{\em JHEP} {\bfseries
  09} (2009) 033}, \href{http://arxiv.org/abs/0905.2607}{{\ttfamily
  arXiv:0905.2607 [hep-th]}}.

\bibitem{Brax:2019koq}
P.~Brax, S.~Fichet, and P.~Tanedo, ``{The Warped Dark Sector},''
  \href{http://dx.doi.org/10.1016/j.physletb.2019.135012}{{\em Phys. Lett. B}
  {\bfseries 798} (2019) 135012},
  \href{http://arxiv.org/abs/1906.02199}{{\ttfamily arXiv:1906.02199
  [hep-ph]}}.

\bibitem{Costantino:2020msc}
A.~Costantino, S.~Fichet, and P.~Tanedo, ``{Effective Field Theory in AdS:
  Continuum Regime, Soft Bombs, and IR Emergence},''
  \href{http://arxiv.org/abs/2002.12335}{{\ttfamily arXiv:2002.12335
  [hep-th]}}.

\bibitem{Bullock:2017xww}
J.~S. Bullock and M.~Boylan-Kolchin, ``{Small-Scale Challenges to the
  $\Lambda$CDM Paradigm},''
  \href{http://dx.doi.org/10.1146/annurev-astro-091916-055313}{{\em Ann. Rev.
  Astron. Astrophys.} {\bfseries 55} (2017) 343--387},
  \href{http://arxiv.org/abs/1707.04256}{{\ttfamily arXiv:1707.04256
  [astro-ph.CO]}}.

\bibitem{Fadeev:2018rfl}
P.~Fadeev, Y.~V. Stadnik, F.~Ficek, M.~G. Kozlov, V.~V. Flambaum, and
  D.~Budker, ``{Revisiting spin-dependent forces mediated by new bosons:
  Potentials in the coordinate-space representation for macroscopic- and
  atomic-scale experiments},''
  \href{http://dx.doi.org/10.1103/PhysRevA.99.022113}{{\em Phys. Rev. A}
  {\bfseries 99} no.~2, (2019) 022113},
  \href{http://arxiv.org/abs/1810.10364}{{\ttfamily arXiv:1810.10364
  [hep-ph]}}.

\bibitem{Fichet:2017bng}
S.~Fichet, ``{Quantum Forces from Dark Matter and Where to Find Them},''
  \href{http://dx.doi.org/10.1103/PhysRevLett.120.131801}{{\em Phys. Rev.
  Lett.} {\bfseries 120} no.~13, (2018) 131801},
  \href{http://arxiv.org/abs/1705.10331}{{\ttfamily arXiv:1705.10331
  [hep-ph]}}.

\bibitem{Costantino:2019ixl}
A.~Costantino, S.~Fichet, and P.~Tanedo, ``{Exotic Spin-Dependent Forces from a
  Hidden Sector},'' \href{http://dx.doi.org/10.1007/JHEP03(2020)148}{{\em JHEP}
  {\bfseries 03} (2020) 148}, \href{http://arxiv.org/abs/1910.02972}{{\ttfamily
  arXiv:1910.02972 [hep-ph]}}.

\bibitem{Katz:2015zba}
A.~Katz, M.~Reece, and A.~Sajjad, ``{Continuum-mediated dark matter–baryon
  scattering},'' \href{http://dx.doi.org/10.1016/j.dark.2016.01.002}{{\em Phys.
  Dark Univ.} {\bfseries 12} (2016) 24--36},
\href{http://arxiv.org/abs/1509.03628}{{\ttfamily arXiv:1509.03628 [hep-ph]}}.

\bibitem{Randall:1999vf}
L.~Randall and R.~Sundrum, ``{An Alternative to Compactification},''
  \href{http://dx.doi.org/10.1103/PhysRevLett.83.4690}{{\em Phys. Rev. Lett.}
  {\bfseries 83} (1999) 4690--4693},
\href{http://arxiv.org/abs/hep-th/9906064}{{\ttfamily arXiv:hep-th/9906064
  [hep-th]}}.

\bibitem{Goldberger:1999uk}
W.~D. Goldberger and M.~B. Wise, ``{Modulus Stabilization with Bulk Fields},''
  \href{http://dx.doi.org/10.1103/PhysRevLett.83.4922}{{\em Phys. Rev. Lett.}
  {\bfseries 83} (1999) 4922--4925},
  \href{http://arxiv.org/abs/hep-ph/9907447}{{\ttfamily arXiv:hep-ph/9907447}}.

\bibitem{Manohar:1983md}
A.~Manohar and H.~Georgi, ``{Chiral Quarks and the Nonrelativistic Quark
  Model},'' \href{http://dx.doi.org/10.1016/0550-3213(84)90231-1}{{\em Nucl.
  Phys. B} {\bfseries 234} (1984) 189--212}.

\bibitem{Georgi:1986kr}
H.~Georgi and L.~Randall, ``{Flavor Conserving CP Violation in Invisible Axion
  Models},'' \href{http://dx.doi.org/10.1016/0550-3213(86)90022-2}{{\em Nucl.
  Phys. B} {\bfseries 276} (1986) 241--252}.

\bibitem{Georgi:1992dw}
H.~Georgi, ``{Generalized Dimensional Analysis},''
  \href{http://dx.doi.org/10.1016/0370-2693(93)91728-6}{{\em Phys. Lett. B}
  {\bfseries 298} (1993) 187--189},
  \href{http://arxiv.org/abs/hep-ph/9207278}{{\ttfamily arXiv:hep-ph/9207278}}.

\bibitem{Luty:1997fk}
M.~A. Luty, ``{Naive Dimensional Analysis and Supersymmetry},''
  \href{http://dx.doi.org/10.1103/PhysRevD.57.1531}{{\em Phys. Rev. D}
  {\bfseries 57} (1998) 1531--1538},
  \href{http://arxiv.org/abs/hep-ph/9706235}{{\ttfamily arXiv:hep-ph/9706235}}.

\bibitem{Jenkins:2013sda}
E.~E. Jenkins, A.~V. Manohar, and M.~Trott, ``{Naive Dimensional Analysis
  Counting of Gauge Theory Amplitudes and Anomalous Dimensions},''
  \href{http://dx.doi.org/10.1016/j.physletb.2013.09.020}{{\em Phys. Lett. B}
  {\bfseries 726} (2013) 697--702},
  \href{http://arxiv.org/abs/1309.0819}{{\ttfamily arXiv:1309.0819 [hep-ph]}}.

\bibitem{Dvali:2008ec}
G.~Dvali and C.~Gomez, ``{Quantum Information and Gravity Cutoff in Theories
  with Species},'' \href{http://dx.doi.org/10.1016/j.physletb.2009.03.024}{{\em
  Phys. Lett. B} {\bfseries 674} (2009) 303--307},
  \href{http://arxiv.org/abs/0812.1940}{{\ttfamily arXiv:0812.1940 [hep-th]}}.

\bibitem{Fichet:2019owx}
S.~Fichet, ``{Braneworld Effective Field Theories --- Holography, Consistency
  and Conformal Effects},''
  \href{http://dx.doi.org/10.1007/JHEP04(2020)016}{{\em JHEP} {\bfseries 04}
  (2020) 016}, \href{http://arxiv.org/abs/1912.12316}{{\ttfamily
  arXiv:1912.12316 [hep-th]}}.

\bibitem{Davoudiasl:2008hx}
H.~Davoudiasl, G.~Perez, and A.~Soni, ``{The Little Randall-Sundrum Model at
  the Large Hadron Collider},''
  \href{http://dx.doi.org/10.1016/j.physletb.2008.05.024}{{\em Phys. Lett. B}
  {\bfseries 665} (2008) 67--71},
  \href{http://arxiv.org/abs/0802.0203}{{\ttfamily arXiv:0802.0203 [hep-ph]}}.

\bibitem{Breitenlohner:1982bm}
P.~Breitenlohner and D.~Z. Freedman, ``{Positive Energy in Anti-de~Sitter
  Backgrounds and Gauged Extended Supergravity},''
  \href{http://dx.doi.org/10.1016/0370-2693(82)90643-8}{{\em Phys. Lett. B}
  {\bfseries 115} (1982) 197--201}.

\bibitem{Breitenlohner:1982jf}
P.~Breitenlohner and D.~Z. Freedman, ``{Stability in Gauged Extended
  Supergravity},'' \href{http://dx.doi.org/10.1016/0003-4916(82)90116-6}{{\em
  Annals Phys.} {\bfseries 144} (1982) 249}.

\bibitem{Maldacena:1997re}
J.~M. Maldacena, ``{The Large N limit of superconformal field theories and
  supergravity},'' \href{http://dx.doi.org/10.1023/A:1026654312961,
  10.4310/ATMP.1998.v2.n2.a1}{{\em Int. J. Theor. Phys.} {\bfseries 38} (1999)
  1113--1133}, \href{http://arxiv.org/abs/hep-th/9711200}{{\ttfamily
  arXiv:hep-th/9711200 [hep-th]}}.
[Adv. Theor. Math. Phys.2,231(1998)].

\bibitem{Gubser:1998bc}
S.~S. Gubser, I.~R. Klebanov, and A.~M. Polyakov, ``{Gauge theory correlators
  from noncritical string theory},''
  \href{http://dx.doi.org/10.1016/S0370-2693(98)00377-3}{{\em Phys. Lett.}
  {\bfseries B428} (1998) 105--114},
\href{http://arxiv.org/abs/hep-th/9802109}{{\ttfamily arXiv:hep-th/9802109
  [hep-th]}}.

\bibitem{Witten:1998qj}
E.~Witten, ``{Anti-de Sitter space and holography},''
  \href{http://dx.doi.org/10.4310/ATMP.1998.v2.n2.a2}{{\em Adv. Theor. Math.
  Phys.} {\bfseries 2} (1998) 253--291},
\href{http://arxiv.org/abs/hep-th/9802150}{{\ttfamily arXiv:hep-th/9802150
  [hep-th]}}.

\bibitem{Freedman:1998bj}
D.~Z. Freedman, S.~D. Mathur, A.~Matusis, and L.~Rastelli, ``{Comments on 4
  point functions in the CFT / AdS correspondence},''
  \href{http://dx.doi.org/10.1016/S0370-2693(99)00229-4}{{\em Phys. Lett. B}
  {\bfseries 452} (1999) 61--68},
  \href{http://arxiv.org/abs/hep-th/9808006}{{\ttfamily arXiv:hep-th/9808006}}.

\bibitem{Liu:1998ty}
H.~Liu and A.~A. Tseytlin, ``{On four point functions in the CFT / AdS
  correspondence},'' \href{http://dx.doi.org/10.1103/PhysRevD.59.086002}{{\em
  Phys. Rev. D} {\bfseries 59} (1999) 086002},
  \href{http://arxiv.org/abs/hep-th/9807097}{{\ttfamily arXiv:hep-th/9807097}}.

\bibitem{Freedman:1998tz}
D.~Z. Freedman, S.~D. Mathur, A.~Matusis, and L.~Rastelli, ``{Correlation
  functions in the CFT(d) / AdS(d+1) correspondence},''
  \href{http://dx.doi.org/10.1016/S0550-3213(99)00053-X}{{\em Nucl. Phys. B}
  {\bfseries 546} (1999) 96--118},
  \href{http://arxiv.org/abs/hep-th/9804058}{{\ttfamily arXiv:hep-th/9804058}}.

\bibitem{DHoker:1999mqo}
E.~D'Hoker, D.~Z. Freedman, and L.~Rastelli, ``{AdS / CFT four point functions:
  How to succeed at z integrals without really trying},''
  \href{http://dx.doi.org/10.1016/S0550-3213(99)00526-X}{{\em Nucl. Phys. B}
  {\bfseries 562} (1999) 395--411},
  \href{http://arxiv.org/abs/hep-th/9905049}{{\ttfamily arXiv:hep-th/9905049}}.

\bibitem{DHoker:1999kzh}
E.~D'Hoker, D.~Z. Freedman, S.~D. Mathur, A.~Matusis, and L.~Rastelli,
  ``{Graviton exchange and complete four point functions in the AdS / CFT
  correspondence},''
  \href{http://dx.doi.org/10.1016/S0550-3213(99)00525-8}{{\em Nucl. Phys. B}
  {\bfseries 562} (1999) 353--394},
  \href{http://arxiv.org/abs/hep-th/9903196}{{\ttfamily arXiv:hep-th/9903196}}.

\bibitem{Aharony:1999ti}
O.~Aharony, S.~S. Gubser, J.~M. Maldacena, H.~Ooguri, and Y.~Oz, ``{Large N
  field theories, string theory and gravity},''
  \href{http://dx.doi.org/10.1016/S0370-1573(99)00083-6}{{\em Phys. Rept.}
  {\bfseries 323} (2000) 183--386},
\href{http://arxiv.org/abs/hep-th/9905111}{{\ttfamily arXiv:hep-th/9905111
  [hep-th]}}.

\bibitem{Zaffaroni:2000vh}
A.~Zaffaroni, ``{Introduction to the AdS-CFT correspondence},''
  \href{http://dx.doi.org/10.1088/0264-9381/17/17/306}{{\em Class. Quant.
  Grav.} {\bfseries 17} (2000) 3571--3597}.

\bibitem{Nastase:2007kj}
H.~Nastase, ``{Introduction to AdS-CFT},''
\href{http://arxiv.org/abs/0712.0689}{{\ttfamily arXiv:0712.0689 [hep-th]}}.

\bibitem{Kap:lecture}
J.~Kaplan, ``{Lectures on AdS/CFT from the Bottom Up}.''.

\bibitem{ArkaniHamed:2000ds}
N.~Arkani-Hamed, M.~Porrati, and L.~Randall, ``{Holography and
  phenomenology},'' \href{http://dx.doi.org/10.1088/1126-6708/2001/08/017}{{\em
  JHEP} {\bfseries 08} (2001) 017},
\href{http://arxiv.org/abs/hep-th/0012148}{{\ttfamily arXiv:hep-th/0012148
  [hep-th]}}.

\bibitem{Creminelli:2001th}
P.~Creminelli, A.~Nicolis, and R.~Rattazzi, ``{Holography and the Electroweak
  Phase Transition},''
  \href{http://dx.doi.org/10.1088/1126-6708/2002/03/051}{{\em JHEP} {\bfseries
  03} (2002) 051}, \href{http://arxiv.org/abs/hep-th/0107141}{{\ttfamily
  arXiv:hep-th/0107141}}.

\bibitem{Hebecker:2001nv}
A.~Hebecker and J.~March-Russell, ``{Randall-Sundrum II Cosmology, AdS / CFT,
  and the Bulk Black Hole},''
  \href{http://dx.doi.org/10.1016/S0550-3213(01)00286-3}{{\em Nucl. Phys.}
  {\bfseries B608} (2001) 375--393},
\href{http://arxiv.org/abs/hep-ph/0103214}{{\ttfamily arXiv:hep-ph/0103214
  [hep-ph]}}.

\bibitem{Langlois:2002ke}
D.~Langlois, L.~Sorbo, and M.~Rodriguez-Martinez, ``{Cosmology of a brane
  radiating gravitons into the extra dimension},''
  \href{http://dx.doi.org/10.1103/PhysRevLett.89.171301}{{\em Phys. Rev. Lett.}
  {\bfseries 89} (2002) 171301},
  \href{http://arxiv.org/abs/hep-th/0206146}{{\ttfamily arXiv:hep-th/0206146}}.

\bibitem{Langlois:2003zb}
D.~Langlois and L.~Sorbo, ``{Bulk gravitons from a cosmological brane},''
  \href{http://dx.doi.org/10.1103/PhysRevD.68.084006}{{\em Phys. Rev. D}
  {\bfseries 68} (2003) 084006},
  \href{http://arxiv.org/abs/hep-th/0306281}{{\ttfamily arXiv:hep-th/0306281}}.

\bibitem{us:DR}
A.~Costantino, S.~Fichet, and F.~Tanedo, ``{Work in progress}.''.

\bibitem{Giddings:2000mu}
S.~B. Giddings, E.~Katz, and L.~Randall, ``{Linearized gravity in brane
  backgrounds},'' \href{http://dx.doi.org/10.1088/1126-6708/2000/03/023}{{\em
  JHEP} {\bfseries 03} (2000) 023},
  \href{http://arxiv.org/abs/hep-th/0002091}{{\ttfamily arXiv:hep-th/0002091}}.

\bibitem{Lee:2020zjt}
J.~G. Lee, E.~G. Adelberger, T.~S. Cook, S.~M. Fleischer, and B.~R. Heckel,
  ``{New Test of the Gravitational $1/r^2$ Law at Separations down to 52
  $\mu$m},'' \href{http://dx.doi.org/10.1103/PhysRevLett.124.101101}{{\em Phys.
  Rev. Lett.} {\bfseries 124} no.~10, (2020) 101101},
  \href{http://arxiv.org/abs/2002.11761}{{\ttfamily arXiv:2002.11761
  [hep-ex]}}.

\bibitem{Brax:2017xho}
P.~Brax, S.~Fichet, and G.~Pignol, ``{Bounding Quantum Dark Forces},''
  \href{http://dx.doi.org/10.1103/PhysRevD.97.115034}{{\em Phys. Rev. D}
  {\bfseries 97} no.~11, (2018) 115034},
  \href{http://arxiv.org/abs/1710.00850}{{\ttfamily arXiv:1710.00850
  [hep-ph]}}.

\bibitem{Kahlhoefer:2017umn}
F.~Kahlhoefer, K.~Schmidt-Hoberg, and S.~Wild, ``{Dark Matter Self-Interactions
  from a General Spin-0 Mediator},''
  \href{http://dx.doi.org/10.1088/1475-7516/2017/08/003}{{\em JCAP} {\bfseries
  08} (2017) 003}, \href{http://arxiv.org/abs/1704.02149}{{\ttfamily
  arXiv:1704.02149 [hep-ph]}}.

\bibitem{Zwicky:2016lka}
R.~Zwicky, \href{http://dx.doi.org/10.3204/DESY-PROC-2016-04/Zwicky}{``{A Brief
  Introduction to Dispersion Relations and Analyticity},''} in {\em {Quantum
  Field Theory at the Limits}: {From Strong Fields to Heavy Quarks}},
  pp.~93--120.
\newblock 2017.
\newblock \href{http://arxiv.org/abs/1610.06090}{{\ttfamily arXiv:1610.06090
  [hep-ph]}}.

\bibitem{Fichet:2019hkg}
S.~Fichet, ``{Opacity and effective field theory in anti--de Sitter
  backgrounds},'' \href{http://dx.doi.org/10.1103/PhysRevD.100.095002}{{\em
  Phys. Rev. D} {\bfseries 100} no.~9, (2019) 095002},
  \href{http://arxiv.org/abs/1905.05779}{{\ttfamily arXiv:1905.05779
  [hep-th]}}.

\bibitem{Costantino:2020vdu}
A.~Costantino and S.~Fichet, ``{Opacity from Loops in AdS},''
  \href{http://arxiv.org/abs/2011.06603}{{\ttfamily arXiv:2011.06603
  [hep-th]}}.

\bibitem{Kaplinghat:2015aga}
M.~Kaplinghat, S.~Tulin, and H.-B. Yu, ``{Dark Matter Halos as Particle
  Colliders: Unified Solution to Small-Scale Structure Puzzles from Dwarfs to
  Clusters},'' \href{http://dx.doi.org/10.1103/PhysRevLett.116.041302}{{\em
  Phys. Rev. Lett.} {\bfseries 116} no.~4, (2016) 041302},
  \href{http://arxiv.org/abs/1508.03339}{{\ttfamily arXiv:1508.03339
  [astro-ph.CO]}}.

\bibitem{Dave:2000ar}
R.~Dave, D.~N. Spergel, P.~J. Steinhardt, and B.~D. Wandelt, ``{Halo properties
  in cosmological simulations of selfinteracting cold dark matter},''
  \href{http://dx.doi.org/10.1086/318417}{{\em Astrophys. J.} {\bfseries 547}
  (2001) 574--589}, \href{http://arxiv.org/abs/astro-ph/0006218}{{\ttfamily
  arXiv:astro-ph/0006218}}.

\bibitem{Sakurai:1167961}
J.~J. Sakurai, {\em {Modern quantum mechanics; rev. ed.}}
\newblock Addison-Wesley, Reading, MA, 1994.
\newblock \url{https://cds.cern.ch/record/1167961}.

\bibitem{Chiron:2016lzn}
D.~Chiron and B.~Marcos, ``{Classical particle scattering for power-law
  two-body potentials},'' \href{http://arxiv.org/abs/1601.00064}{{\ttfamily
  arXiv:1601.00064 [cond-mat.stat-mech]}}.

\bibitem{Khrapak:2003kjw}
S.~A. Khrapak, A.~V. Ivlev, G.~E. Morfill, and S.~K. Zhdanov, ``{Scattering in
  the Attractive Yukawa Potential in the Limit of Strong Interaction},''
  \href{http://dx.doi.org/10.1103/PhysRevLett.90.225002}{{\em Phys. Rev. Lett.}
  {\bfseries 90} no.~22, (2003) 225002}.

\bibitem{Cyburt:2015mya}
R.~H. Cyburt, B.~D. Fields, K.~A. Olive, and T.-H. Yeh, ``{Big Bang
  Nucleosynthesis: 2015},''
  \href{http://dx.doi.org/10.1103/RevModPhys.88.015004}{{\em Rev. Mod. Phys.}
  {\bfseries 88} (2016) 015004},
  \href{http://arxiv.org/abs/1505.01076}{{\ttfamily arXiv:1505.01076
  [astro-ph.CO]}}.

\bibitem{Hisano:2004ds}
J.~Hisano, S.~Matsumoto, M.~M. Nojiri, and O.~Saito, ``{Non-Perturbative Effect
  on Dark Matter Annihilation and Gamma Ray Signature from Galactic Center},''
  \href{http://dx.doi.org/10.1103/PhysRevD.71.063528}{{\em Phys. Rev. D}
  {\bfseries 71} (2005) 063528},
  \href{http://arxiv.org/abs/hep-ph/0412403}{{\ttfamily arXiv:hep-ph/0412403}}.

\bibitem{ArkaniHamed:2008qn}
N.~Arkani-Hamed, D.~P. Finkbeiner, T.~R. Slatyer, and N.~Weiner, ``{A Theory of
  Dark Matter},'' \href{http://dx.doi.org/10.1103/PhysRevD.79.015014}{{\em
  Phys. Rev. D} {\bfseries 79} (2009) 015014},
  \href{http://arxiv.org/abs/0810.0713}{{\ttfamily arXiv:0810.0713 [hep-ph]}}.

\bibitem{Lattanzi:2008qa}
M.~Lattanzi and J.~I. Silk, ``{Can the WIMP annihilation boost factor be
  boosted by the Sommerfeld enhancement?},''
  \href{http://dx.doi.org/10.1103/PhysRevD.79.083523}{{\em Phys. Rev. D}
  {\bfseries 79} (2009) 083523},
  \href{http://arxiv.org/abs/0812.0360}{{\ttfamily arXiv:0812.0360
  [astro-ph]}}.

\bibitem{Iengo:2009ni}
R.~Iengo, ``{Sommerfeld Enhancement: General Results from Field Theory
  Diagrams},'' \href{http://dx.doi.org/10.1088/1126-6708/2009/05/024}{{\em
  JHEP} {\bfseries 05} (2009) 024},
  \href{http://arxiv.org/abs/0902.0688}{{\ttfamily arXiv:0902.0688 [hep-ph]}}.

\bibitem{Iengo:2009xf}
R.~Iengo, ``{Sommerfeld enhancement for a Yukawa potential},''
  \href{http://arxiv.org/abs/0903.0317}{{\ttfamily arXiv:0903.0317 [hep-ph]}}.

\bibitem{Cassel:2009wt}
S.~Cassel, ``{Sommerfeld factor for arbitrary partial wave processes},''
  \href{http://dx.doi.org/10.1088/0954-3899/37/10/105009}{{\em J. Phys. G}
  {\bfseries 37} (2010) 105009},
  \href{http://arxiv.org/abs/0903.5307}{{\ttfamily arXiv:0903.5307 [hep-ph]}}.

\bibitem{Hannestad:2010zt}
S.~Hannestad and T.~Tram, ``{Sommerfeld Enhancement of DM Annihilation:
  Resonance Structure, Freeze-Out and CMB Spectral Bound},''
  \href{http://dx.doi.org/10.1088/1475-7516/2011/01/016}{{\em JCAP} {\bfseries
  01} (2011) 016}, \href{http://arxiv.org/abs/1008.1511}{{\ttfamily
  arXiv:1008.1511 [astro-ph.CO]}}.

\bibitem{Bellazzini:2013foa}
B.~Bellazzini, M.~Cliche, and P.~Tanedo, ``{Effective theory of
  self-interacting dark matter},''
  \href{http://dx.doi.org/10.1103/PhysRevD.88.083506}{{\em Phys. Rev. D}
  {\bfseries 88} no.~8, (2013) 083506},
  \href{http://arxiv.org/abs/1307.1129}{{\ttfamily arXiv:1307.1129 [hep-ph]}}.

\bibitem{Klebanov:1999tb}
I.~R. Klebanov and E.~Witten, ``{AdS / CFT correspondence and symmetry
  breaking},'' \href{http://dx.doi.org/10.1016/S0550-3213(99)00387-9}{{\em
  Nucl. Phys. B} {\bfseries 556} (1999) 89--114},
  \href{http://arxiv.org/abs/hep-th/9905104}{{\ttfamily arXiv:hep-th/9905104}}.

\bibitem{gross1999relativistic}
F.~Gross, {\em Relativistic Quantum Mechanics and Field Theory}.
\newblock A Wiley-Intersience publication. Wiley, 1999.

\bibitem{Petraki:2016cnz}
K.~Petraki, M.~Postma, and J.~de~Vries, ``{Radiative bound-state-formation
  cross-sections for dark matter interacting via a Yukawa potential},''
  \href{http://dx.doi.org/10.1007/JHEP04(2017)077}{{\em JHEP} {\bfseries 04}
  (2017) 077}, \href{http://arxiv.org/abs/1611.01394}{{\ttfamily
  arXiv:1611.01394 [hep-ph]}}.

\bibitem{Landau1976Mechanics}
L.~D. Landau and E.~M. Lifshitz, {\em Mechanics, Third Edition: Volume 1
  (Course of Theoretical Physics)}.
\newblock Butterworth-Heinemann, 3~ed., Jan., 1976.
\newblock \url{http://www.worldcat.org/isbn/0750628960}.

\bibitem{Buckley:2009in}
M.~R. Buckley and P.~J. Fox, ``{Dark Matter Self-Interactions and Light Force
  Carriers},'' \href{http://dx.doi.org/10.1103/PhysRevD.81.083522}{{\em Phys.
  Rev.} {\bfseries D81} (2010) 083522},
\href{http://arxiv.org/abs/0911.3898}{{\ttfamily arXiv:0911.3898 [hep-ph]}}.

\end{thebibliography}\endgroup

\end{document}